\documentclass[useAMS]{mn2e}

\usepackage{graphicx,color,soul}
\usepackage{latexsym}
\usepackage{multirow}
\usepackage{amsmath,amssymb}  

\title[Jets driven by relativistic radiation hydrodynamics]{Evolution of jets driven by relativistic radiation hydrodynamics as long and low-luminosity GRBs}

\author[F. J. Rivera-Paleo and F. S. Guzm\'an]{F. J. Rivera-Paleo  \thanks{E-mail:friverap@ifm.umich.mx (FJRP)} and F. S. Guzm\'an \thanks{E-mail:guzman@ifm.umich.mx (FSG)}   \\ 
Laboratorio de Inteligencia Artificial y Superc\'omputo,
Instituto de F\'{\i}sica y Matem\'{a}ticas, Universidad
              Michoacana de San Nicol\'as de Hidalgo. \\ Edificio C-3, Cd.
              Universitaria, 58040 Morelia, Michoac\'{a}n,
              M\'{e}xico.\\
			  }

\begin{document}

\date{\today}

\pagerange{\pageref{firstpage}--\pageref{lastpage}} \pubyear{201X}

\maketitle

\label{firstpage}

\begin{abstract}
We present numerical simulations of jets modelled with relativistic radiation hydrodynamics (RRH), which evolve across two environments: i) a stratified surrounding medium and ii) a 16TI progenitor model. We consider opacities consistent with 
various processes of interaction between the fluid and radiation, specifically  free-free, bound-free, bound-bound, and electron scattering. We explore various initial conditions, with different radiation energy densities of the beam in hydrodynamical and radiation-pressure-dominated scenarios, considering only highly relativistic jets.  In order to investigate the impact of the radiation field on the evolution of the jets, we compare our results with purely hydrodynamical jets. Comparing among jets driven by an RRH, we find that radiation-pressure-dominated jets propagate slightly faster than gas pressure dominated ones. Finally, we construct the luminosity light curves (LCs) associated with the two  cases. The construction of LCs uses the fluxes of the radiation field that is fully coupled to the hydrodynamics equations during the evolution. The main properties of the jets propagating on the stratified surrounding medium are that the LCs show the same order  of magnitude as the gamma-ray luminosity of typical Long gamma-ray Bursts $10^{50}-10^{54}$ erg/s, and the difference between the radiation and gas temperatures is of nearly one order of magnitude. The properties of jets breaking out from the progenitor star model are that the LCs are of the order  of magnitude of low-luminosity GRBs $10^{46}-10^{49}$ erg/s, and in this scenario, the difference between the gas and radiation temperature is of four orders of magnitude, which is a case far from thermal equilibrium.
\end{abstract}


%\pacs{95.30.Lz,4.25.D-,04.25.Dm,95.30.Sf, 98.80.Jk}
%95.30.Lz Hydrodynamics in astrophysical applications
%04.20.-q	% Classical general relativity
%4.25.D-   % Numerical relativity
%04.30.Nk % Wave propagation and interactions
%04.25.Dm, % numerical relativity
%04.25.dg %numerical relativistic studies of Black Holes
%Dark energy, 95.36.+x
%Dark matter, 95.35.+d
%galactic Accretion and accretion disks, 98.62.Mw
%classical black holes, 04.70.Bw
%astrophysics Gravitation , 95.30.Sf
%Relativistic astrophysics, 95.30.Sf, 98.80.Jk

\maketitle

% --------------------------------------------
% ----->     INTRODUCTION     <-----
% --------------------------------------------

\section{Introduction}
\label{sec:introduction}
There is observational evidence that long gamma-ray bursts (LGRBs) are produced after the death of massive stars \cite{Woosley,Galama,Stanek,Hjorth,WoosleyBloom}, whose spectrum agrees with those of Type Ic supernovae (SNe). Although the LGRBs have been identified spectroscopically with SNe, many of these events show smaller luminosity  than those of standard LGRBs. These events are called low-luminosity GRBs (LLGRBs). The emission mechanisms and the surrounding medium density profiles that make the difference between LGRBs and LLGRBs are still a matter of debate. As a consequence of this, several authors have developed numerical models for the jet propagation in a stratified surrounding medium applied to LGRBs and massive star models applied to LLGRBs. 

In this context, several numerical studies have been done. For instance, the evolution of jets within a surrounding medium using different approximations has been studied in  \cite{Aloy,DeColle,MacFadyen,Meliani_2007,Meliani_2010,Mizuta,MizutaAloy,Morsony,MizutaII,Nagakura,Lazzati,Lopez,ZhangI,ZhangII} using pure relativistic  hydrodynamics. More elaborate models   include the use of ideal magnetohydrodynamics \cite{Bromberg,Obergaulinger} or radiation hydrodynamics simulations \cite{Van_2010,Van_2011,Vlasis,Cuesta,DeColleII}. However, in these later works the radiation is not coupled with hydrodynamics during the evolution of the system, but the radiation is analysed by post-processing the hydrodynamical variables.

From a theoretical point of view, in the RRH simulations there are two key variables at modelling the system, these are: the rest-mass density of the fluid and the radiation energy density of the radiation field, which in turn has an impact on the optical depth of the fluid and the radiation pressure \cite{Mihalas}. In accordance with the later, the radiative transport does not significantly affect the fluid dynamics as long as the optical depth is in the optically thin regime or the fluid pressure dominates. In this case, the radiation attached to the system in a post-processing of the numerical simulations is a good approach because emitted photons can be assumed to escape with no further interactions. Otherwise, in the optically thick regime or in the radiation-pressure-domination scenario, the effect of radiative transport cannot be negligible because the photons carry significant momentum and energy that affect the dynamics of the fluid around. 

In this case, the radiation and fluid feedback each other's dynamics, and we need to solve the equations of radiative transport during the evolution at the same time as Euler's equations. In view of the above, and under the assumption that the radiation carried by jets goes from an optically thick region to an optically thin region, coupling the radiation field and the fluid during the evolution, is expected to be a better approximation than post-processing or pure hydrodynamics.
Therefore, in this paper, we assume a model of the jet, where the fluid is coupled to the radiation field, and the evolution is dictated by the RRH equations. In our numerical simulations, we assume, initially, that the radiation field and matter are in local thermal equilibrium (LTE) and evolve according to the RRH equations. However, after initial time, the system loses the LTE. We explore various initial conditions for the jet, with different radiation energy density and a fixed but high Lorentz factor. 

In order to describe GRB events in a more realistic astrophysical scenario,  we study the dynamics of the jet when it propagates in two different environments. In the first one, the jets propagate in a stratified external medium with  rest mass toy density profile $\rho\sim r^{-2}$, which is associated with LGBRs \cite{Mignone}. In the second one, the jet propagates within its progenitor star. For the later, we consider a pre-supernova 16TI model as the progenitor star that is considered to be a progenitor of LLGBRs. In each of the models, we construct the LCs associated with particular processes of interaction between the fluid and radiation, specifically, free-free, bound-free, bound-bound, and electron scattering opacities.

The paper is organized as follows. In section \ref{sec:RRH}, we describe the system of the RRH equations. In section \ref{sec:Opacities}, we describe the opacities used in our numerical simulations. In section \ref{sec:IntSetup}, we describe the initial conditions for the parameters of our simulations. In section \ref{sec:GRBJet}, we study the evolution of jets on a stratified medium and illustrate how the LCs can be associated to LGRBs. Also, we discuss the implications of the radiation field in the evolution of the system. In section \ref{sec:LLGRBs}, we present the evolution of jets starting from inside a progenitor and their relation to LLGRBs. Finally, in section \ref{sec:conclusions}, we discuss our results.  

% --------------------------------------------
% ---------->     SECTION     <----------
% --------------------------------------------

\section{Equations of evolution and numerical methods}
\label{sec:RRH}
As mentioned before, the model assumed for the jet corresponds to a fluid interacting with a radiation field. This implies the need to solve the equations coupling such a system in order to capture the back-reaction of one of the components on to the other. One important advantage of considering the radiation field is that the construction of LCs is natural because one is constantly calculating the variables  of the radiation field. The RRH equations governing the evolution of this system are form \cite{Farris,Fragile}: 

\begin{eqnarray}
\nabla_\alpha(\rho u^\alpha) &=&0,\\
\nabla_\alpha T^{\alpha\beta}_\text{m} &=&G^\beta_\text{r},\\
\nabla_\alpha T^{\alpha\beta}_\text{r} &=&-G^\beta_\text{r},
\label{eq:rad-hyd}
\end{eqnarray}

\noindent where $T^{\alpha\beta}_m$ is the stress-energy tensor of a perfect fluid 

\begin{equation}
T^{\alpha\beta}_m = \rho hu^\alpha u^\beta + Pg^{\alpha\beta},
\end{equation}

\noindent where $g^{\alpha\beta}$ is the metric of the space-time, $u^\alpha$ is the four-velocity of fluid elements, $\rho$, $h=1+\epsilon+P/\rho$, $\epsilon$ and $P$ are the rest-mass density, specific enthalpy, specific internal energy, and the thermal pressure, respectively. The thermal pressure is related to $\rho$ and $\epsilon$ through a gamma-law equation of state $P=\rho\epsilon(\Gamma-1)$, where $\Gamma$ is the adiabatic index of the fluid. Here, $T^{\alpha\beta}_\text{r}$ is the stress-energy tensor that describes the radiation field and is given by

\begin{equation}
T^{\alpha\beta}_\text{r} = (E_\text{r}+P_\text{r})u^\alpha u^\beta +F^\alpha_\text{r} u^\beta +u^\alpha F^\beta_\text{r} + P_\text{r}g^{\alpha\beta},
\end{equation}

\noindent where $E_\text{r}$, $F^\alpha_\text{r}$, and $P_\text{r}$ are the radiated energy density,  radiated flux, and  radiation pressure, respectively, measured in the comoving reference frame. The source term $G^{\alpha}_\text{r}$ is the radiation four-force that describes the interaction between the fluid and the radiation field. Among the various regimes of coupling between radiation and fluid, we choose the `grey-body' approximation, which technically means that the radiation field variables do not depend on its frequency \cite{Mihalas}. In this case the radiative four-force is given by \cite{Farris}:

\begin{equation}
G^\alpha_\text{r} = \chi^\text{t}(E_\text{r} - 4\pi B)u^\alpha + (\chi^\text{t} + \chi^\text{s})F^\alpha_\text{r},
\label{eq:source}
\end{equation}

\noindent with $\chi^\text{t}$ and $\chi^\text{s}$ the coefficients of thermal and scattering opacities, respectively. Finally $B=\frac{1}{4\pi}a_\text{r}T_\text{fluid}^4$, is the Planck function, $T_\text{fluid}$ the temperature of the fluid and $a_\text{r}$ the radiation constant. 

The above set of equations is completed with a closure relation that identifies the second moment of radiation with one of the lower order moments. The simplest approach is the Eddington approximation, which assumes a nearly isotropic radiation field and in the fluid frame shows a pressure tensor with the form $P^{ij}_\text{r} = \frac{1}{3}E_\text{r}\delta^{ij}$ \cite{Mihalas}. This assumption is valid only in the optically thick regime within the diffusion limit. The radiation field in the optically thin regime requires a more general assumption. A scheme that allows a description of the radiation field in both optically thick and thin regimes is the M1 \cite{Levermore,Dubroca,Gonzalez}. The M1 closure provides a better approximation than Eddington's to the radiation field because it describes the diffusion limit as well as the free-streaming limit, where the radiative energy is transported at the speed of light. Explicitly, this closure relation is given by

\begin{equation}
P^{ij}_\text{r} = \left(\frac{1-\zeta}{2}\delta^{ij} + \frac{3\zeta-1}{2}\frac{f^if^j}{|f|^2}\right) E_\text{r},
\end{equation}  

\noindent where $f^i=F_\text{r}^i/cE_\text{r}$ is the reduced radiative flux and $\zeta=\frac{3+4f^if_i}{5+2\sqrt{4-3f^if_i}}$ is the Eddington factor \cite{Levermore}. This closure relation recovers the two regimes of radiative transfer. In the optically thick regime $F_\text{r}^i\approx 0$, $f^i=0$, and $\zeta=1/3$ that correspond to Eddington's approximation. On the other hand, in the optically thin regime $F_\text{r}^i= cE_\text{r}$, $f^i=1$, and $\zeta=1$ that correspond to the free-streaming limit.

The fluid temperature is estimated taking into account contributions of both, baryons and radiation pressure. An approximate expression for the total pressure is written as \cite{Cuesta}

\begin{equation}
P_\text{t} = \frac{k_B}{\mu m_\text{p}}\rho T_\text{fluid} + (1 - e^{-\tau})\zeta(T) a_\text{r}T_\text{rad}^4,
\label{eq:tem}
\end{equation}

\noindent where $k_B$ is the Boltzmann constant, $\mu$ the mean molecular weight, and $m_\text{p}$ the mass of the proton, $\tau=\int(\rho\chi^\text{t}+\rho\chi^\text{s})d\text{s}$ is the total optical depth. Finally, $T_\text{rad} = (E_\text{r}/a_\text{r} )^{1/4}$, is the temperature of radiation. Here, $\tau$ depends on the temperature only, if any of the opacity coefficients does. The path to integrate $\tau$ in our simulations is straight lines parallel to the $z$-axis. In general, the Eddington factor depends on the temperature $\zeta=\zeta(T_\text{fluid})$. When the fluid and radiation are in LTE, that is  $T_\text{fluid}=T_\text{rad}$, the temperature approximately obeys a fourth degree equation similar to (\ref{eq:tem}) (see \cite{Cuesta}).

We programmed a code that solves the 3D RRH equations above, together with the M1-closure relation, using the following numerical methods. First, the RRH equations are written in flux balance form $\partial_t\textbf{U}+\partial_{i}\textbf{F}^{i}=\textbf{S}$, where  $\textbf{U}$ is the vector of conserved variables, $\textbf{F}^i$ are the fluxes, and  $\textbf{S}$ the sources \cite{ZanottiI,Roeding}. Based on this structure of the equations, we apply high-resolution shock capturing methods that use a finite volume discretization, the HLLE flux formula, and the minmod-slope limiter. For the evolution, we use the Method of Lines, with an explicit-implicit Runge-Kutta (IMEX RK) time integrator with second-order accuracy, as done in \cite{Panchos}. In order to benefit from efficient parallelization and standard I/O, we mounted our code on the Cactus frame \cite{Cactus}, using the Carpet driver \cite{Carpet}, and in all the cases, we use an unigrid discretization. In the appendices, we present canonical tests showing that our implementation works properly.

% --------------------------------------------
% ---------->     SECTION     <----------
% --------------------------------------------

\section{Opacities}
\label{sec:Opacities}

An essential ingredient in our analysis is the use of appropriate opacities, because their values determine the radiative processes  associated with the GRB emission. When temperature is of the order of $\sim 10^9 K$, the energy of photons becomes an appreciable fraction of the electron rest mass, photons may be scattered only on some electrons, and the electron-scattering opacity is given by \cite{Buchler}:

\begin{equation}
\chi^\text{s} = 0.2(1 + X)\left[1 + 2.7\times 10^{11}\frac{\rho}{T^2}\right]^{-1}\left[1 + \left(\frac{T}{4.5\times 10^8}\right)^{0.86}\right]^{-1}.
\label{eq:ScOp}
\end{equation}

\noindent Moreover, at these high temperatures $(\sim 10^9 K)$ and low densities, a primary source of opacity comes from the creation of pairs. On the other hand, at intermediate temperatures  $(<10^6 K)$ and low densities $(\sim 1 \text{gr}/\text{cm}^3)$,  bound-free opacity may be dominant. Finally, at sufficiently low temperatures and densities, bound-bound absorption in the ultraviolet (UV) and far UV dominate the opacity. This effect is relevant in the low-density regions, both in the stratified model and the progenitor model used later, specially in the wind-like structure surrounding the progenitor star. The opacity due to free-free, bound-free and bound-bound emission can be roughly approximated with the so-called Kramers formula \cite{George,Hayashi,Schwarzschild}:

\begin{eqnarray}
\kappa^\text{ff} &\simeq& 3.8\times10^{22}(1+X)(X+Y+Z)\rho T^{-7/2} \text{cm}^2/\text{gr}, \\
\kappa^\text{bf} &\simeq& 4.3\times10^{25}Z(1+X)\rho T^{-7/2} \text{cm}^2/\text{gr}, \\
\kappa^\text{bb} &\simeq& 10^{25}Z\rho T^{-7/2} \text{cm}^2/\text{gr}.
\label{eq:opacities}
\end{eqnarray}

\noindent The total coefficients of thermal opacity may be approximated over a wide range of temperatures, $10^4 \leq T \leq 10^9 K$ with the sum of free-free, bound-free, and bound-bound coefficients $(\chi^\text{t} = \kappa^\text{ff} +\kappa^\text{bf} +\kappa^\text{bb})$. In all opacities, $X$, $Y$, and $Z$ represent the mass fractions of hydrogen, helium, and elements heavier than helium, respectively. 
 A more accurate treatment of  opacities consists in using the mean thermal Gaunt factor in the free-free opacity, and the mean Gaunt and guillotine factors in the bound-free opacity \cite{Sutherland,Van_2015,Hayashi,Schwarzschild}, or directly calculating the Rosseland mean opacities by using the metallicity and the microphysics involved in the processes \cite{opal}.

% --------------------------------------------
% ---------->     SECTION     <----------
% --------------------------------------------

\section{Simulations setup}
\label{sec:IntSetup}

The GRB jet model that we study here, is produced by the injection of a relativistic beam evolving through a fluid at rest, starting from a nozzle with radius $r_\text{b}$ and velocity $v_\text{b}$. The process is characterized by the ratio between the density of the beam (subindex b) and that of the medium (subindex m)  $\eta=\rho_\text{b}/\rho_\text{m}$, as well as by the ratio between their pressures $K = P_\text{b}/P_\text{m} $. The relativistic Mach number of the beam is $M_\text{b}={\cal M}_\text{b}W_\text{b}\sqrt{1-c_\text{s}^2}$, where ${\cal M}_\text{b}$, $W_\text{b}$, and $c_\text{s}^2$ are the Newtonian Mach number, Lorentz factor, and speed of sound, respectively. Outflow boundary conditions are used at the boundaries, except inside the nozzle radius, where the values of the variables are kept constant in time during the time window in which we inject the beam.
 
In order to study the jet interaction with the surrounding medium,  we assume the radiation field of the beam starts in an optically thick region and propagates towards an optically thin region \cite{Pe'er,Giannios}. This fact allows one to assume LTE between the fluid and the radiation field initially. This assumption is satisfied by the external medium density $\rho_\text{m}$ and pressure profiles $P_\text{m}$ that decrease with distance. We use density and pressure profiles described by the following power law \cite{Mignone,DeColle}:

\begin{equation}
\rho_\text{m}=\rho_0\left(\frac{r_\text{b}}{r}\right)^{2}, \ \ P_\text{m}= P_0\left(\frac{r_\text{b}}{r}\right)^{2},\label{eq:stratified}
\end{equation}

where the parameters of the surrounding medium are $\rho_0=10^5 \text{g} \text{cm}^{-3}$ and $P_0 =  10^{22} \text{g} \text{cm}^{-1} \text{s}^{-2}$. This is a simplified model for the propagation of a relativistic jet through a collapsing, non-rotating massive star with $0.1$ times solar metallicity \cite{Mignone}.

Other interesting properties of GRBs are the luminosity and total injected energy $L_\text{j}$ and $E_\text{j}$, respectively. The injected jet luminosity is given by the flux of the momentum density equation times the surface of injection, $A_\text{b}$, which in an optically thick regime is

\begin{equation}
L_\text{j} \simeq  \left[ \left(\rho_\text{b}  h_\text{b} + 4/3  E_\text{r,b} \right)  W_\text{b}^2  v_{z,\text{b}} + F_\text{r}^z  W_\text{b}  \left(v_{z,\text{b}} +1\right) \right] A_\text{b},
\end{equation}

\noindent 
where $h_\text{b}$, $E_\text{r,b}$, $W_\text{b}$, and $v_{z,\text{b}}$ are the specific enthalpy, radiated energy density, Lorentz factor, and velocity of the beam. Finally, the total injected energy is approximately

\begin{equation}
E_\text{j} \simeq  L_\text{j}  t_\text{inj},
\end{equation}

\noindent where $t_\text{inj}$ is the injection time.

We explore various initial conditions for a highly relativistic jet, with different radiation energy densities, in both gas and radiation-pressure-dominated scenarios. In order to determine whether a jet is radiation or matter dominated initially, we  measure the ratio between the effective inertia of the radiation field $(4/3 E_\text{r,b}W^2_\text{b}\text{b})$ and the effective inertia in the purely hydrodynamical case $(\rho_\text{b} h_\text{b} W^2_\text{b})$, as well as the ratio between radiation  and gas pressures. These are, respectively,

\begin{eqnarray}
g_\text{1,b} =  \frac{4/3 E_\text{r,b}}{\rho_\text{b} h_\text{b}}, \\
g_\text{2,b} =  \frac{1/3 E_\text{r,b}}{P_\text{b}}.
\end{eqnarray}

A very important part of our simulations' diagnostics is the LC curve that we calculate  by directly integrating the radiation fluxes in the laboratory frame $\vec{F^\prime}_\text{r}$, on a given surface

\begin{equation}
L=\int \vec{F^\prime}_\text{r}\cdot\hat{n} dA.
\label{eq:luminosity}
\end{equation}

\noindent We use the second-order trapezoidal rule to calculate the integral.
The relation between laboratory  and comoving frames of the radiation moments is given by a Lorentz transformation \cite{Mihalas,Park}:

\begin{eqnarray}
\nonumber
E^\prime_\text{r}       &=&  W^2\left( E_\text{r} + 2v_iF^i_\text{r} +v_iv_jP^{ij}_\text{r}\right), \\
\nonumber 
F^{\prime i}_\text{r} &=&  W^2v^iE_\text{r} + \left[\delta^i_j +\left(\frac{W-1}{v^2} + W\right)v^iv_j\right]F^j_\text{r} \\
\nonumber
                   & & +Wv_j\left(\delta^i_k + \frac{W-1}{v^2}v^iv_k\right)P^{jk}_\text{r},\\
\nonumber
P^{\prime ij}_\text{r}  &=&  W^2v^iv^jE_\text{r} + W\left(v^i\delta^j_k + v^j\delta^i_k -2\frac{W-1}{v^2}v^iv^jv_k\right)F^k_\text{r} \\
\nonumber
                   & & +\left(\delta^i_k +\frac{W-1}{v^2}v^iv_k\right)\left(\delta^j_k +\frac{W-1}{v^2}v^jv_l\right)P^{kl}_\text{r},
\nonumber
\end{eqnarray}

\noindent where the primed variables are those measured by an observer in the laboratory frame. As a detecting surface in the laboratory frame, we choose $A$ to be a plane included in the numerical domain. The surface is located in a region, where the optically thin regime holds. We calculate this luminosity in two different planes, the first one perpendicular to $\hat{z}$, whereas the second one is a plane perpendicular to $(\hat{x}+\hat{z})/\sqrt{2}$. 

In order to evaluate the radiation LC seen by a distant observer, we need to compute quantities in an observer frame that take into account the cosmological effects induced by the redshift at which the source is located. We consider a distant observer whose line of sight makes an angle $\theta$ with respect to the jet axis. We define the time at which the observer sees the radiation coming from a fluid element located at a distance $z_i$ at time $t$ (both measured in a laboratory frame) as $t_\text{det}=t-z_i\cos\theta / c$. Assuming that the emitting source is located at redshift $z$, the time measured in the observer's frame is $t_\text{obs}= t_\text{det}(1 + z)$ \cite{Cuesta,Chucho}.
On the other hand, the total luminosity in the observer's frame is given by $L_\text{obs}(t_\text{obs}) = (1 + z)D^2L$, where $D = [W(1-v_z\cos\theta/c)]^{-1}$ is the Doppler factor \cite{dermer}. In our calculations we assume a generic $z=1$ and $\theta = 0$ for the perpendicular detector $\hat{n}=\hat{z}$ and $\theta=\pi/4$ for the inclined detector $\hat{n}=(\hat{x}+\hat{z})/\sqrt{2}$.

% --------------------------------------------
% ---------->     SECTION     <----------
% --------------------------------------------

\section{Different jet models}
\label{sec:GRBJet}

The  aim of this paper is to study the evolution of jets and their LCs across interesting surrounding media. The first case assumes the medium density has the toy density profile (\ref{eq:stratified}), and we  analyse various scenarios. The parameters of the cases studied are summarized in Table \ref{t1}. In these models, we choose the density and pressure ratios to be initially $\eta=0.01$ and $K = 0.01$,  a nozzle radius of $r_\text{b}= 8\times10^8 \text{cm}$ and a beam Lorentz factor $W_\text{b} = 10$. We inject the jet during a finite time $t_\text{inj}=12 \text{s}$, which is a lapse consistent with the amount of total energy of a generic GRB.

In all these simulations, the jet propagates along the $z$-direction and the numerical parameters have been standardized such that we use a resolution of $\Delta x=\Delta y=\Delta z=1\times10^{8}$ cm, in a numerical domain of $[-1,1]\times[-1,1]\times[0,5]\times 10^{10}\text{cm}$. This resolution is enough to contain eight zones per beam radius, which is a recommended resolution to properly resolve the internal structure and the jet/external medium interaction \cite{Aloy}. 

Model 1 corresponds to a purely hydrodynamical jet and will serve as reference to learn how much the radiation field affects the dynamics of the evolution. This is important, because in the ideal case one would like to avoid the use of post-processing to include the radiation effects that do not back react into the hydrodynamics, and these hydrodynamical jets show how much of the dynamics one loses when the radiation is not considered during the evolution.

\begin{figure}%[htb]
\includegraphics[width=8cm]{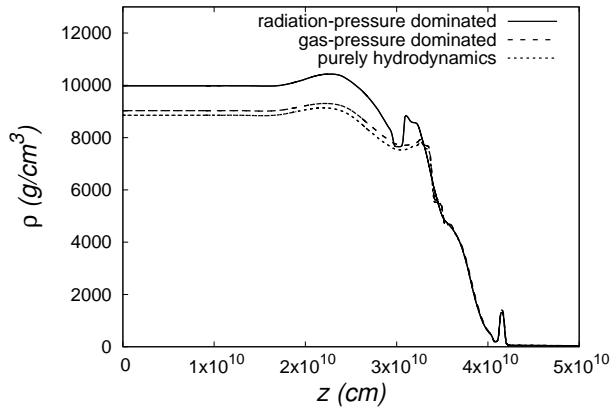}
\caption{Rest-mass density  along $z$-axis of  models 1, 2, and 3 in Table \ref{t1} at $t=11 \text{s}$. We remind that model 1 is purely hydrodynamical, model 2 is a case where hydrodynamical pressure dominates, and model 3 is radiation pressure dominated. In all the cases, the Lorentz factor is $W_\text{b}=10$.} \label{fig:ultra}
\end{figure}

Even though the jets propagate through a stratified surrounding medium, they are collimated along their entire length and present a morphology similar to that of jets propagating on a constant surrounding medium \cite{Marti,MartiII,AloyII,Hughes}. The properties are pretty much same among models 1, 2, and 3, namely the beam, bow shock, contact discontinuity, cocoon, back-flow, and internal shocks. However, quantitatively there are  differences among these models. In Fig. \ref{fig:ultra}, we show the rest-mass density profile along the $z$-axis of the purely hydrodynamical jet (model 1), the gas-pressure-dominated jet (model 2), and the radiation-pressure-dominated jet (model 3) at time $t=11$ s in the laboratory frame. From Fig. \ref{fig:ultra}, we can distinguish a shocked region, which is in front of the jet's head, the profile is nearly steady, and the radiation field does not contribute significantly  because in the three cases the profile is similar. Behind the jet's head, the radiation field plays a more important role because there the rest-mass density is higher for the radiation dominated case by a 10\%.

In the following subsections, we will analyse in more detail the implications of the interaction of the radiation field with the fluid. We explore two cases in which the radiation photon field is coupled with the fluid. 

\begin{table}%[!hbt]
\begin{center}
\begin{tabular}{c  c  c  c  c  c }
\hline
\hline
Model & $L_\text{j}(\text{erg}/\text{s})$ & $E_\text{j}(\text{erg})$   & $E_\text{r,b}\left(\frac{\text{erg}}{\text{cm}^3}\right)$ & $g_\text{1,b}$ $(g_\text{2,b})$  \\
\hline
 1    &$2.4\times 10^{51}$        &$2.88\times 10^{52}$  &$0$                                                   &   $0$ $(0)$    \\
\hline
 2    &$3.18\times 10^{51}$       &$3.81\times 10^{52}$  &$1\times10^{20}$                                      &   $0.33$  $(0.33)$  \\
\hline
 3    &$8.22\times 10^{52}$       &$9.8\times 10^{53}$   &$1\times10^{22}$                                      &   $33.3$   $(33.3)$\\
\hline
\hline
\end{tabular}
\caption{\label{t1} Parameters of the jet models. In all cases, we use the opacities that emulate the free-free, bound-free, bound-bound, and electron-scattering precesses, adiabatic index $\Gamma=4/3$, and Lorentz factor $W_\text{b} = 10$.}
\end{center}
\end{table}

\begin{figure}%[htb]
\includegraphics[width=7cm]{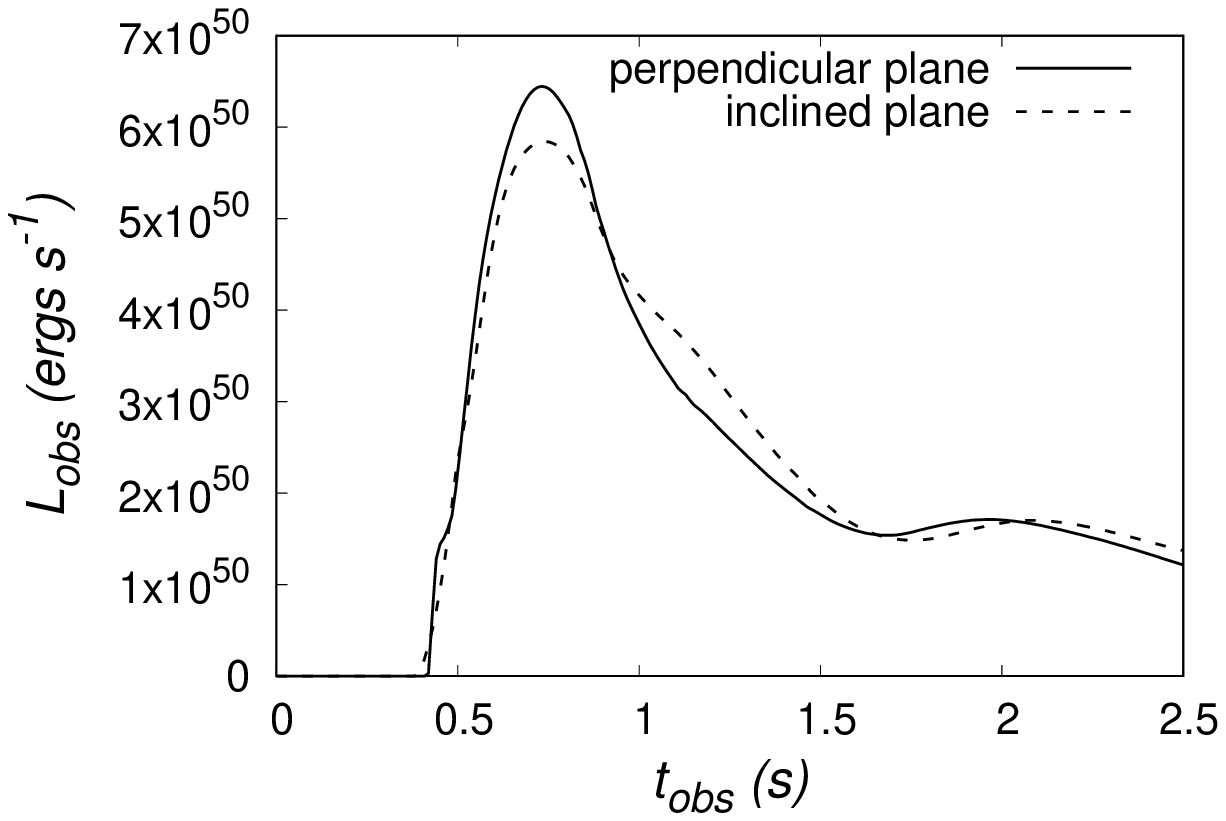}
\includegraphics[width=7cm]{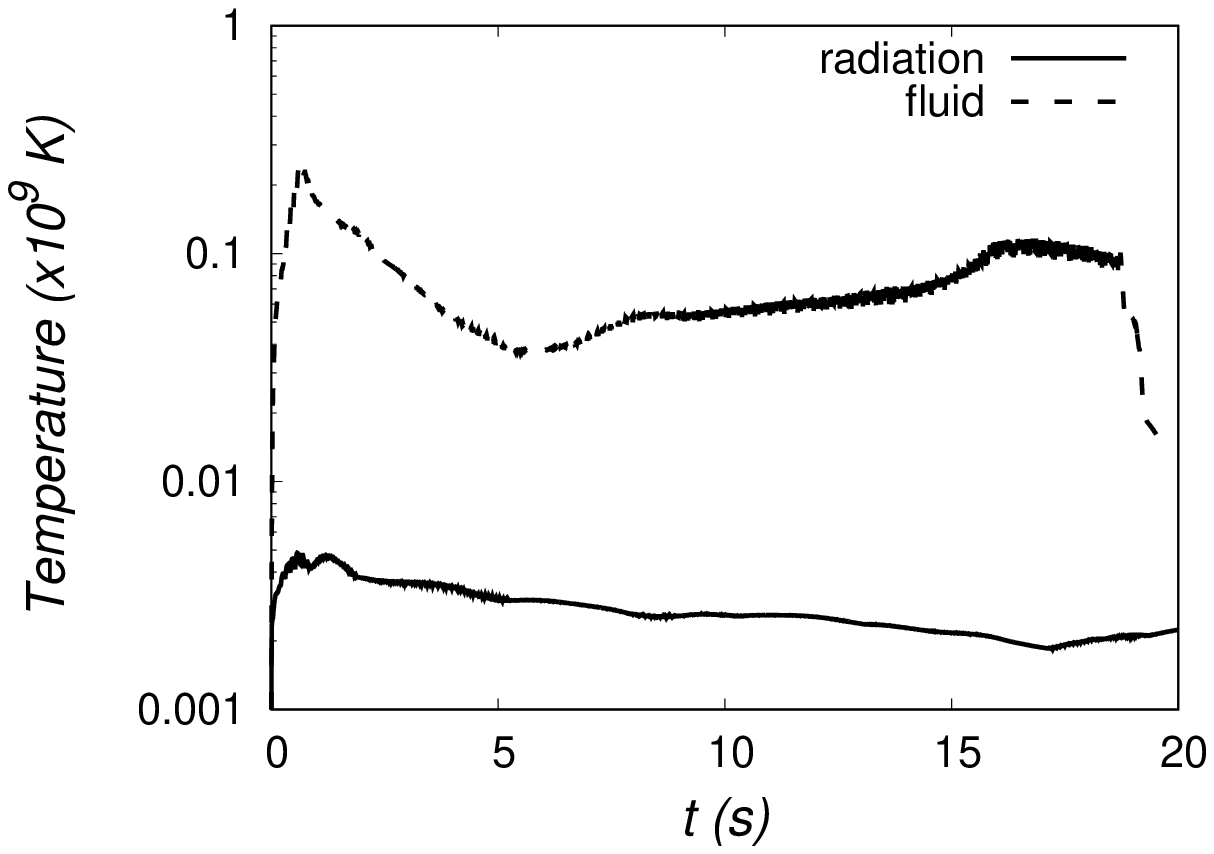}
\caption{Model 2. Top: The LCs  measured in two different planes transformed to the observer's frame. The solid and dotted line are obtained measuring the flux through  the perpendicular and inclined planes, respectively. Bottom: Maximum of  radiation and fluid temperatures, which is located behind the working surface of the jet. The solid and dotted lines correspond to the radiation and fluid temperatures, respectively, measured in the laboratory frame.} \label{fig:ultra-gaspress:a}
\end{figure}

\begin{figure}%[htb]
\includegraphics[width=0.4\textwidth]{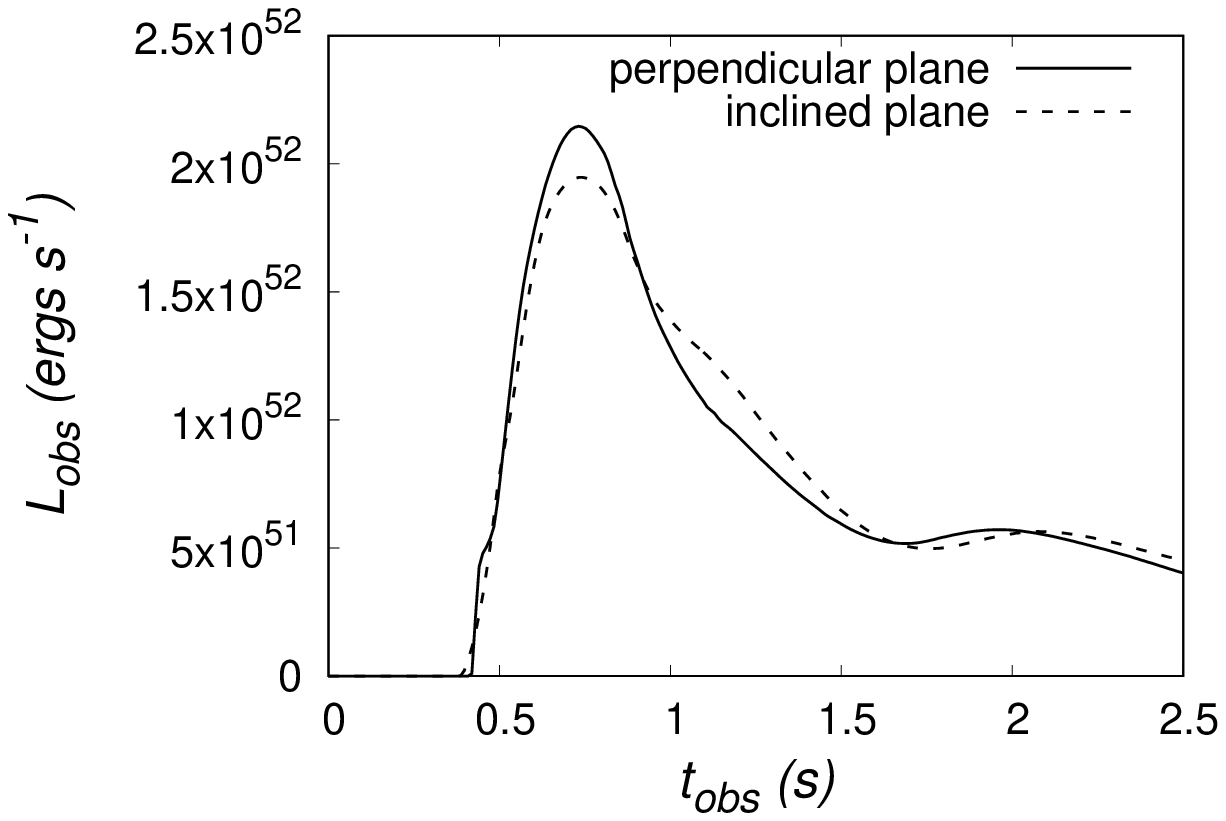}
\includegraphics[width=0.4\textwidth]{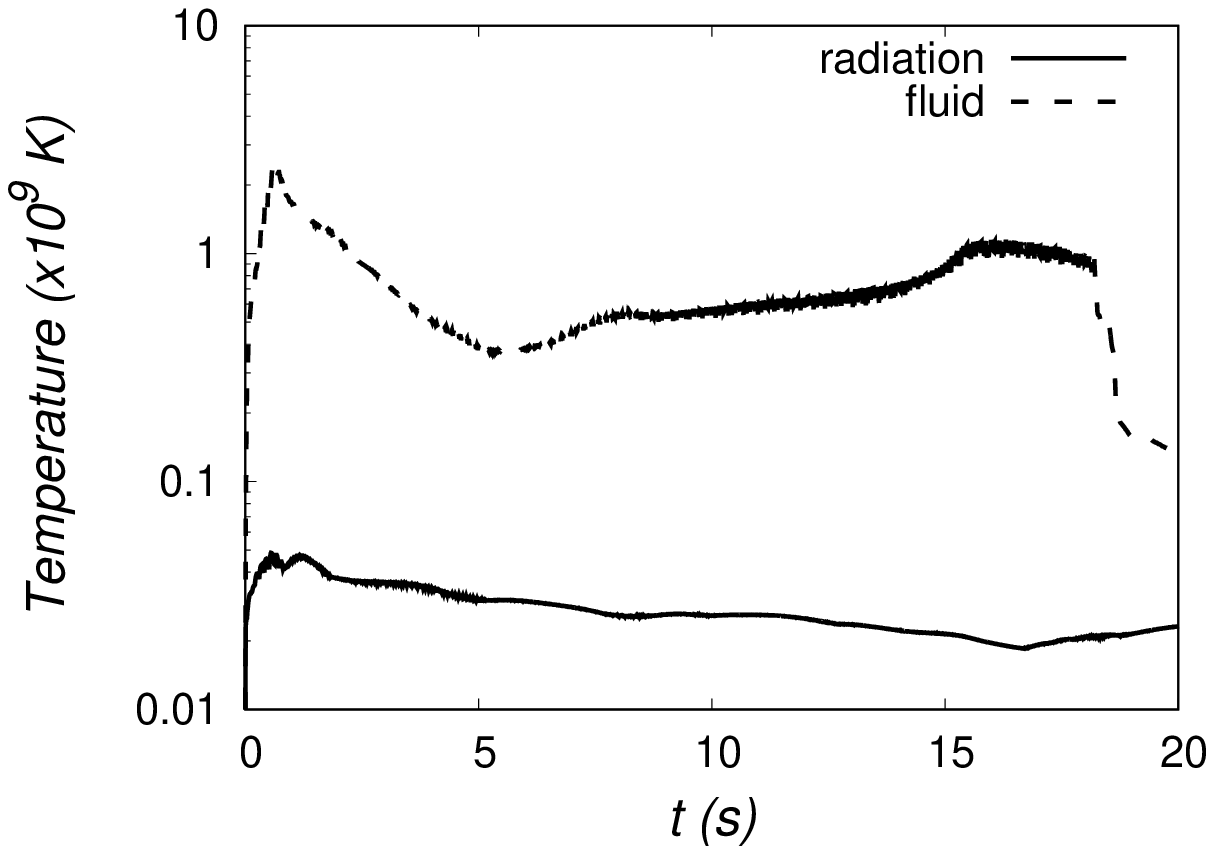}
\caption{Model 3. Top: The LCs measured in two different planes transformed to the observer's frame.  The solid and dotted line corresponds to the perpendicular and inclined planes. Bottom: Maximum of both radiation and fluid temperatures, which is located behind the working surface of the jet. The solid and dotted line corresponds to the radiation and fluid temperature, respectively,  measured in the laboratory frame.} \label{fig:ultra-radpress:a}
\end{figure}

\begin{figure}%[htb]
\includegraphics[width=0.4\textwidth]{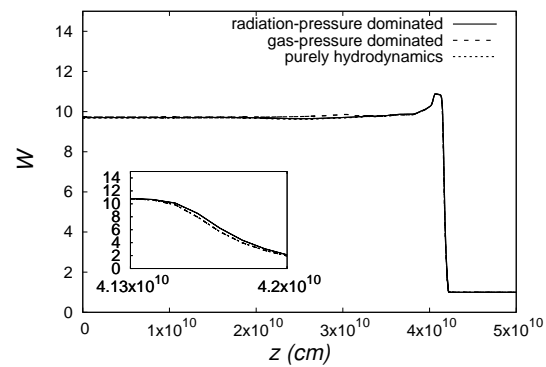}
\caption{Lorentz factor for jet models 1, 2 and 3 at $t=11 \ {\rm s}$. The front shock moves faster in the radiation-pressure-dominated case. The inset shows a zoom in at the front shock region.} \label{fig:ultra-LF}
\end{figure}

\begin{figure*}%[htb]
\includegraphics[width=0.235\textwidth,height=0.35\textheight]{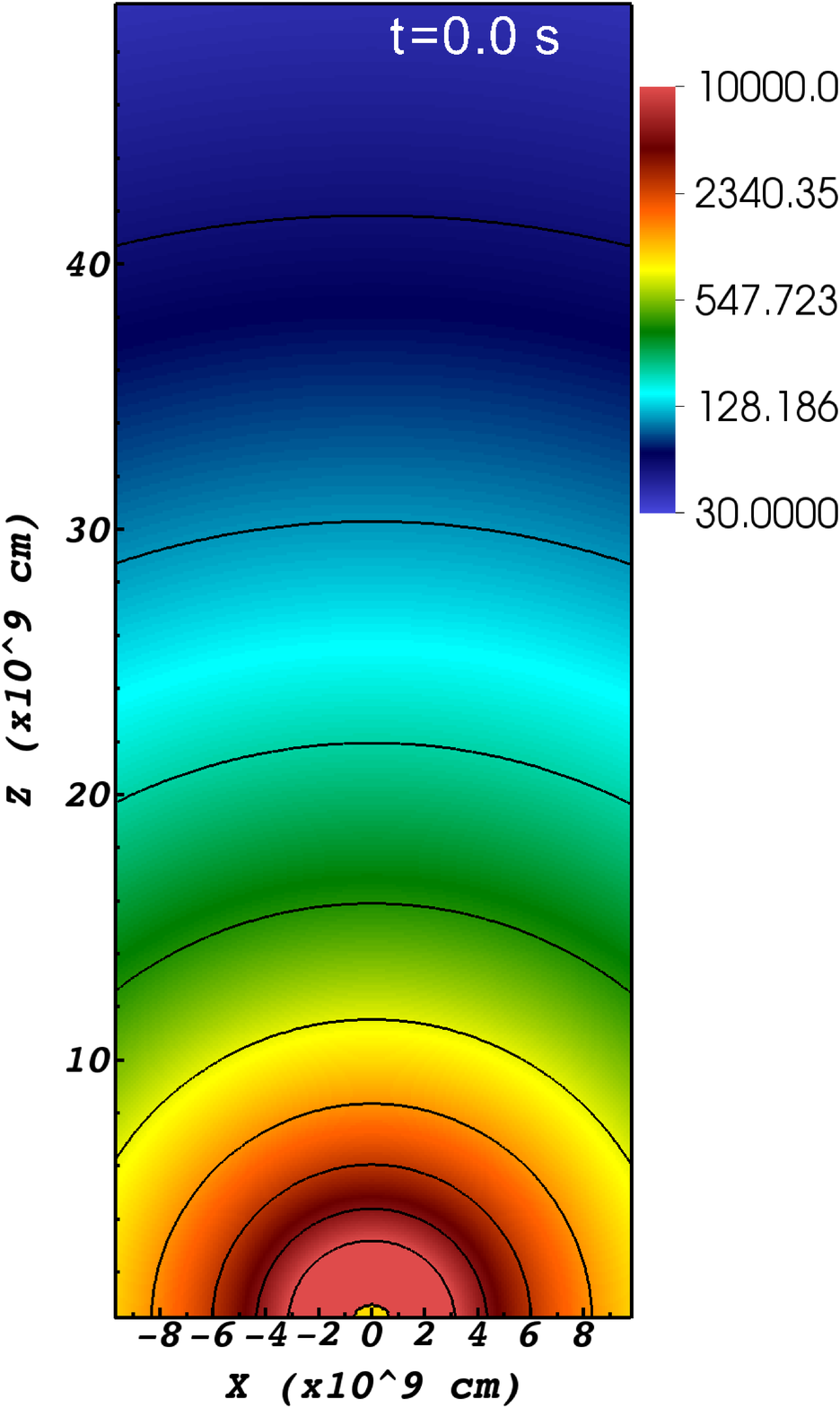}
\includegraphics[width=0.235\textwidth,height=0.35\textheight]{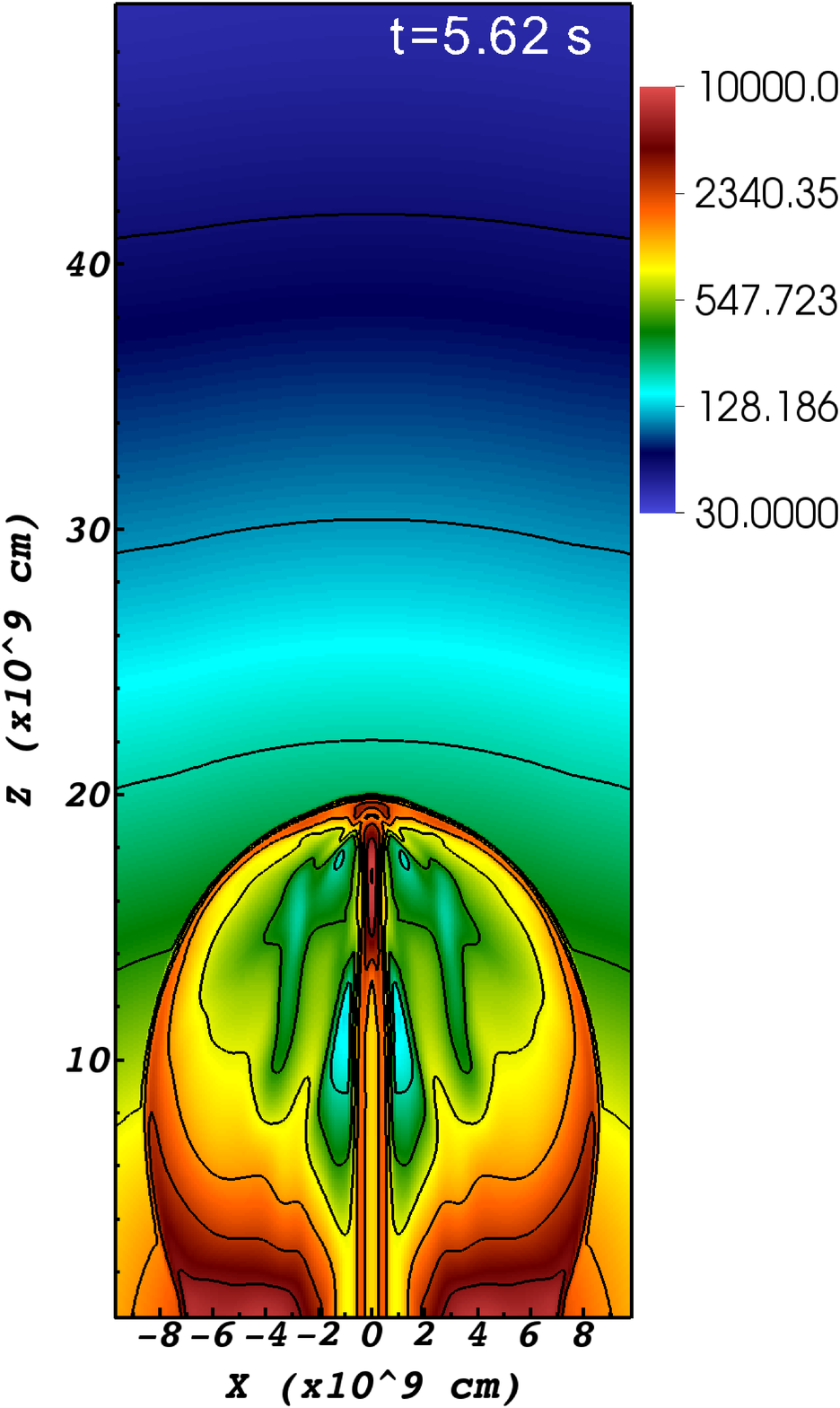}
\includegraphics[width=0.235\textwidth,height=0.35\textheight]{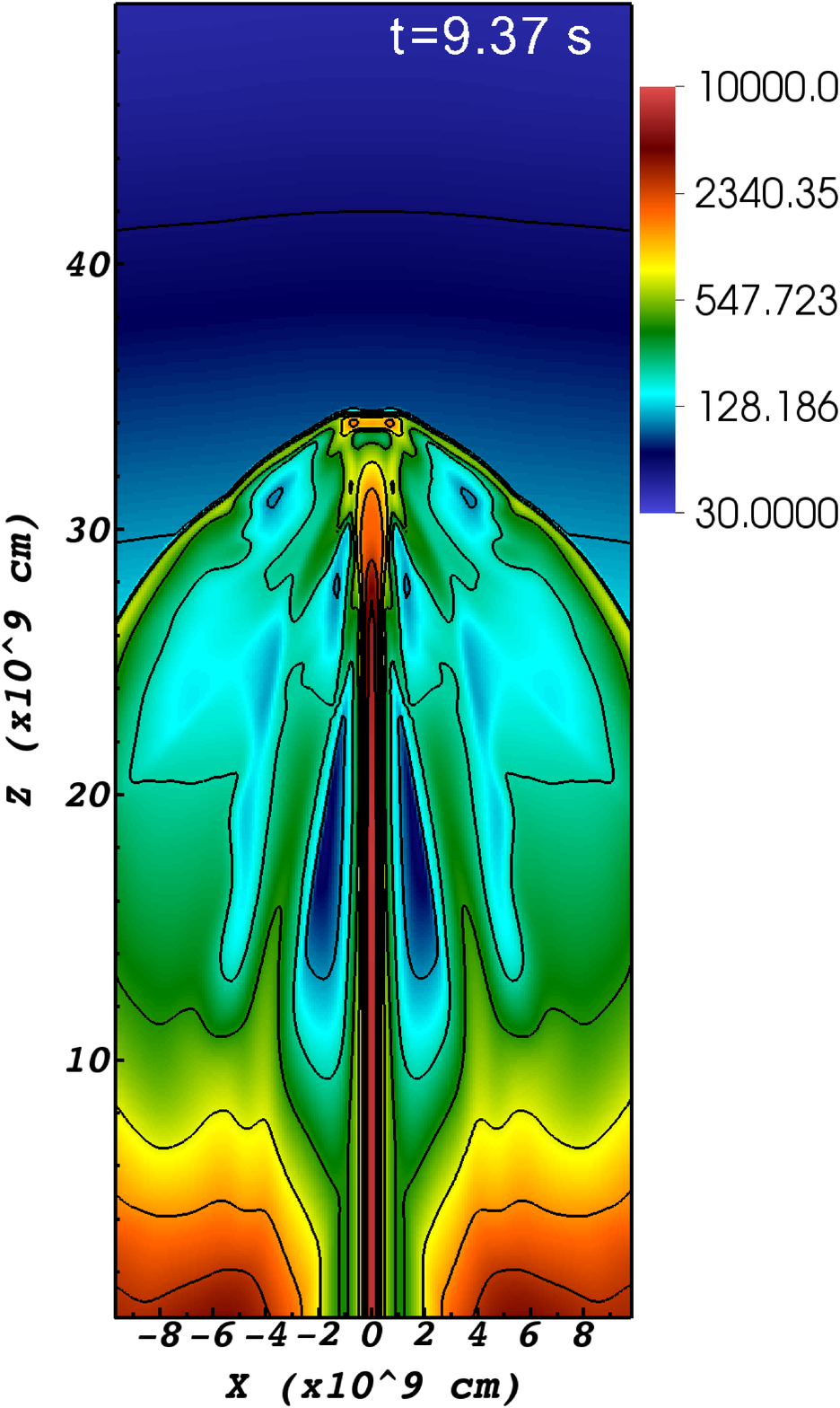}
\includegraphics[width=0.235\textwidth,height=0.35\textheight]{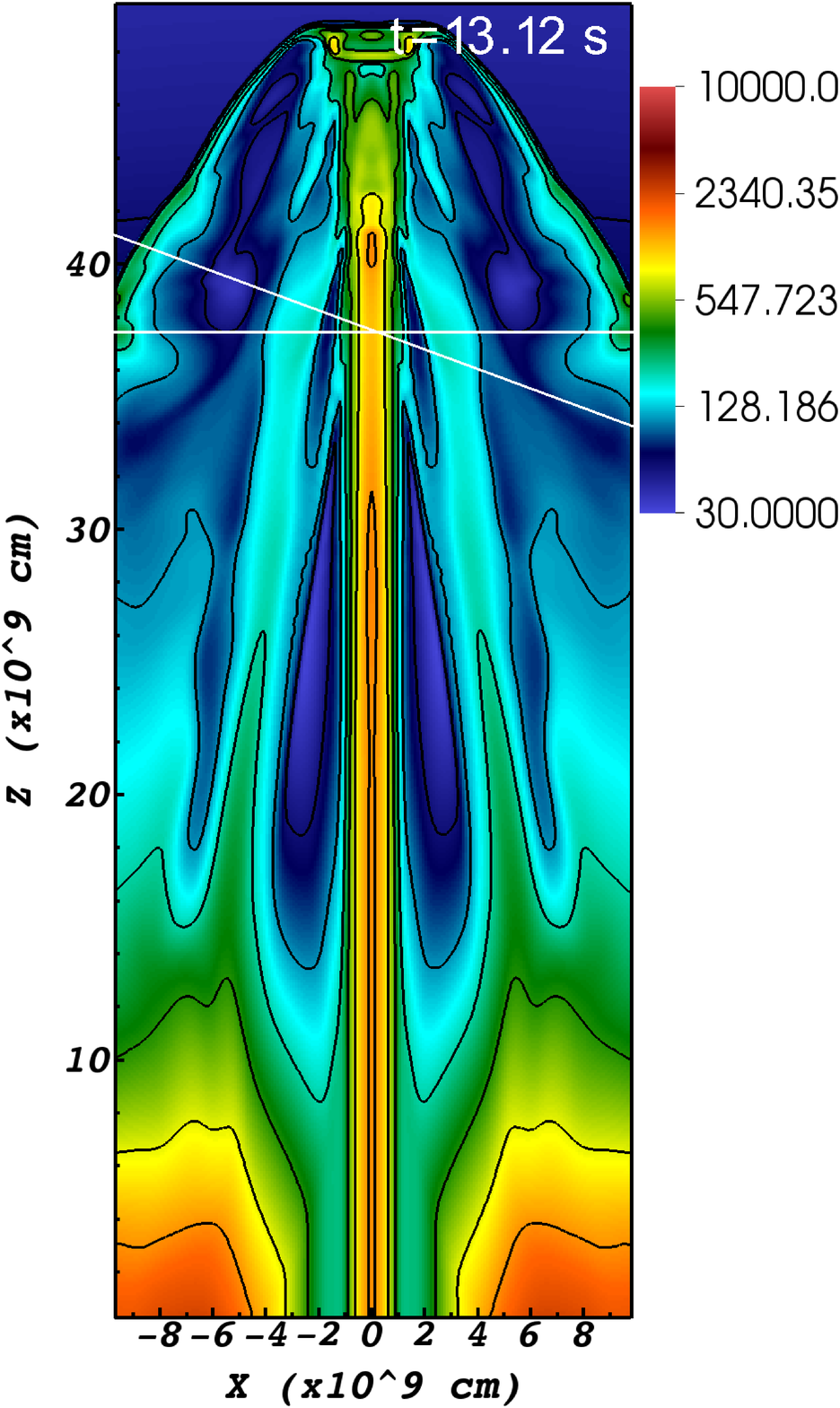}
\includegraphics[width=0.235\textwidth,height=0.35\textheight]{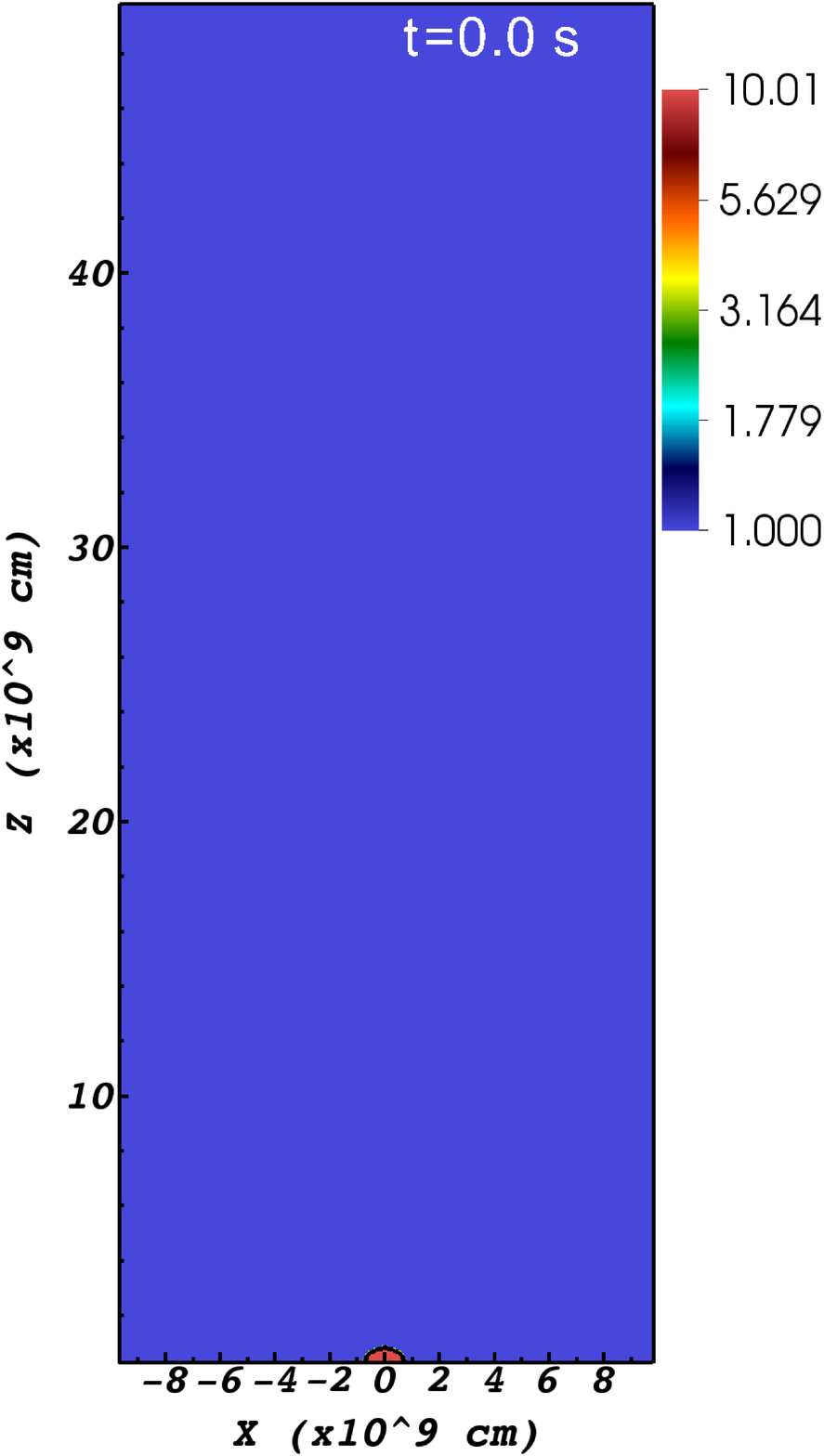}
\includegraphics[width=0.235\textwidth,height=0.35\textheight]{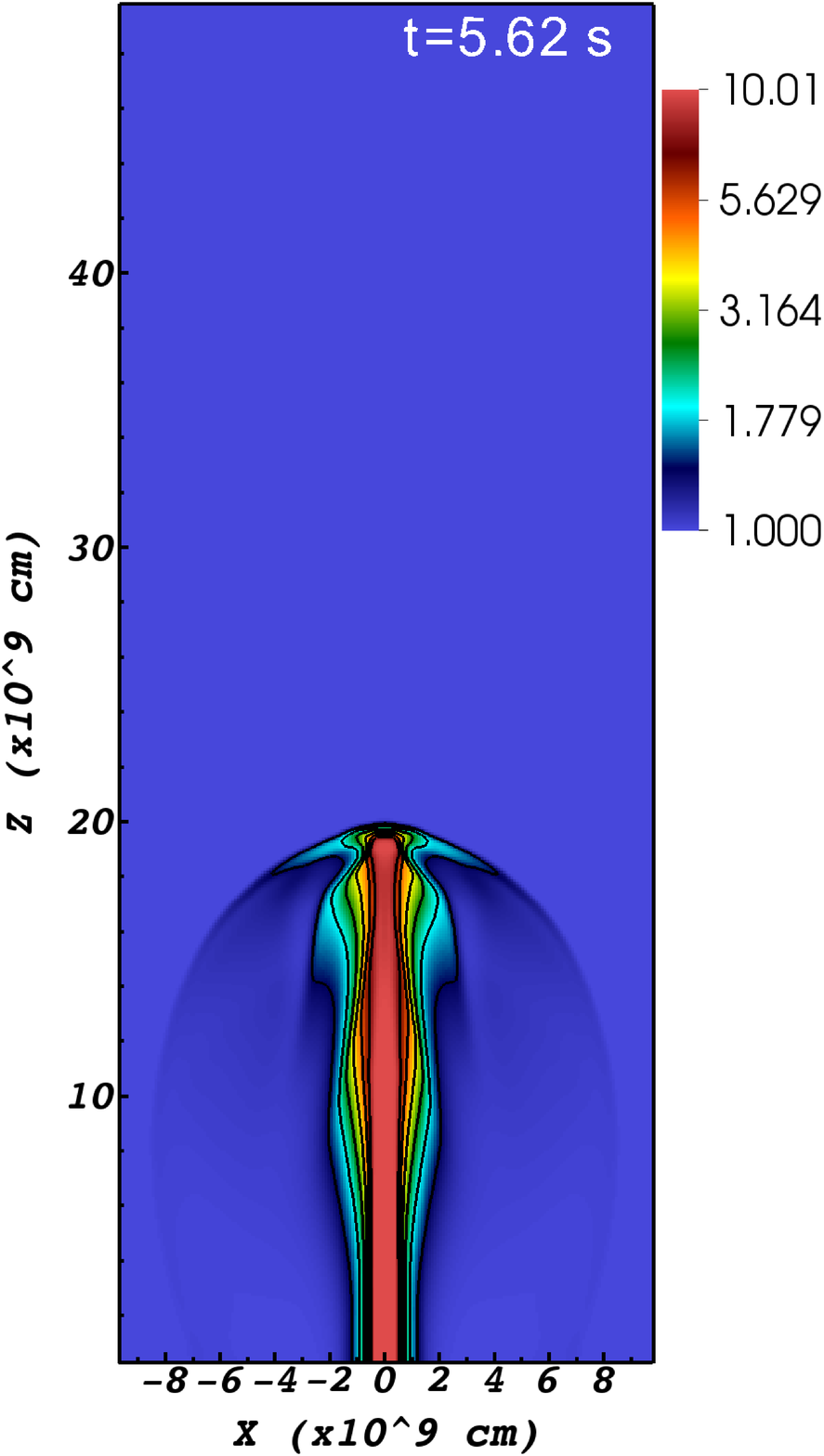}
\includegraphics[width=0.235\textwidth,height=0.35\textheight]{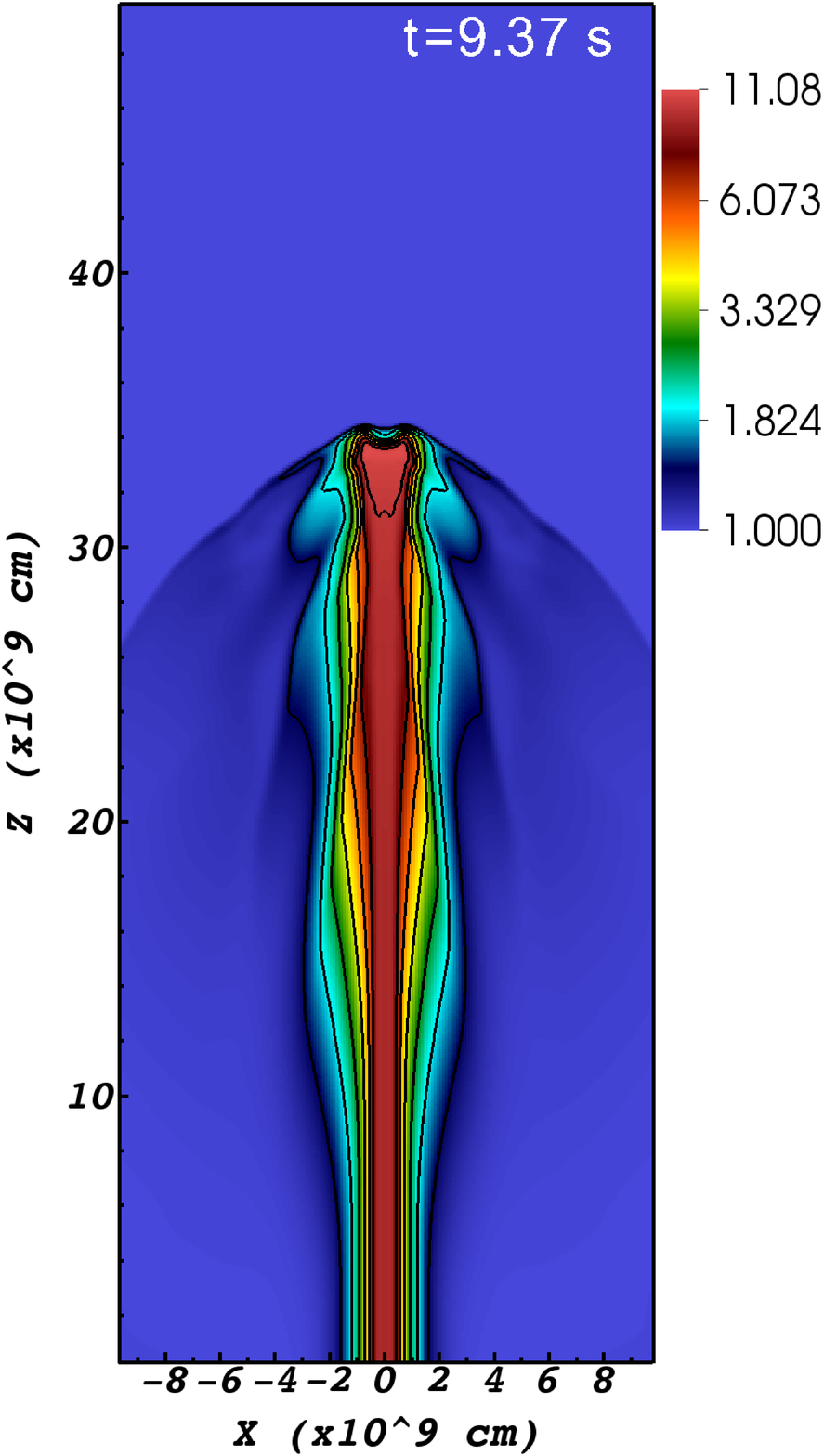}
\includegraphics[width=0.235\textwidth,height=0.35\textheight]{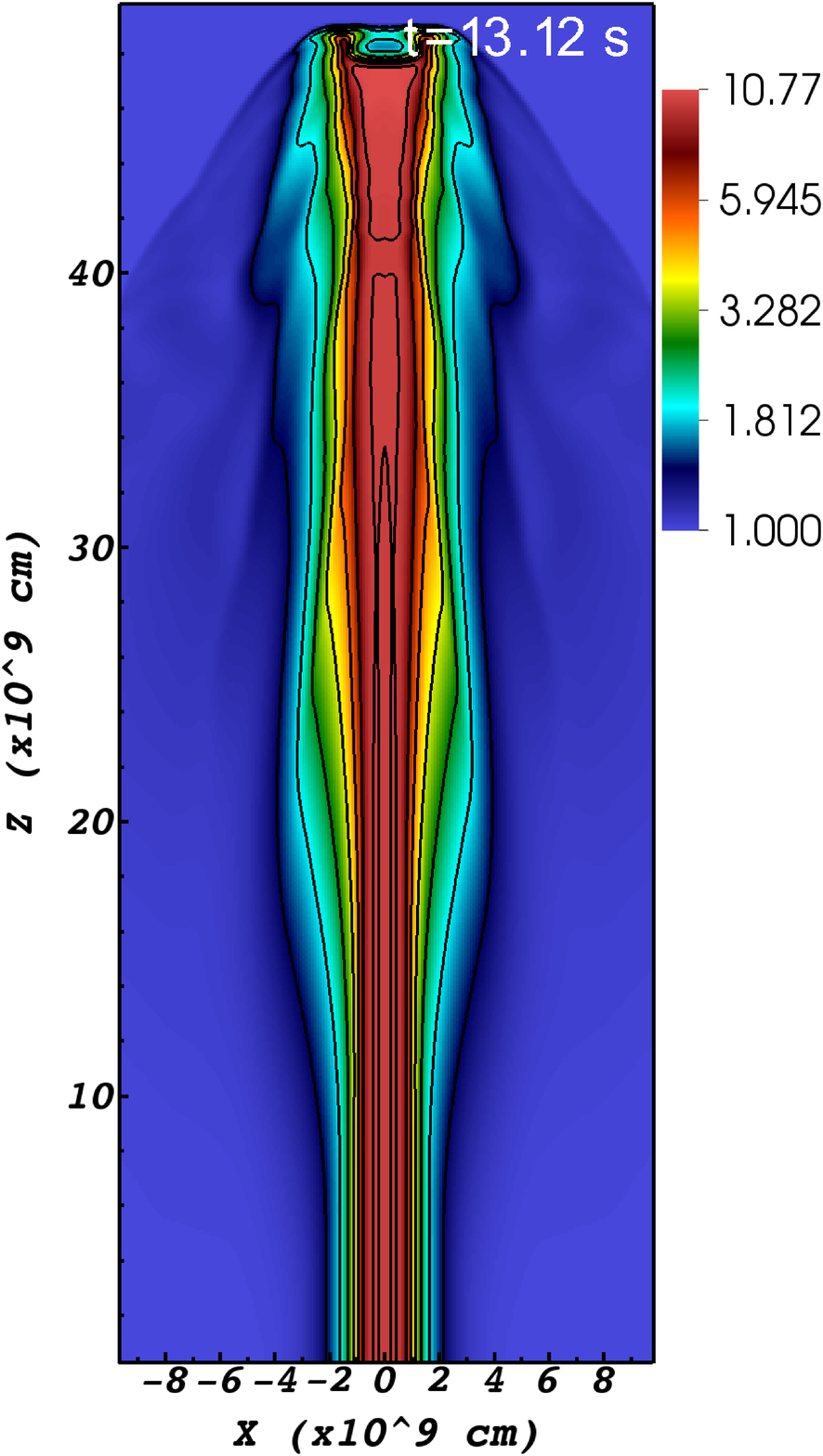}
\caption{Model 3. We show the rest-mass density (top) and Lorentz factor (bottom) at different times.  The jet was injected during $12 \ {\rm s}$ with a Lorentz factor $W_\text{b}=10$. The white lines in the top right-hand panel indicate the position of both detectors.} \label{fig:pl-10radrho}
\end{figure*}

\begin{figure*}%[htb]
\includegraphics[width=0.235\textwidth,height=0.35\textheight]{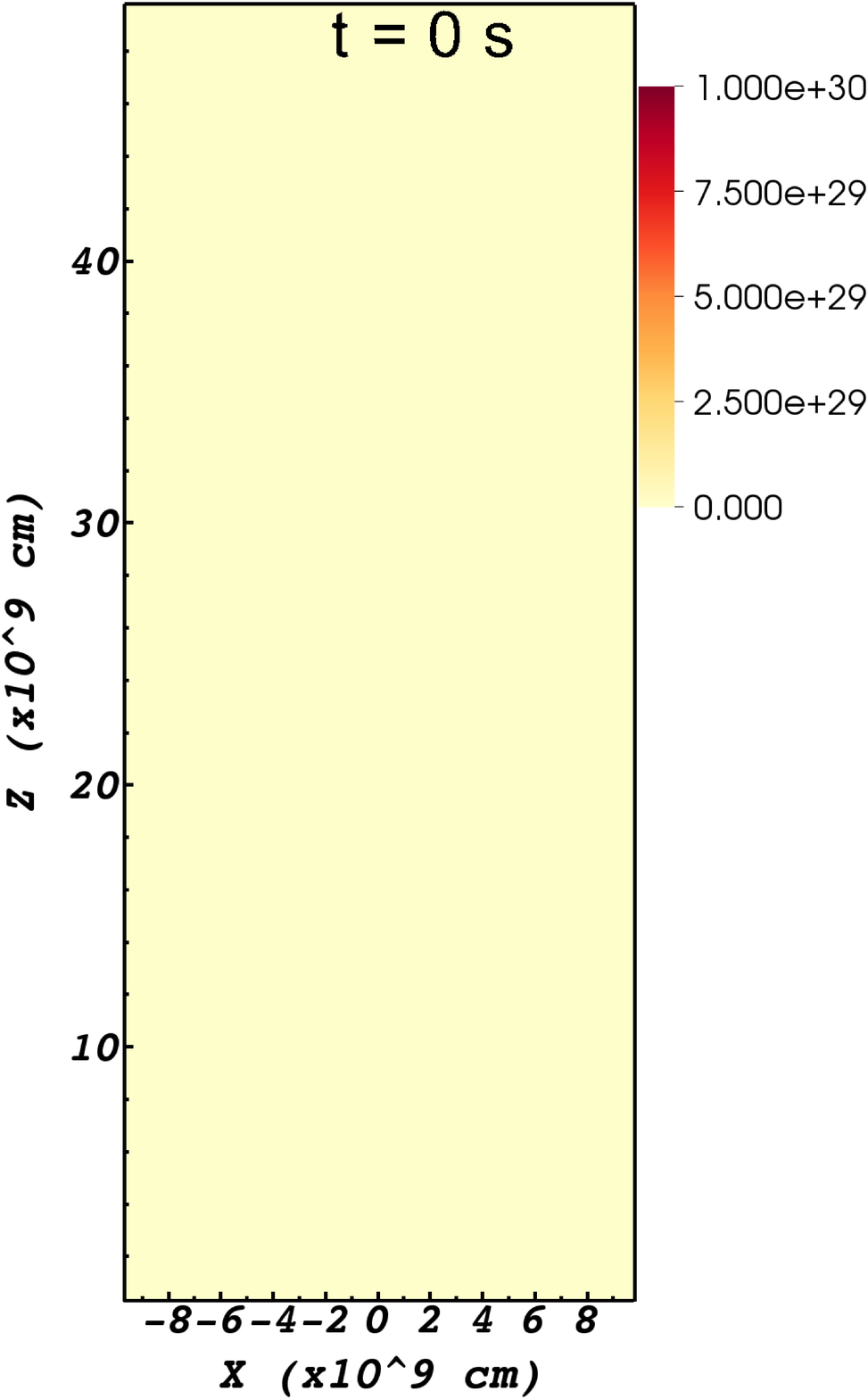}
\includegraphics[width=0.235\textwidth,height=0.35\textheight]{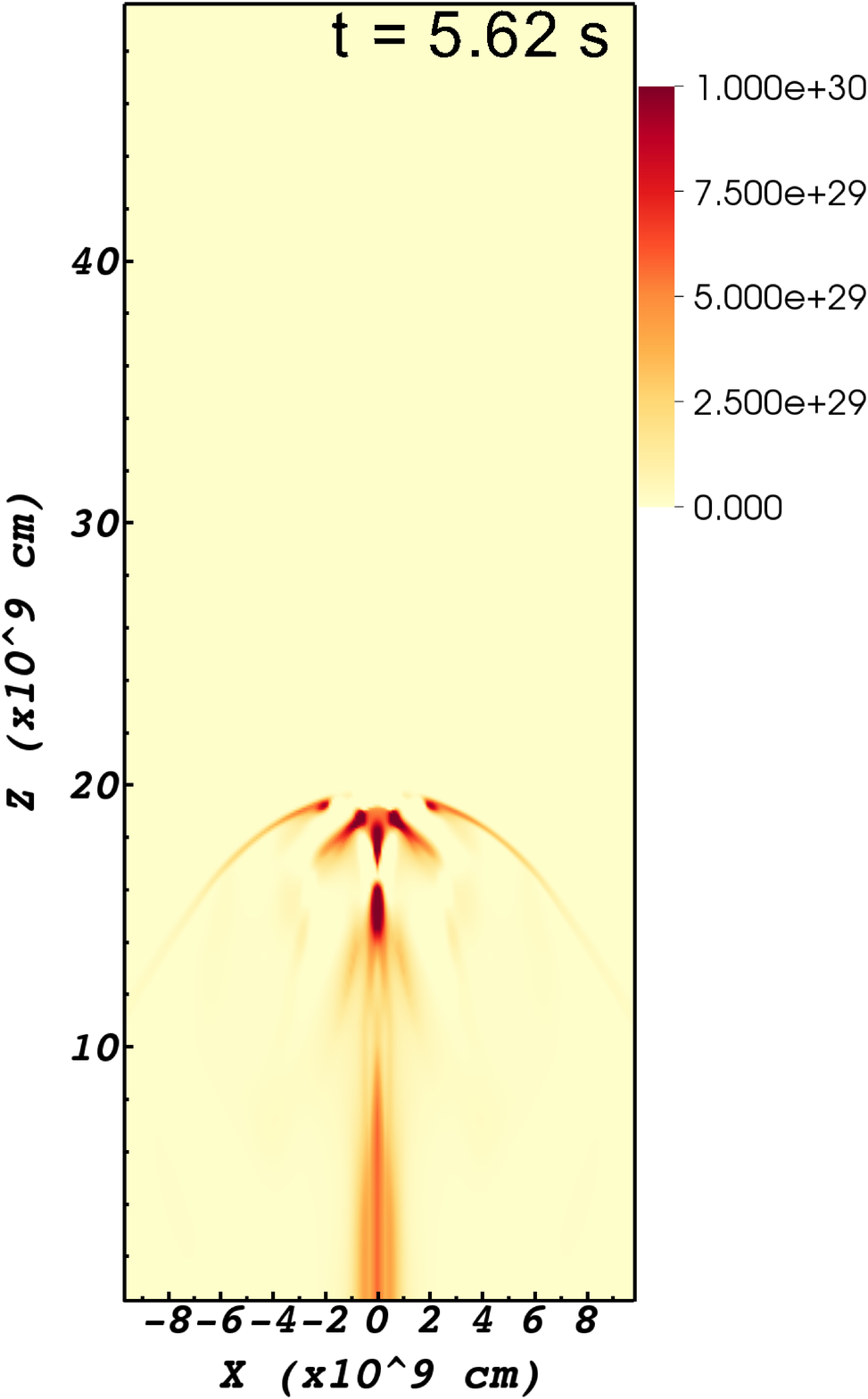}
\includegraphics[width=0.235\textwidth,height=0.35\textheight]{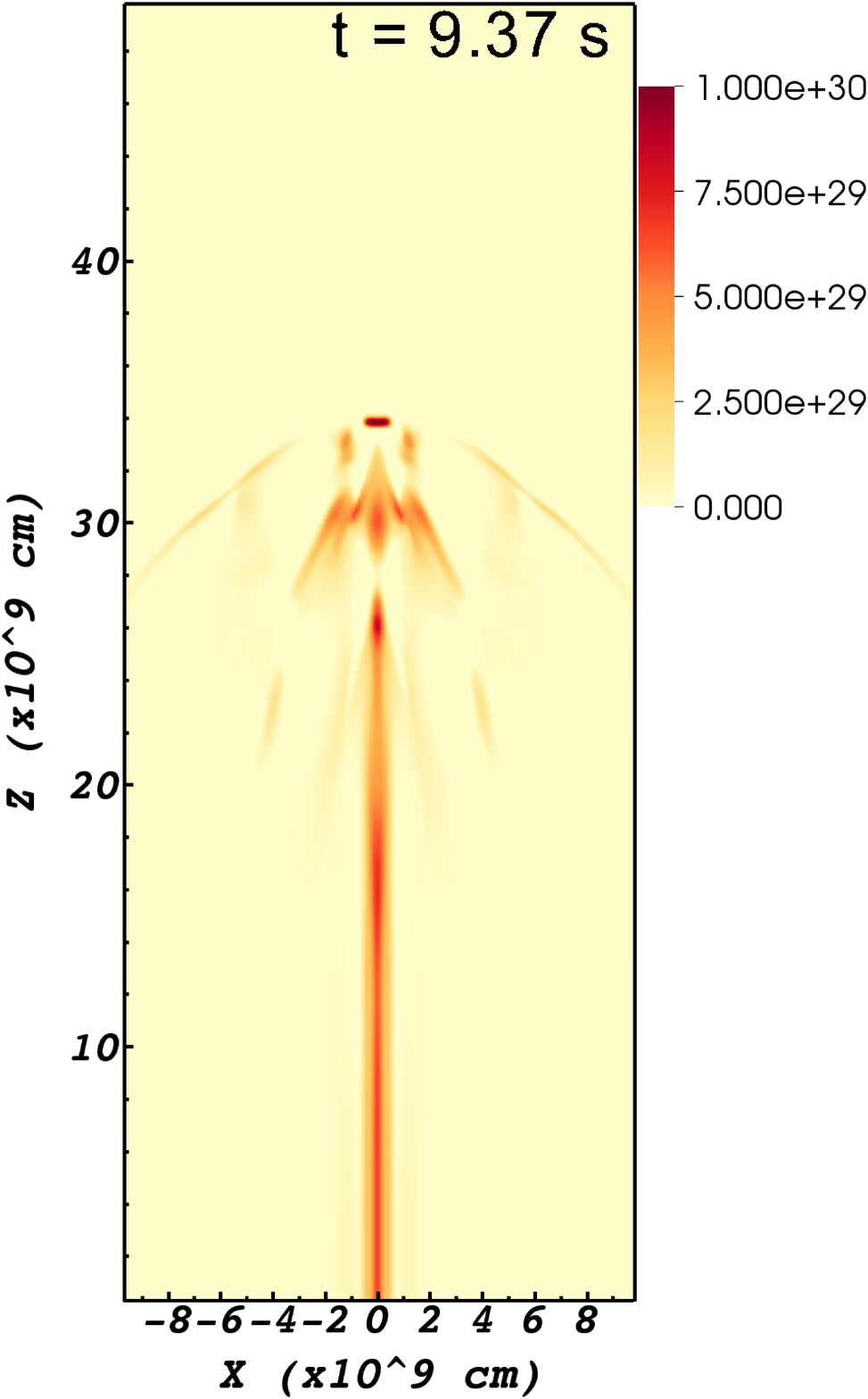}
\includegraphics[width=0.235\textwidth,height=0.35\textheight]{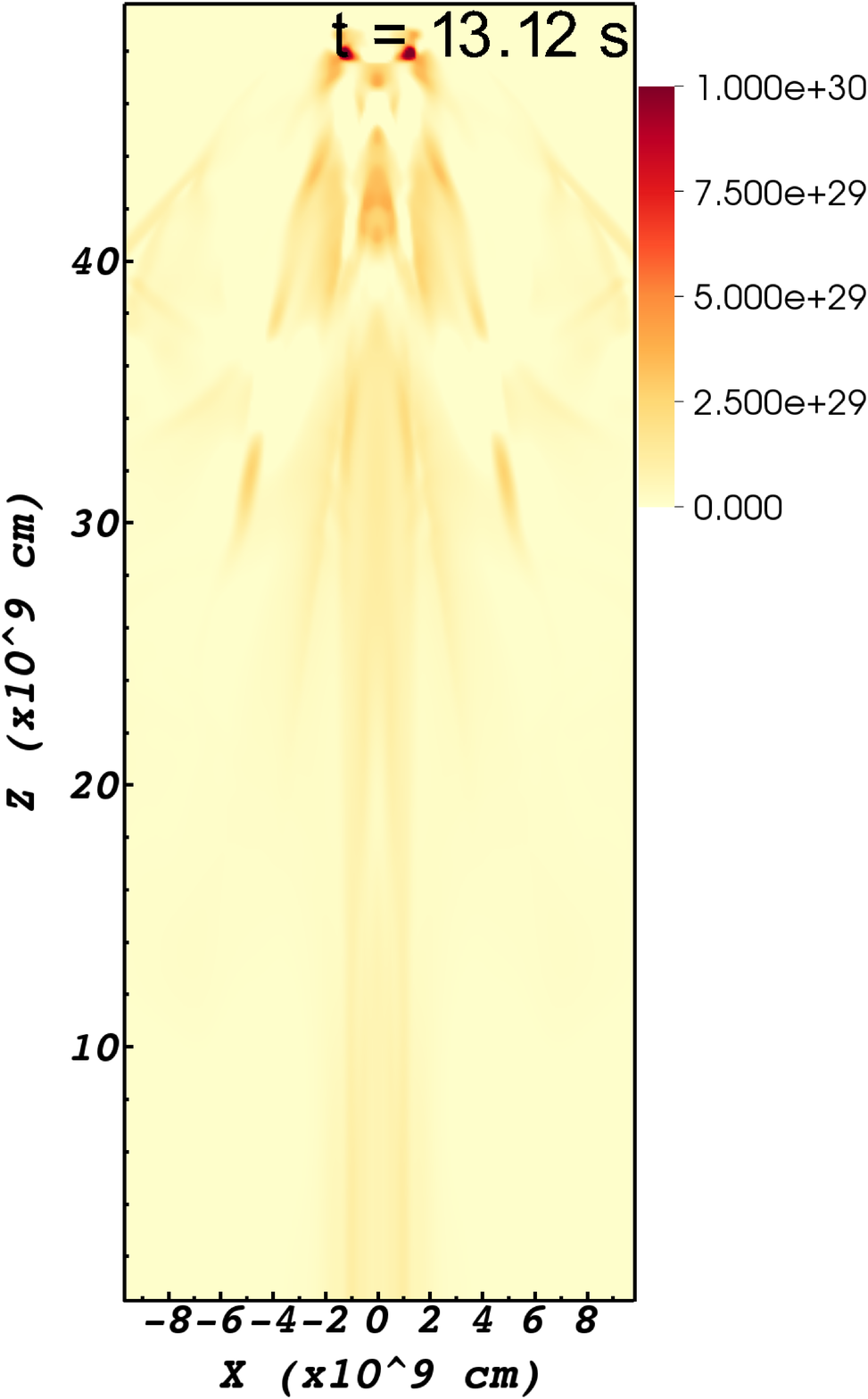}
\caption{Snapshots of the $z-$component of the radiation flux at different times for jet of model 3.} \label{fig:pl-10radfr}
\end{figure*}

% -----> Subsection <-----
\subsection{Gas-pressure-dominated case}
\label{subsec:gpu}

This  is model 2 of Table \ref{t1} and in Fig. \ref{fig:ultra-gaspress:a}, we show the LCs measured in the two different planes mentioned above for comparison. In the top right-hand panel of Fig. \ref{fig:pl-10radrho}, we indicate the surface of the two detectors with white lines, used to calculate the LC using (\ref{eq:luminosity}). From now on we will refer to these planes as perpendicular and inclined planes where the LC is measured. 

Also, in Fig. \ref{fig:ultra-gaspress:a}, we show the maximum of the temperature of the fluid and radiation for this model 2, which is found to be right  behind the bow shock. This figure also illustrates that the initial thermal equilibrium is lost,  and the matter temperature is bigger than radiation temperature. The difference is of one order of magnitude, which indicates how close to equilibrium the gas and radiation are in this case.

Morphologically, the three highly relativistic jets in Table \ref{t1} have a similar structure. The high pressure  at the cocoon compacts the jet and some material is convected backwards. It is known that the shock reflected backwards from the contact discontinuity modifies the structure of the jet head and influences the further propagation into the surrounding medium \cite{Massaglia}. In this context, the radiation pressure helps to push material towards the contact discontinuity by making the internal shock behind the jet head a little bit different. From Fig. \ref{fig:ultra},  the rest-mass density of the purely hydrodynamical case (model 1) and the gas-pressure-dominated case (model 2) are very similar. Quantitatively though, the maximum of the density of model 2 is slightly higher than that of model 1 by a 2\% only in the region behind the front shock.

% -----> Subsection <-----
\subsection{Radiation-pressure-dominated case}
\label{subsec:rpu}

This is model 3 in Table \ref{t1}. In Fig. \ref{fig:ultra-radpress:a}, we show the LCs for this model in the two  detectors. The difference with respect to its gas-pressure-dominated counterpart is the amplitude of the LC, which is of the order of $\sim 10^{52} \ \text{erg}/\text{s}$, two orders of magnitude bigger than that of model 2. Also, in Fig. \ref{fig:ultra-radpress:a}, we show the maximum of the radiation and fluid temperatures. The fluid temperature is bigger than the radiation temperature throughout evolution by approximately one order of magnitude.

With respect to the morphology, in Fig. \ref{fig:pl-10radrho}, we show the evolution of the rest-mass density and Lorentz factor corresponding to  model $3$. At $t \sim 5 \text{s}$, we can see the basic properties of the jet, namely a collimation shock in the beam, a bow shock, and the reverse shock produced by the interaction between the jet/external medium and the formation of a cocoon. At this time, the Lorentz factor of the  beam remains similar to its initial value, whereas the Lorentz factor of the head and cocoon is smaller. Later on, the jet starts to propagate in a very dilute medium, and consequently the pressure of the cocoon drops and the jet begins to expand laterally into the circumstellar matter. We can see a snapshot of this behaviour at $t \sim 9 \text{s}$. Also, at this time, the beam's Lorentz factor grows. When the head of the jet reaches the boundary along the $z-\text{axis}$ at $t \sim 13.12 \text{s}$, the Lorentz factor of the beam starts to decrease because the jet is not being injected anymore. We also can see an indication of a vortical flow formed behind the head of the jet similar to a Kelvin-Helmholtz instability that could be better resolved using higher resolution. Notice that a higher resolution would help to capture finer structures such as in Fig. \ref{fig:Jets}, where snapshot of the rest-mass density of model 3 using three different resolutions at $t \sim 9.37$ s is shown, and the vortical flow behind the jet head is better resolved with the high resolution. Another example is the jets evolved with high resolution  in \cite{Meliani_2010} can capture the fine-scale instabilities at the head of the jet.

Comparing this case with its counterpart models, gas pressure dominated and purely hydrodynamics, in Fig. \ref{fig:ultra}, we can see a difference along the polar axis of the jet. In this model, we  see that the rest-mass density of the jet is bigger than that of models 1 and 2. Quantitatively, at time $t = 11 \text{s}$, the difference between the maximum rest mass density of model 3 with respect to its hydrodynamical version is of the order of $\sim 10\%$. 

In order to learn the effect of radiation pressure in the case when the radiation pressure is dominant, we compare the Lorentz factor profile of model 3 with the gas-pressure dominated  (model 2) and the purely hydrodynamical  (model 1) cases, at the same time $t=11 \ {\rm s}$. This is shown in Fig. \ref{fig:ultra-LF}, where we can see that the radiation-pressure-dominated jet propagates slightly faster than the gas-pressure dominated and the purely hydrodynamic models. This illustrates the influence of  radiation pressure in the dynamics of the jet.

\begin{figure}%[htb]
\includegraphics[width=0.4\textwidth]{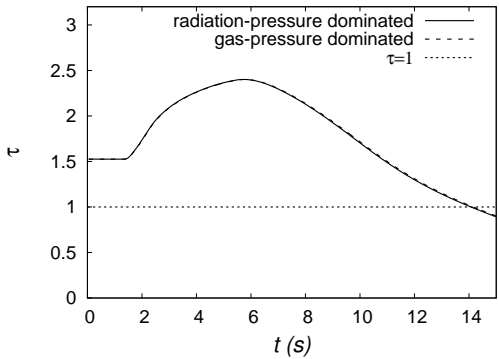}
\caption{Time evolution of optical depth $\tau$ for models 2 and 3, integrated in the laboratory frame along the $z$-axis. The solid and dotted lines correspond to the models 2 and 3 respectively. We also plot the line $\tau=1$ as a threshold of optically thick and thin.} \label{fig:tao}
\end{figure}

 Another important parameter that helps to appreciate the role of the radiative effects incorporated in models 2 and 3 is the optical depth $\tau$. In Fig. \ref{fig:tao}, we see the evolution of $\tau$, integrated all along the $z-$axis. In this figure, we can see that the optically thick ($\tau >1$) and optically thin ($\tau <1$) regimes are very close to each other in both models. This suggests that the thermal radiation field does not significantly affect the morphology of the system, which explains why the outcomes are so similar despite the injection of radiation energy $E_\text{r}$ in model 3.

To finalize this section, in Fig. \ref{fig:pl-10radfr}, we show the $z$ component of the radiative flux during the whole evolution. Because we are assuming that the jet starts in an optically thick regime, the radiative flux in comoving frame at $t = 0$ is equal to zero. At $t\sim 5 \text{s}$, we can see that 
the highest radiative flux takes place in the head of the jet, collimation, oblique, and blow shock. At $t \sim 9 \text{s}$, the radiative flux decreases in the blow shock. Finally, at $t\sim 13 \text{s}$, we can see how the radiative flux in the jet is dissipating because the matter and radiation was switched off at  time $t = 12 \text{s}$.

% --------------------------------------------
% ---------->     SECTION     <----------
% --------------------------------------------
\section{Application to models of Low-Luminosity GRBs}
\label{sec:LLGRBs}

Now, we  study the scenario of jet propagation within a progenitor star, which has been applied to model  Low-luminosity GRBs \cite{BrombergI,Mizuta,DeColleII,Senno,Geng}. In order to study the LLGRBs, we evolve a jet propagating through its progenitor star assuming the 16TI progenitor density profile \cite{WoosleyHeger}. This model consists of a pre-supernova star with radius of $R=4\times10^{10}$ cm, $13.95$ solar masses, and $1\%$ of solar metallicity. From $10^9$ cm to $6\times10^9$ cm the density falls quickly as a power law $\sim r^{-1.5}$, from this point to a radius of $4\times10^{10}$ cm, it decays exponentially. Finally, the surrounding medium from the surface to $1.8\times10^{11}$ cm, the density falls off like $\sim r^{-2}$.

\begin{figure}%[htb]
\includegraphics[width=0.4\textwidth]{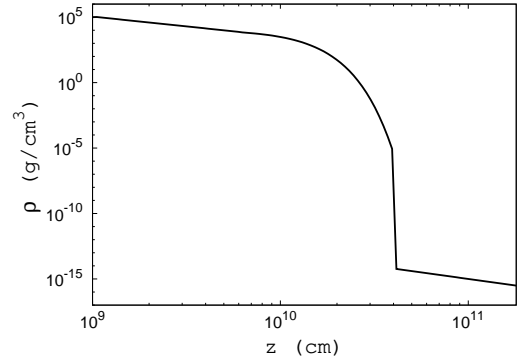}
\caption{Rest-mass density profile of the 16TI progenitor model used in the numerical simulations. This model describes a pre-supernova star with radius of $R=4\times10^{10}$ cm and mass $13.95M_{\odot}$.} \label{fig:16TI}
\end{figure}

In Fig. \ref{fig:16TI}, we show the initial rest-mass density profile of this progenitor model. The rest-mass density and pressure of the beam are $10^3\text{g}~\text{cm}^{-3}$ and $10^{19}\text{g}~ \text{cm}^{-1} \text{s}^{-2}$, respectively. The radius of the jet is $r_\text{b}=2\times10^9 \text{cm}$, and the ratio between their pressures is $K = 0.01$. In our evolution, we do not consider the gravitational field of the progenitor because the jet is moving with high enough velocity such that the dominant effects are due to the interaction of the jet with the matter of the star.

{\it Set up of jets.} Initially, we launched the jet at a distance of $10^9$ cm from the centre of  star in the $z$ direction, as done in \cite{Mizuta}. Unlike in the previous scenario, where the jet was injected during $12$ s, in this case the jet is injected  during $t_\text{inj}=20$ s. 

We carried out  simulations with a Lorentz factor $W_\text{b}=10$ and consider the case, where the gas and radiation pressure are dominant. The specific values for these parameters are in Table \ref{t2}. These models have been standardized with a  resolution of $1.25\times10^8 \text{cm}$ for the numerical domain.

\begin{table}%[!hbt]
\begin{center}
\begin{tabular}{c  c  c  c  c  c }
\hline
\hline
Model & $L_j(\text{erg}/\text{s})$ & $E_j(\text{erg})$   & $E_{r,b}\left(\text{erg}\text{cm}^{-3}\right)$ & $g_\text{1,b}$ $(g_\text{2,b})$  \\
\hline
 4    &$4.97\times 10^{50}$       &$9.94\times 10^{51}$   &$1\times10^{19}$                                     & $0.33$ $(0.33)$\\
\hline
 5    &$1.28\times 10^{52}$       &$2.58\times 10^{53}$   &$1\times10^{21}$                                     & $33.3$  $(33.3)$\\
\hline
\hline
\hline
\hline
\end{tabular}
\caption{\label{t2} Parameters of the  jets that we evolve on the progenitor model. We use the opacities that emulate the  free-free, bound-free, bound-bound and electron-scattering opacities with adiabatic index $\Gamma=4/3$.}
\end{center}
\end{table}

\begin{figure*}%[htb]
\includegraphics[width=0.235\textwidth,height=0.35\textheight]{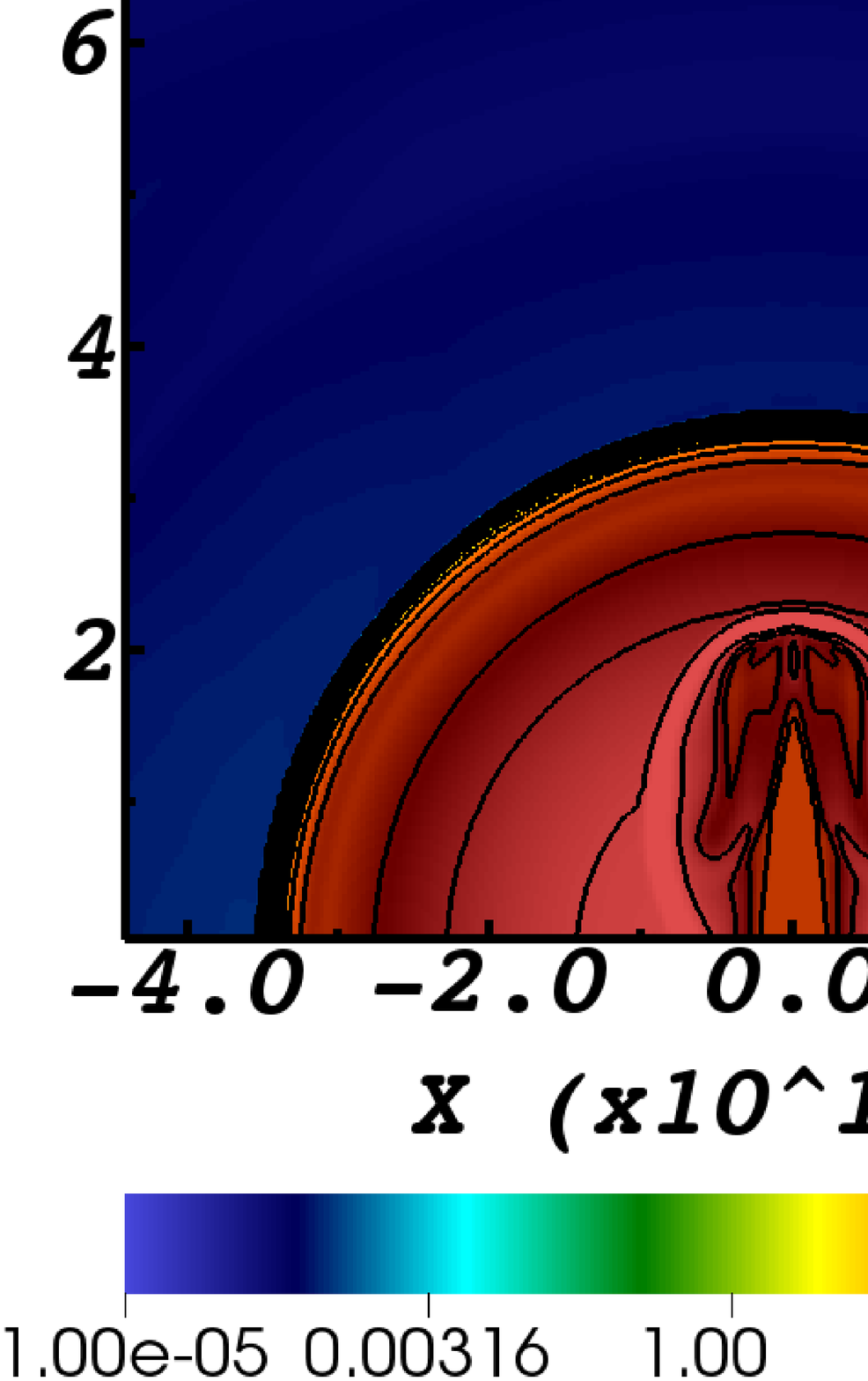}
\includegraphics[width=0.235\textwidth,height=0.35\textheight]{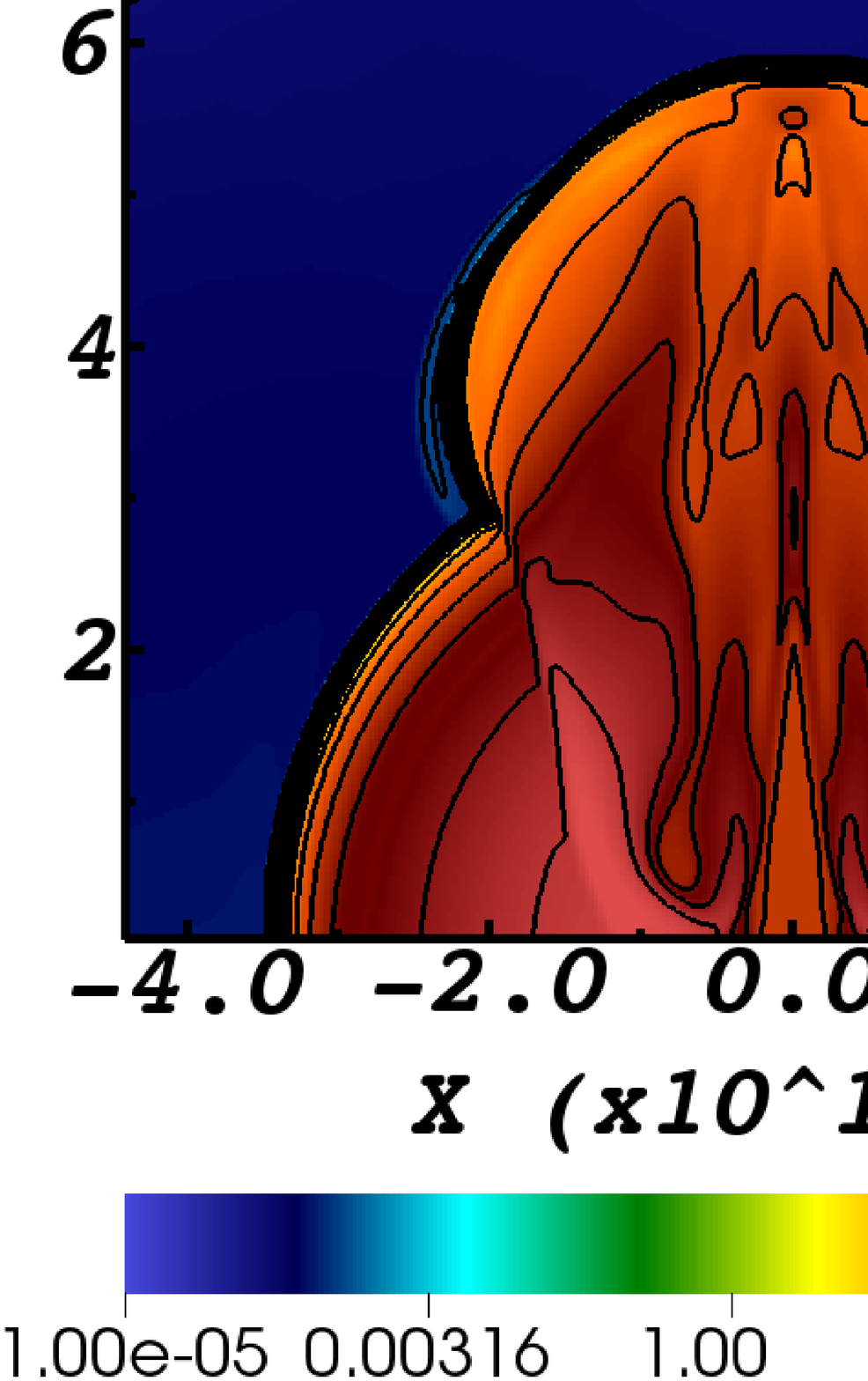}
\includegraphics[width=0.235\textwidth,height=0.35\textheight]{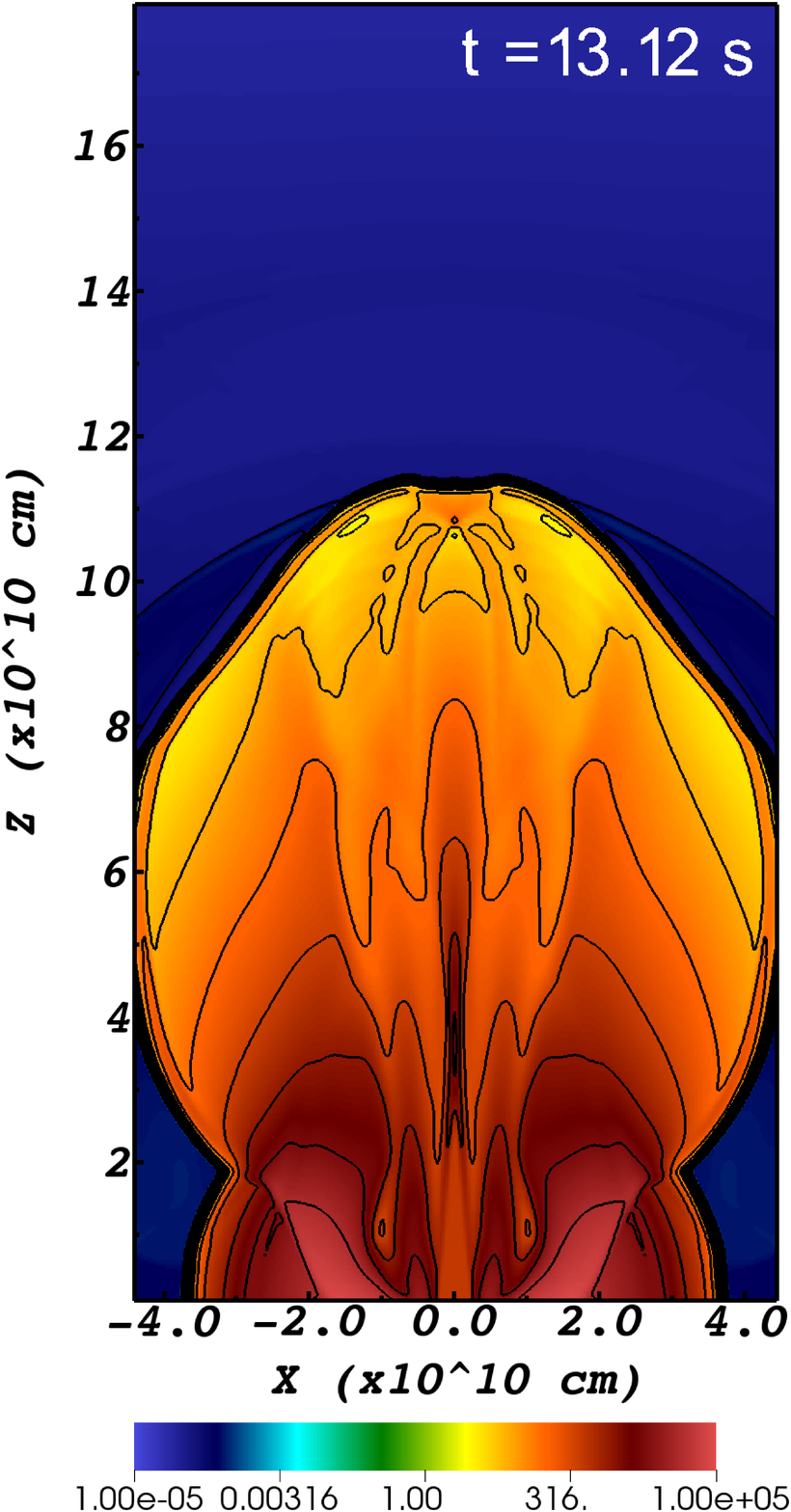}
\includegraphics[width=0.235\textwidth,height=0.35\textheight]{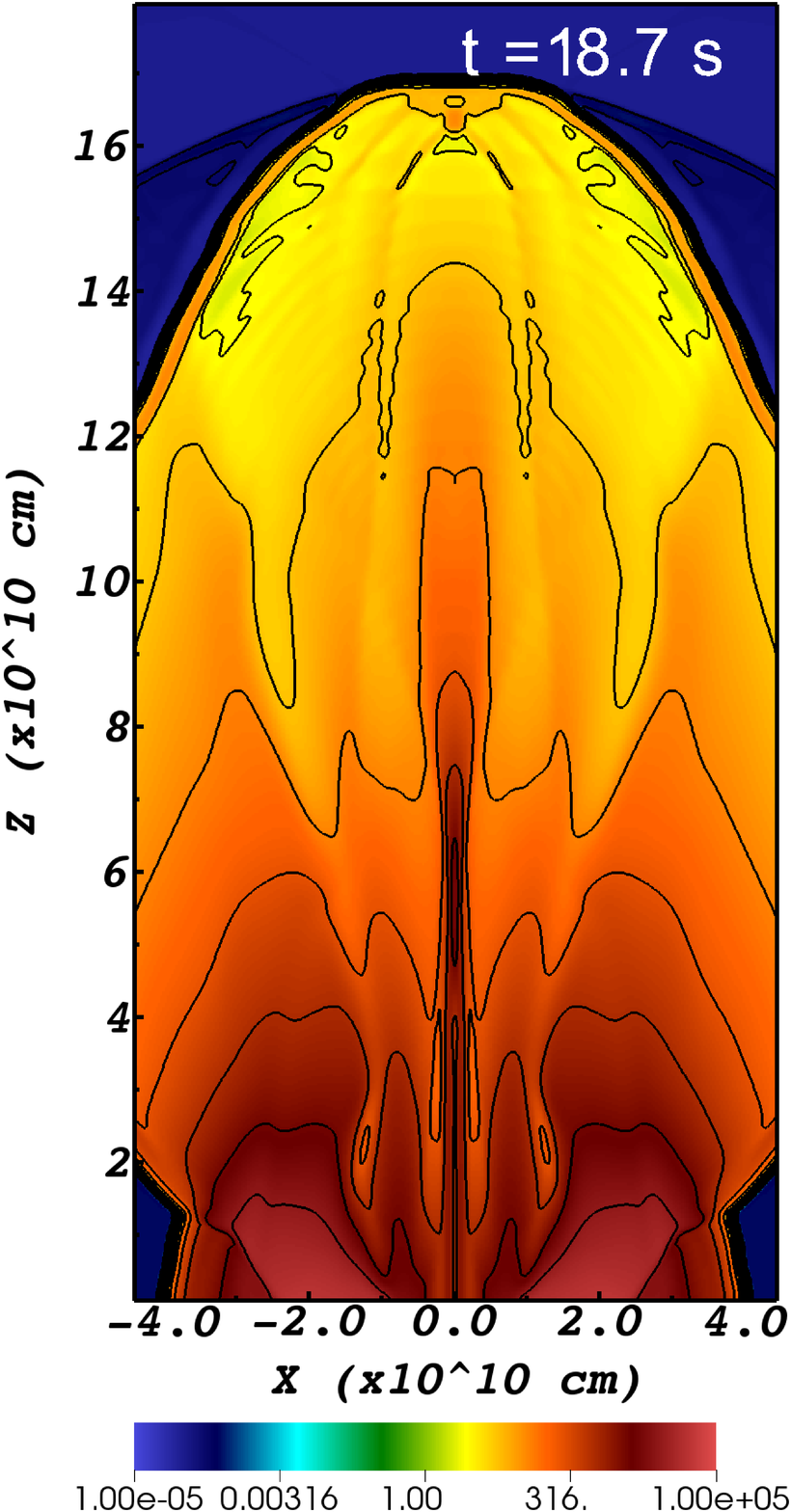}
\includegraphics[width=0.235\textwidth,height=0.35\textheight]{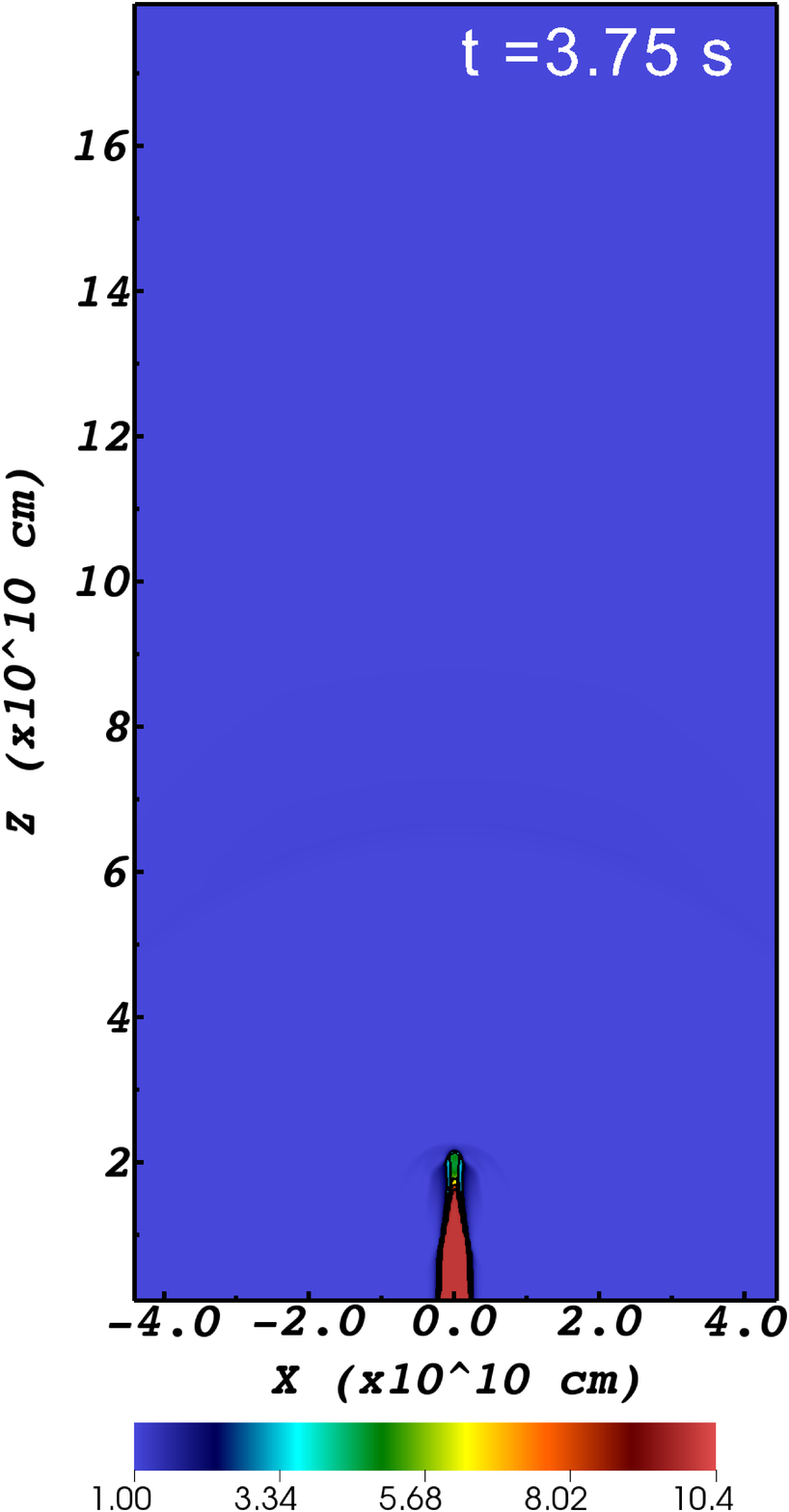}
\includegraphics[width=0.235\textwidth,height=0.35\textheight]{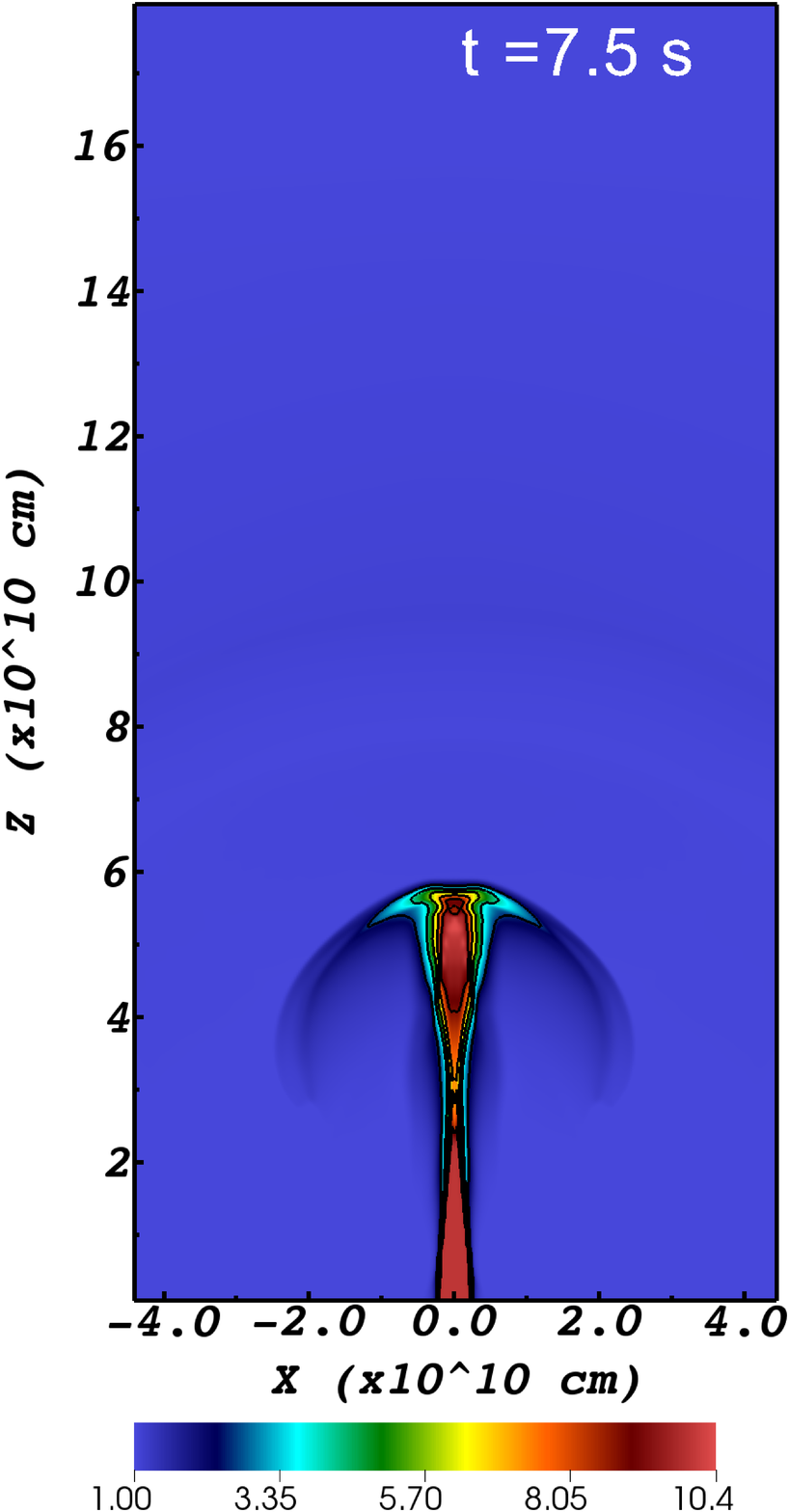}
\includegraphics[width=0.235\textwidth,height=0.35\textheight]{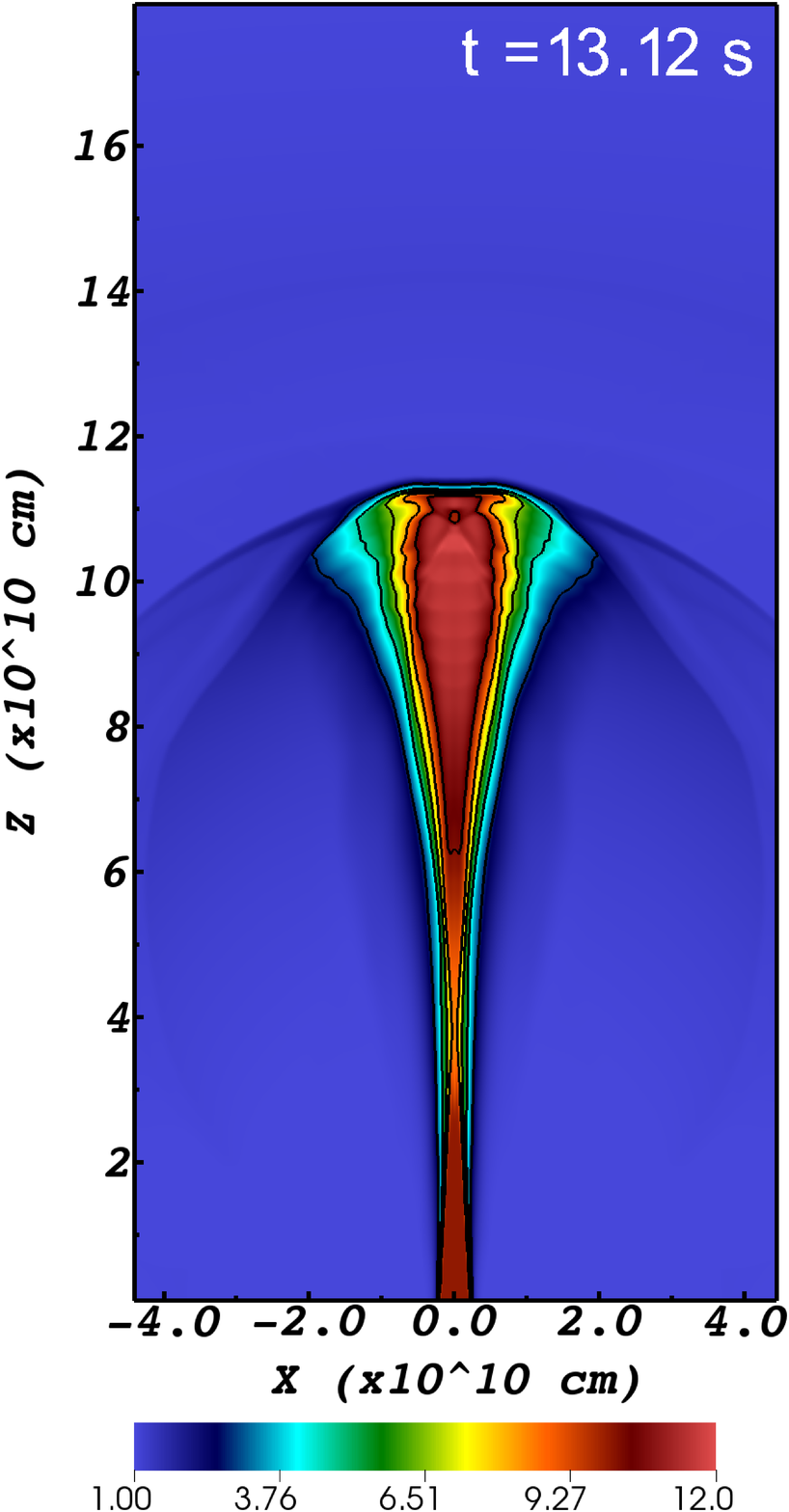}
\includegraphics[width=0.235\textwidth,height=0.35\textheight]{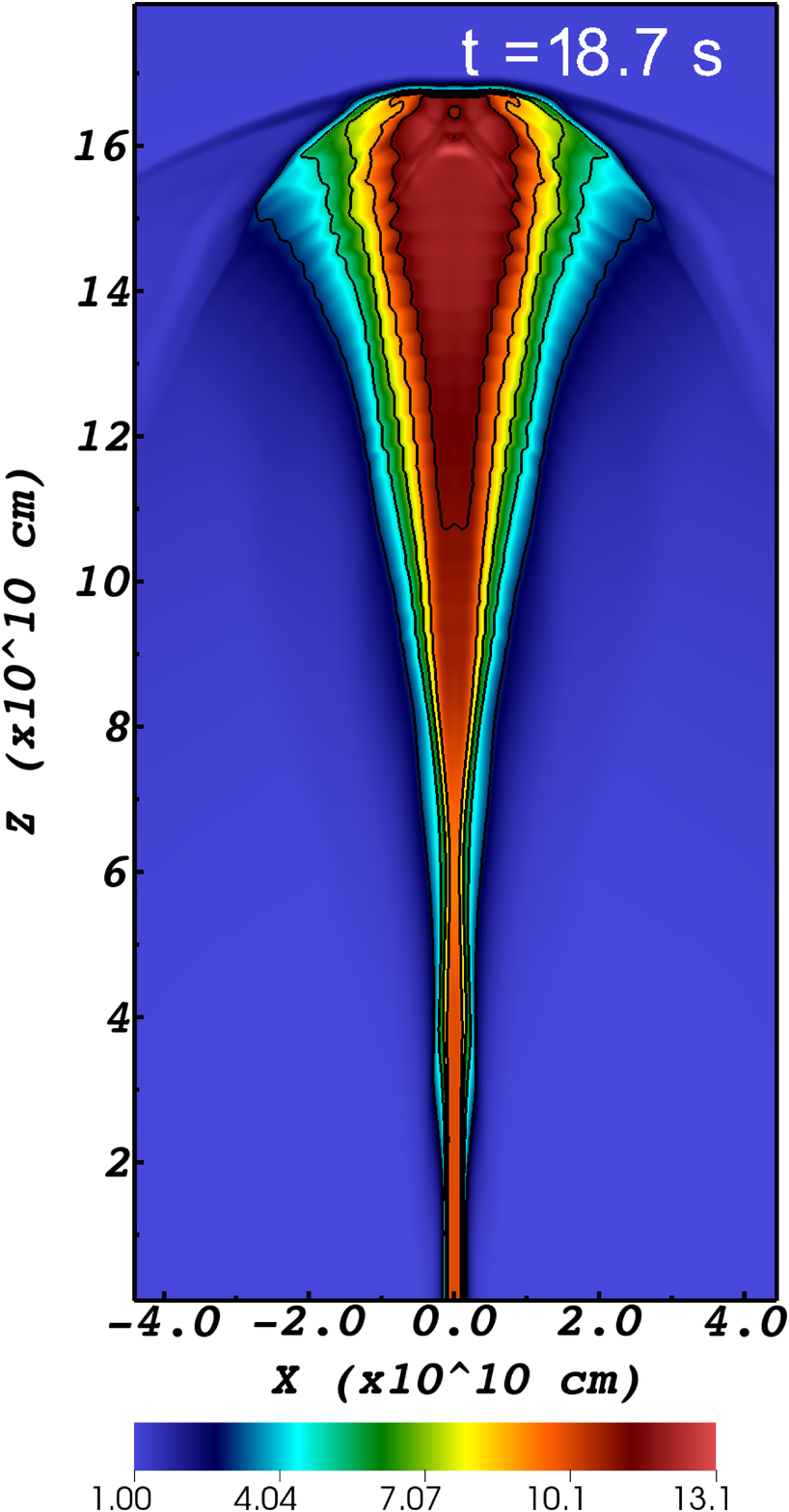}
\caption{Model 4. Snapshots of the rest-mass density (top) and the Lorentz factor (bottom) at different times. This is  the gas-pressure-dominated case.} \label{fig:16ti-10gasrho}
\end{figure*}

In the dynamical evolution of model $4$, there are two important phases shown in Fig. \ref{fig:16ti-10gasrho}. In the first phase, when the jet is propagating through the star, the  Lorentz factor of the beam around the nozzle begins to grow up as expected because the nozzle is continuously injecting energy  to this region, whereas the head of the jet propagates with smaller velocity due to the  interaction with the stellar envelope. Also, as a consequence of the interaction between the jet and stellar envelope a reverse shock is formed, which interacts with the jet and when the beam crosses the reverse shock a cocoon is created, and the beam is deflected sideways. We show a snapshot of the rest-mass density and Lorentz factor in Fig. \ref{fig:16ti-10gasrho} at $t=3 \ {\rm s}$ in the laboratory frame. At this time, the jet is confined by the pressure in the cocoon.

The second phase of the evolution starts when the jet breaks out the progenitor star. In  Fig. \ref{fig:16ti-10gasrho}, we show the rest mass density and Lorentz factor at various times. In particular at time $t\sim 7 \ {\rm s}$ after the breakout, the jet expands into a rarefied medium, the gas pressure is smaller than inside of the star, as a result of this, the cocoon starts to expand laterally.  Later on, at $t=13 \ {\rm s}$, we can see a stratified cocoon close to the  head, which allows the advance of the head. At this time, the jet propagates with high velocity whereas the cocoon expands with a slower velocity. Likewise, we can see that after crossing the collimation shock, which is located around $5\times 10^{10}~\text{cm}$, the jet is not  confined enough as to keep the cylindrical radius fixed and it starts a lateral expansion.  Finally, at $t\sim 18 \text{s}$ the jet continues to expand laterally whereas the collimation shock is larger than at previous times.

\begin{figure*}%[htb]
\includegraphics[width=0.235\textwidth,height=0.35\textheight]{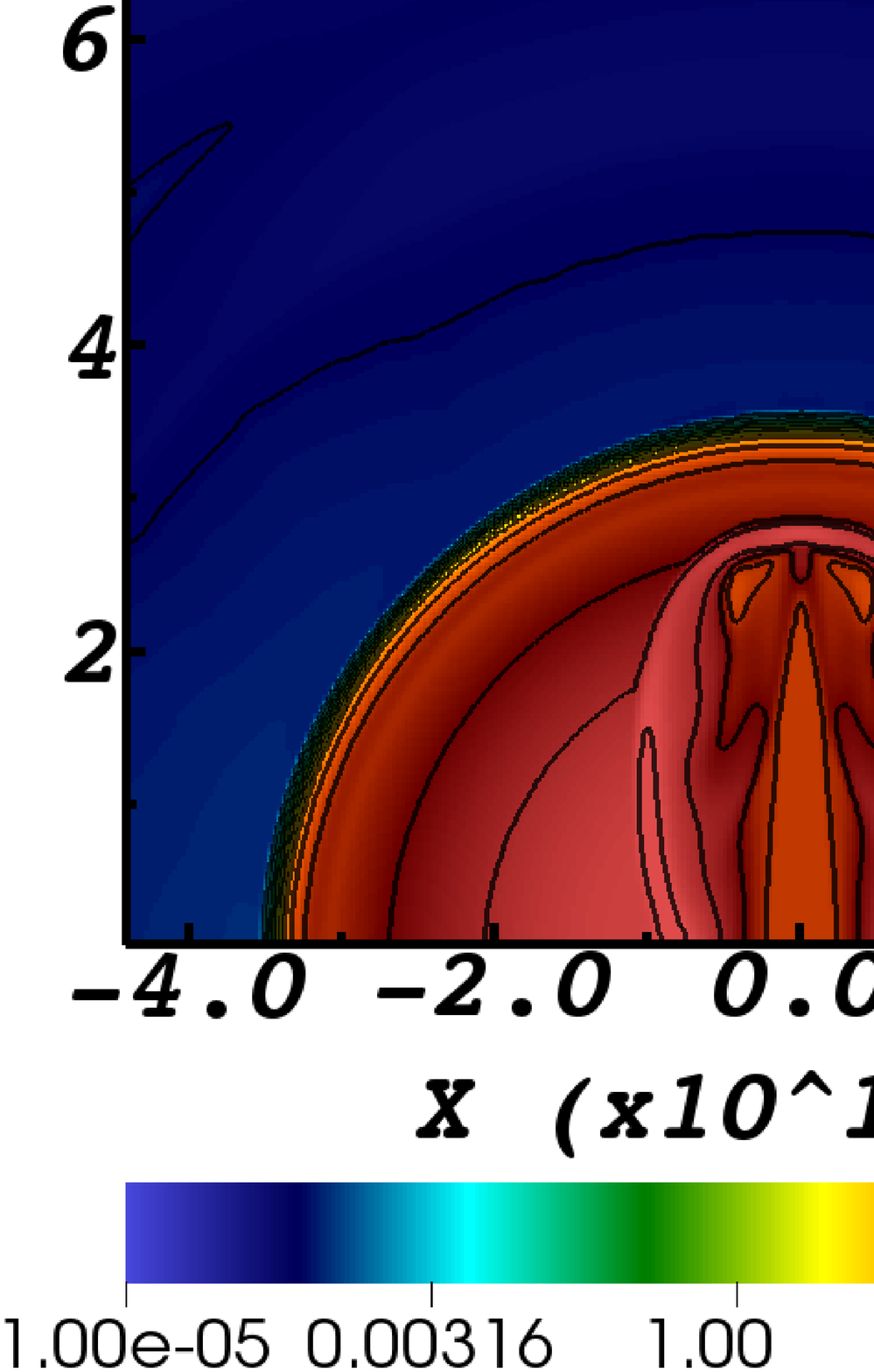}
\includegraphics[width=0.235\textwidth,height=0.35\textheight]{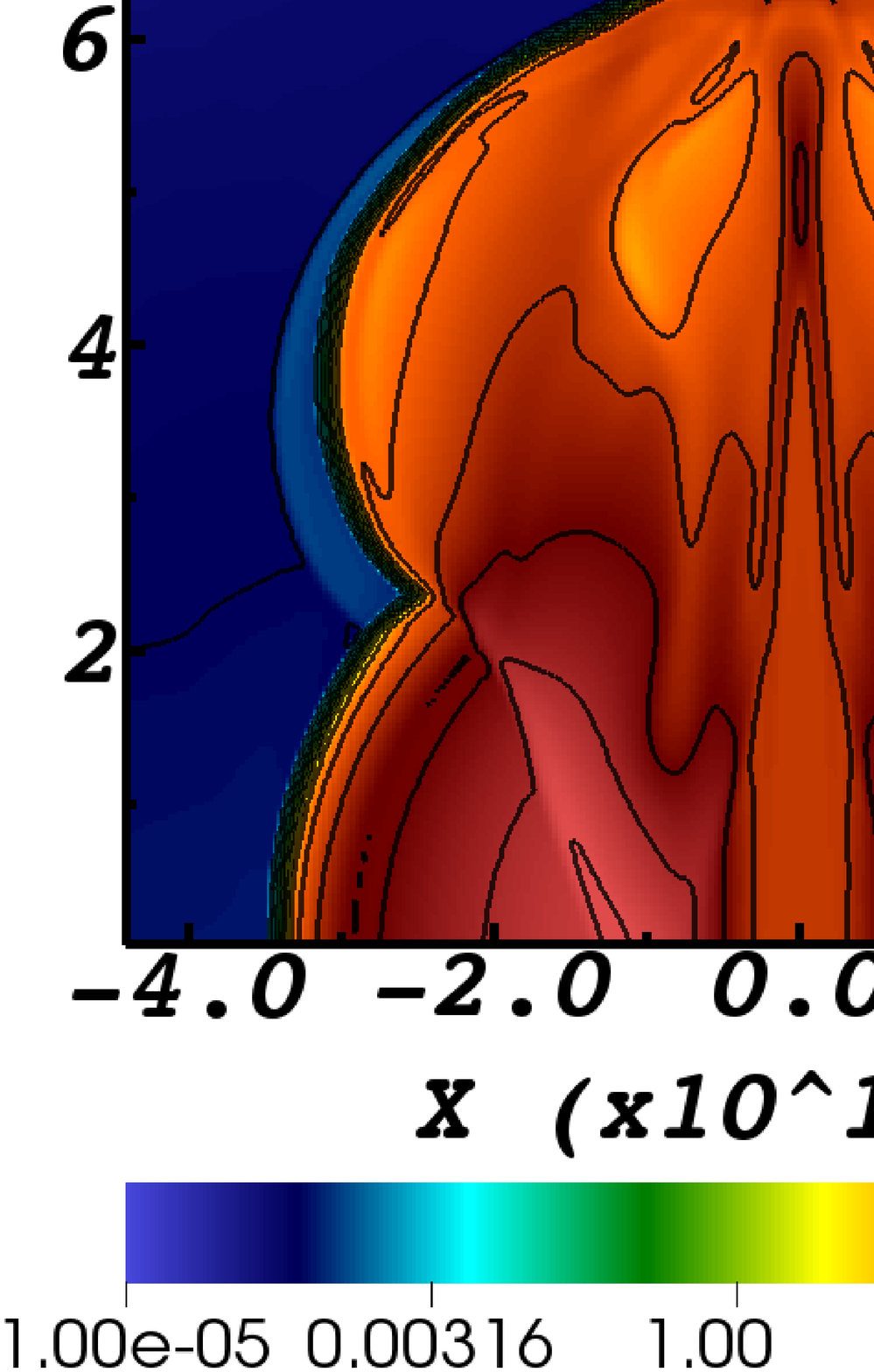}
\includegraphics[width=0.235\textwidth,height=0.35\textheight]{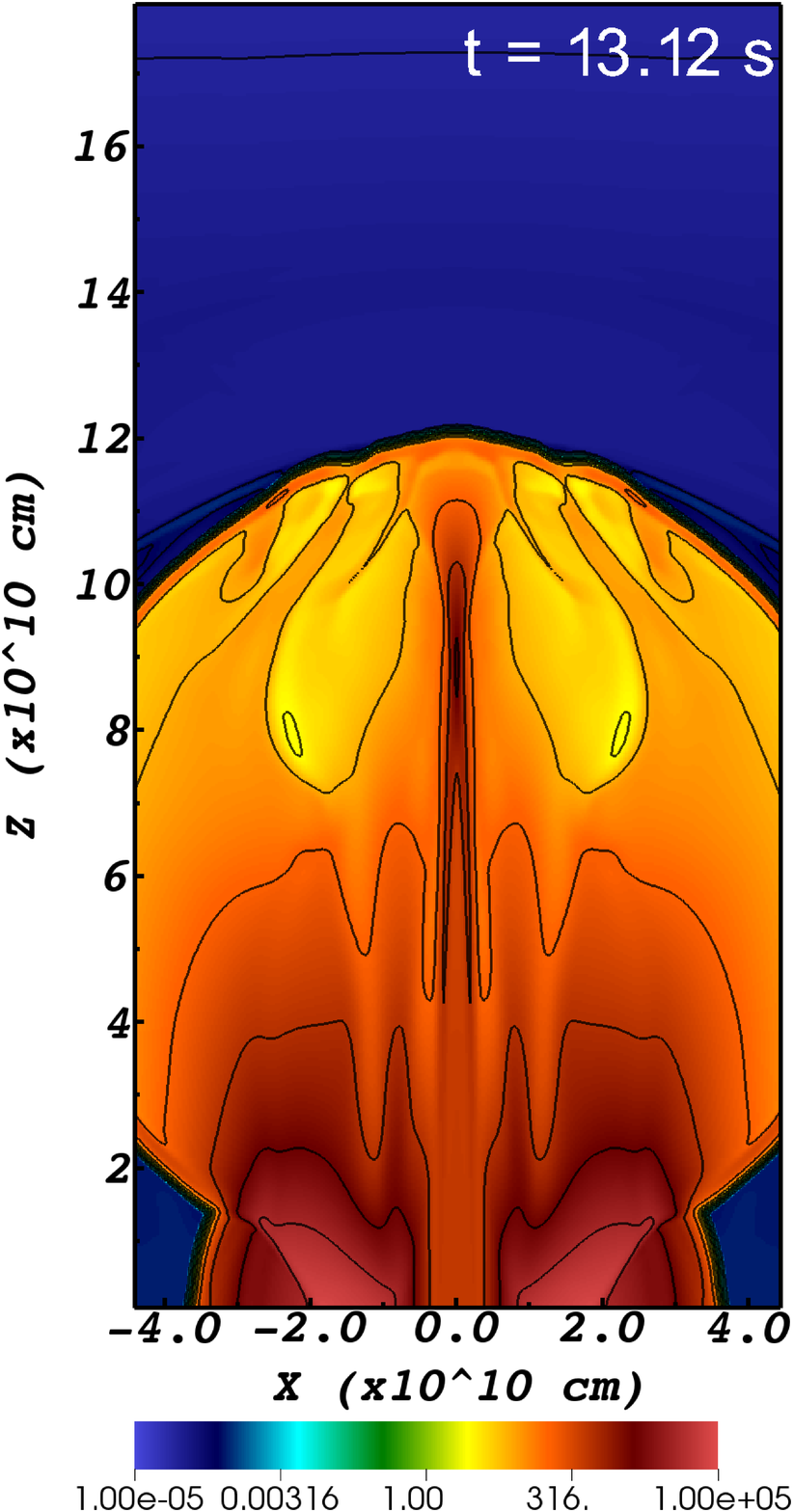}
\includegraphics[width=0.235\textwidth,height=0.35\textheight]{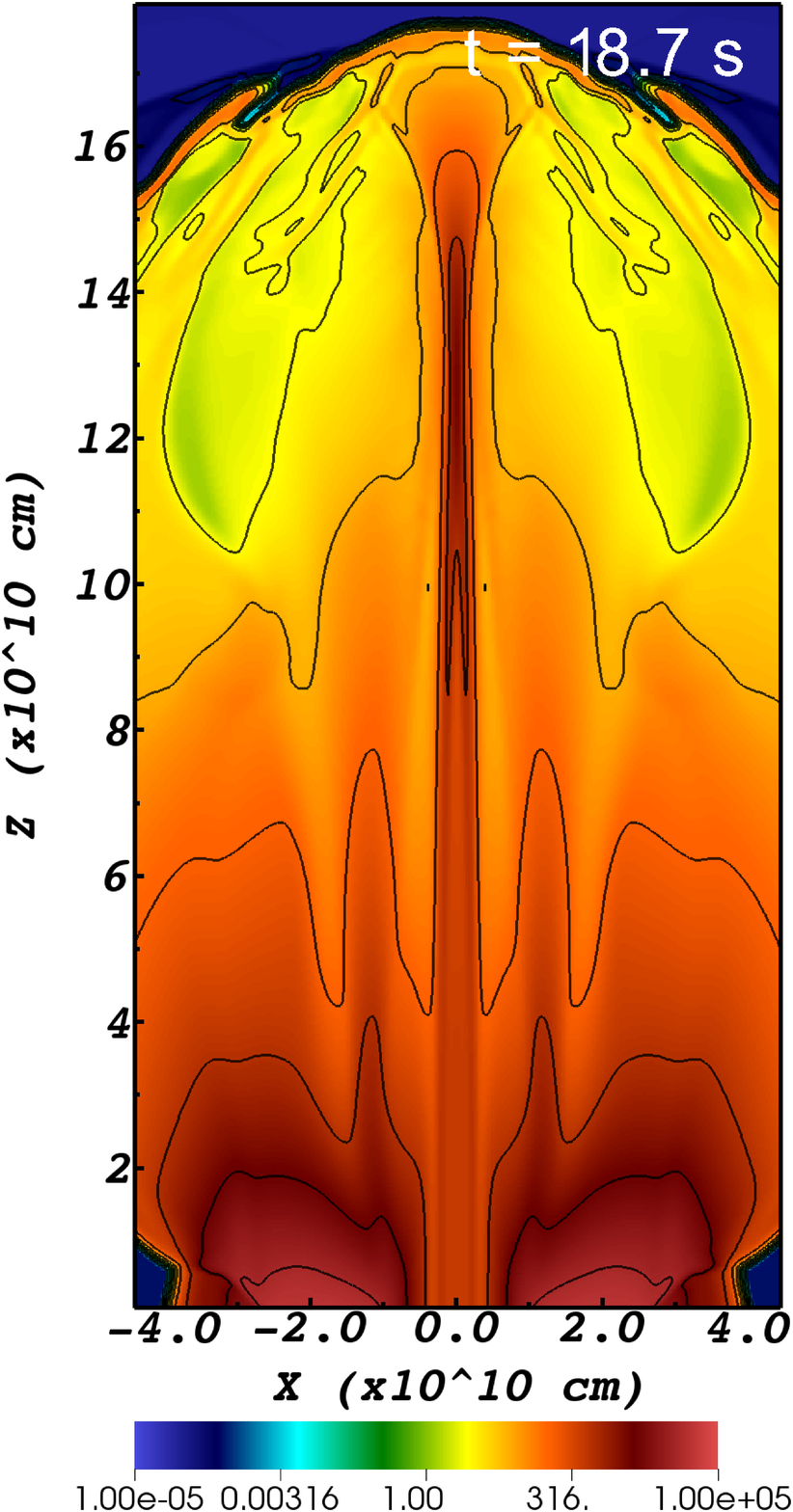}
\includegraphics[width=0.235\textwidth,height=0.35\textheight]{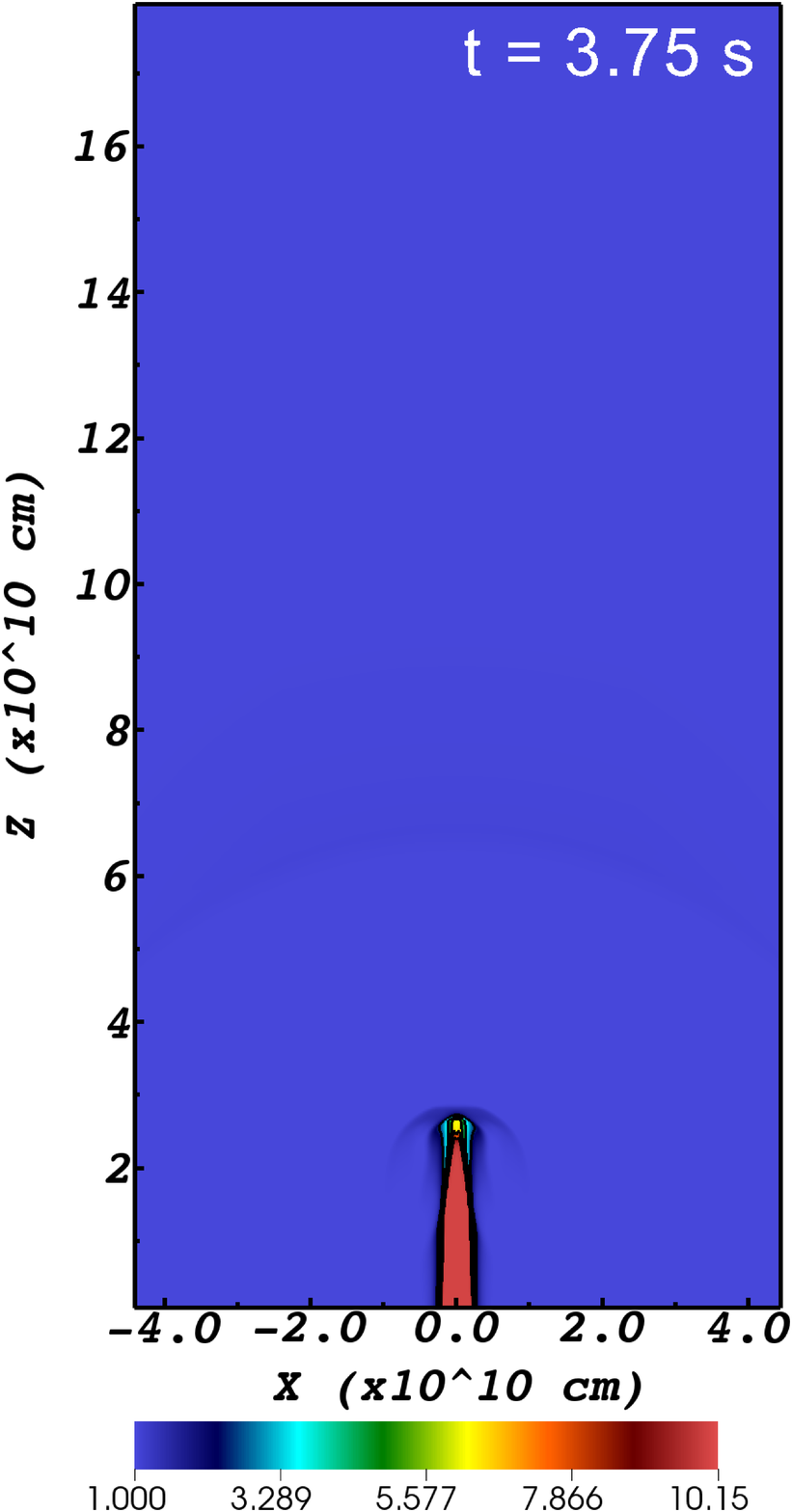}
\includegraphics[width=0.235\textwidth,height=0.35\textheight]{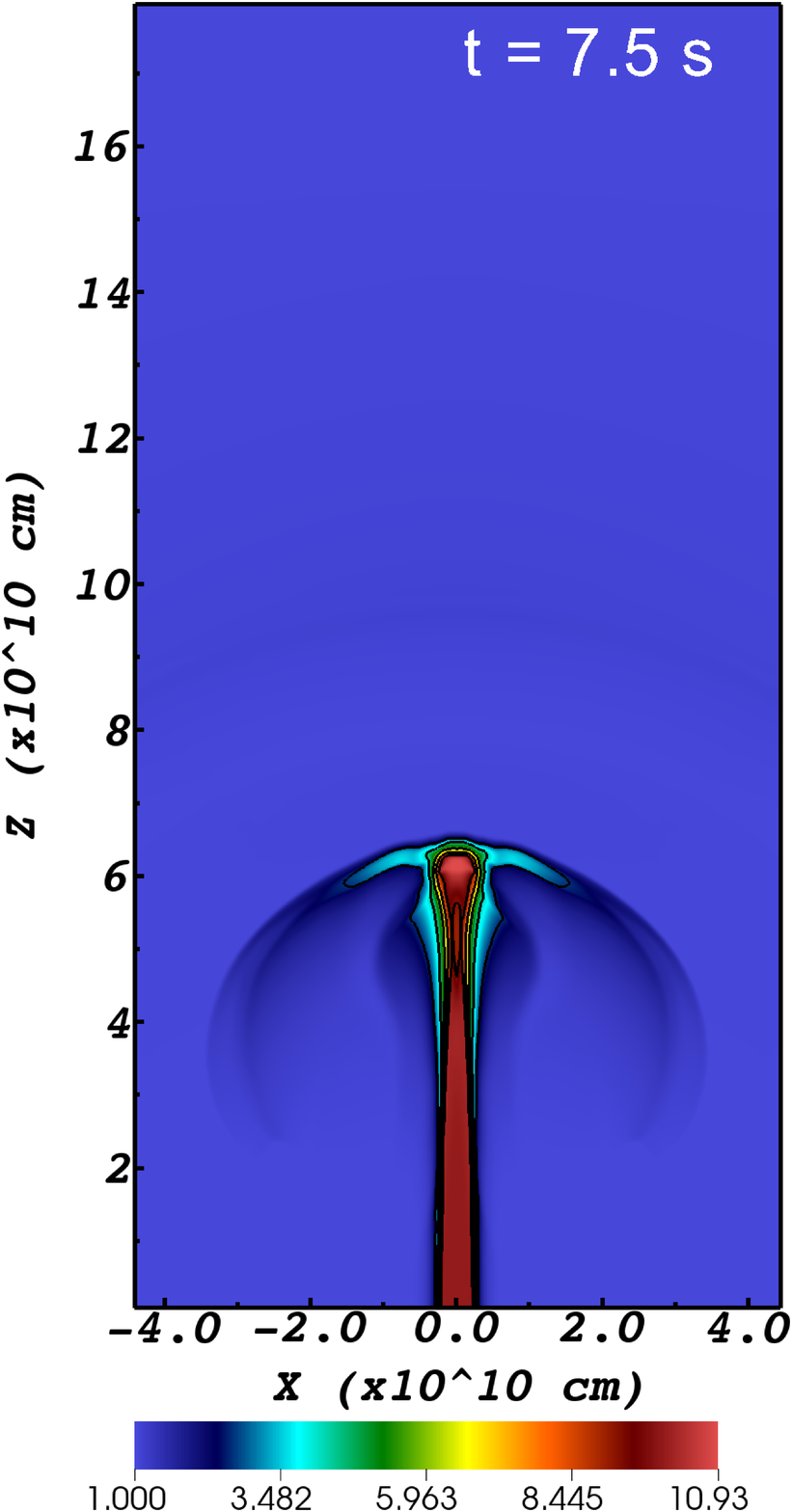}
\includegraphics[width=0.235\textwidth,height=0.35\textheight]{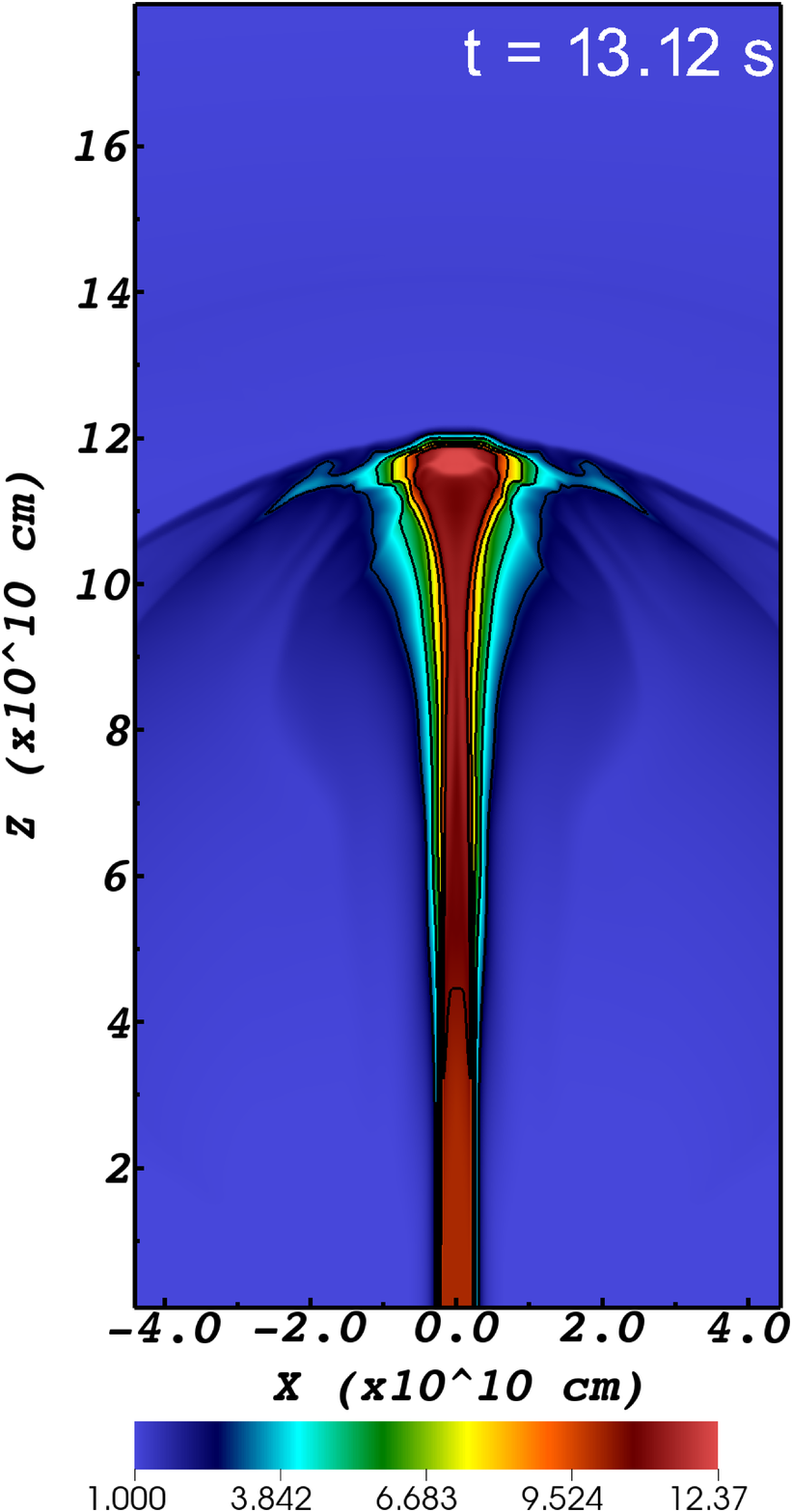}
\includegraphics[width=0.235\textwidth,height=0.35\textheight]{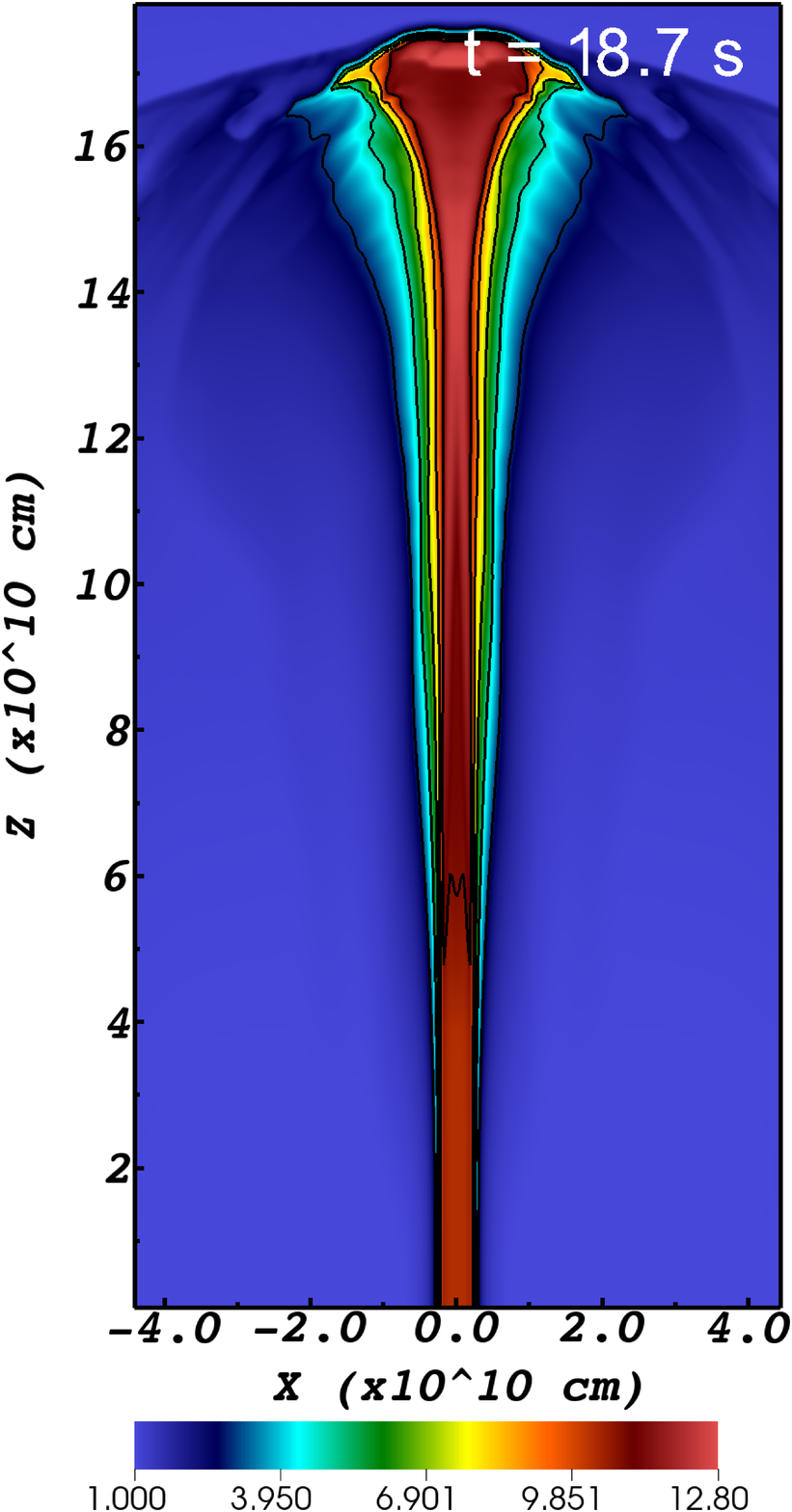}
\caption{Model 5. Snapshots of the rest mass density (top) and the Lorentz factor (bottom) at different times. This is  the radiation-pressure-dominated case.} \label{fig:16ti-10radrho}
\end{figure*}

\begin{figure}%[htb]
\includegraphics[width=0.4\textwidth]{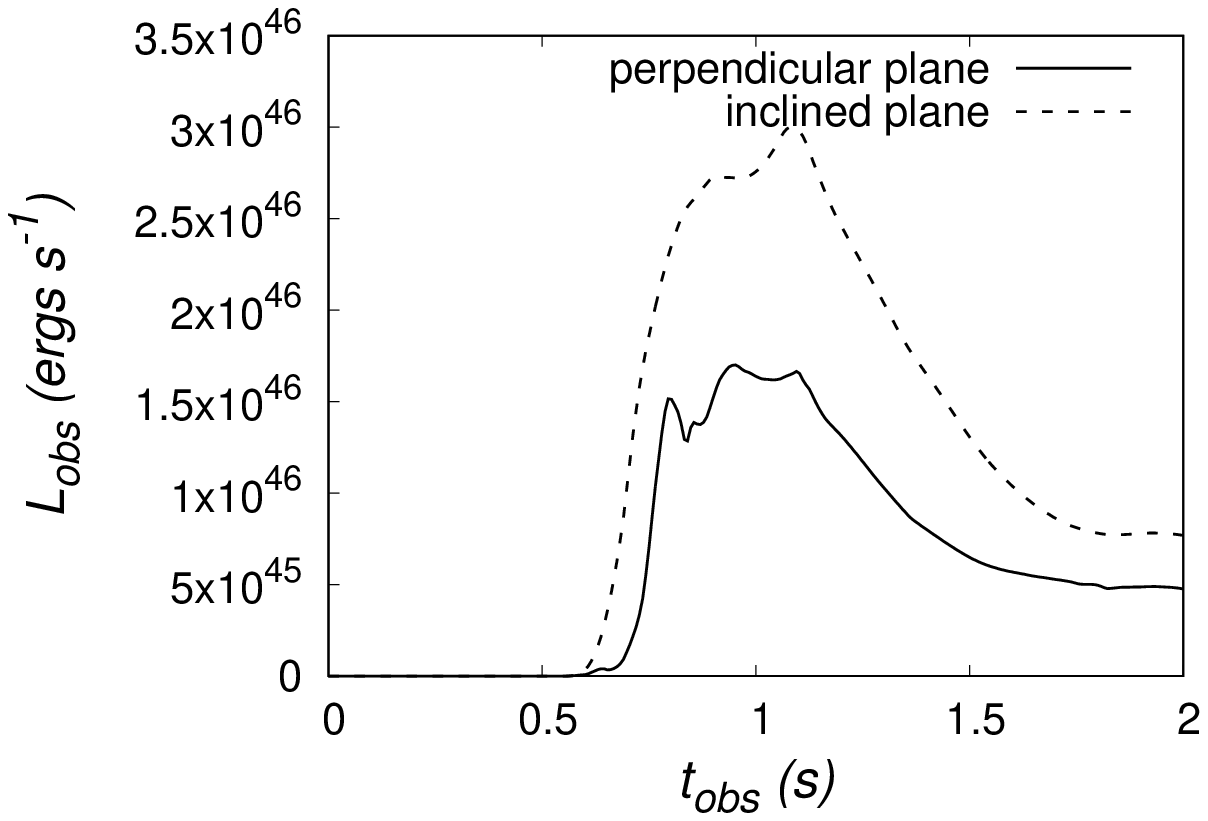}
\includegraphics[width=0.4\textwidth]{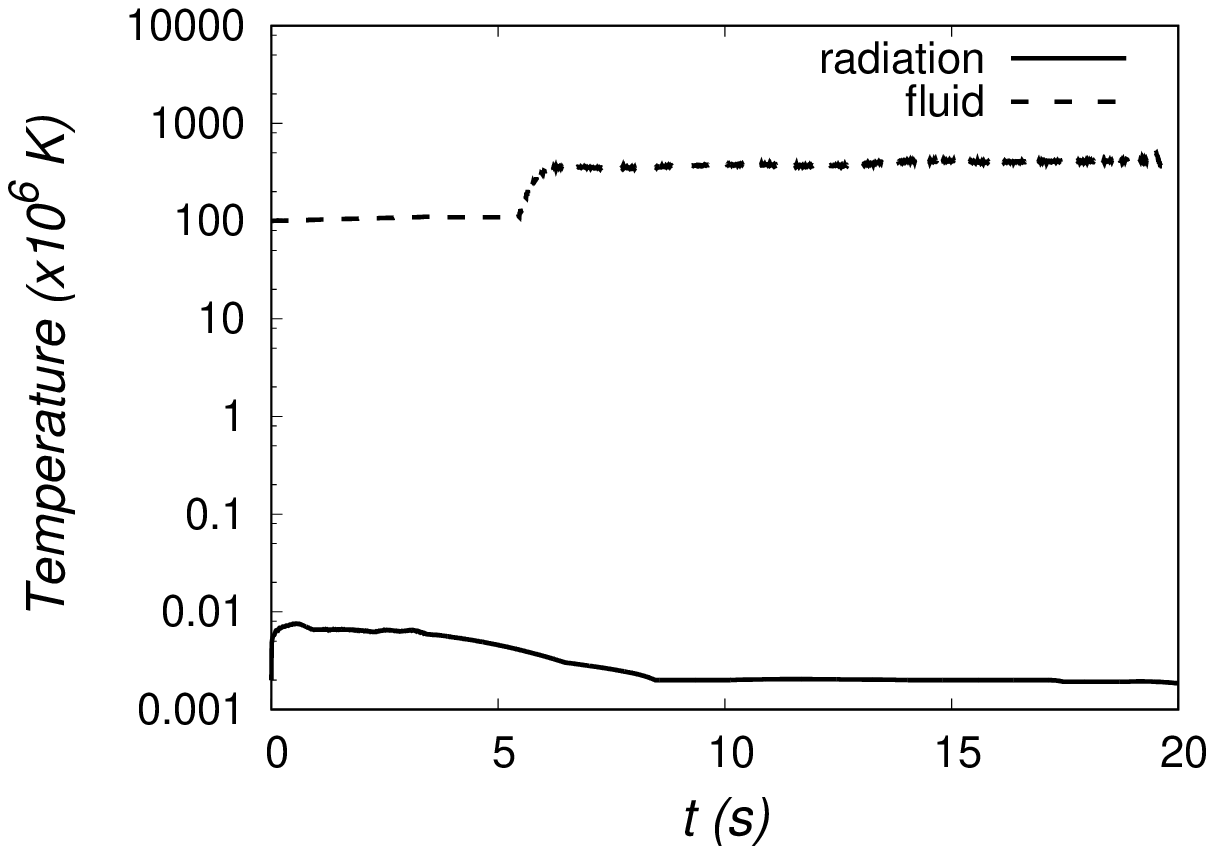}
\caption{Model 4. Top: The LCs measured in two different planes transformed to the observer's frame. The solid and dotted lines correspond to the perpendicular and inclined detector planes. Bottom: Maximum of both radiation and fluid temperatures. The solid and dotted lines correspond to the radiation and fluid temperatures, respectively, measured in the laboratory frame.} \label{fig:16ti-gas:lum}
\end{figure}

The LC  and  the difference of the gas and radiation temperatures for this model 4 are shown in Fig. \ref{fig:16ti-gas:lum}. The first peak in the luminosity measured by a perpendicular plane is produced by the breakout of the jet from the stellar surface. The main peak is due to the one produced by the material near the working surface when it crosses the detector location. The main characteristic of this luminosity curve is that the amplitude of the LC lies within the LLGRBs range ($10^{46} - 10^{49} \ {\rm erg ~s}^{-1}$). The radiation and fluid temperatures have a  behaviour similar to those measured in the previous section for LGRBs, that is, the fluid temperature is higher than the radiation temperature. Nevertheless, the fluid temperature is of the order of $10^8 \ {\rm K}$, whereas the radiation temperature is of the order of $10^4 \ {\rm K}$. Unlike in the jets evolving on the stratified medium from the previous section, in this case, the gas and radiation are not as close to thermal equilibrium, showing a difference in temperature, including four orders of magnitude. It is worth noticing that the opacities used in our simulations are appropriate for this range of temperature.

\begin{figure}%[htb]
\includegraphics[width=0.4\textwidth]{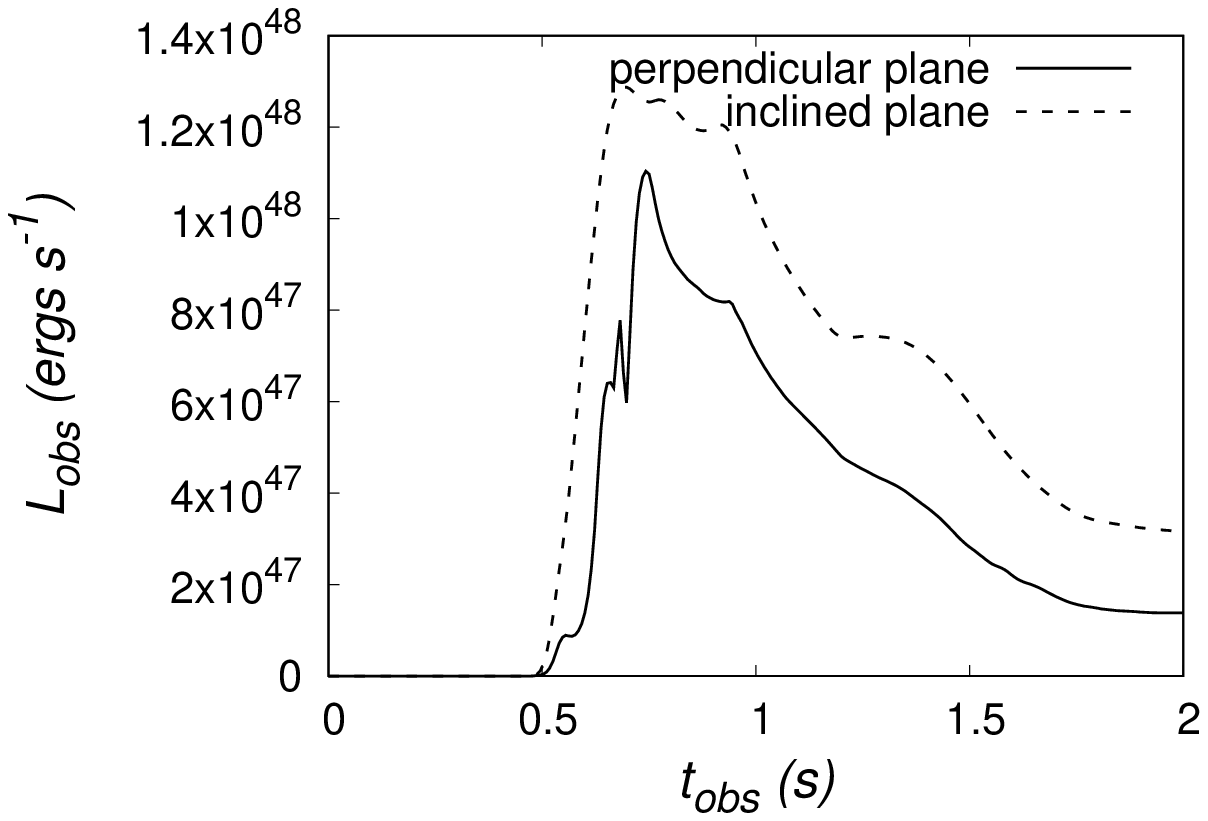}
\includegraphics[width=0.4\textwidth]{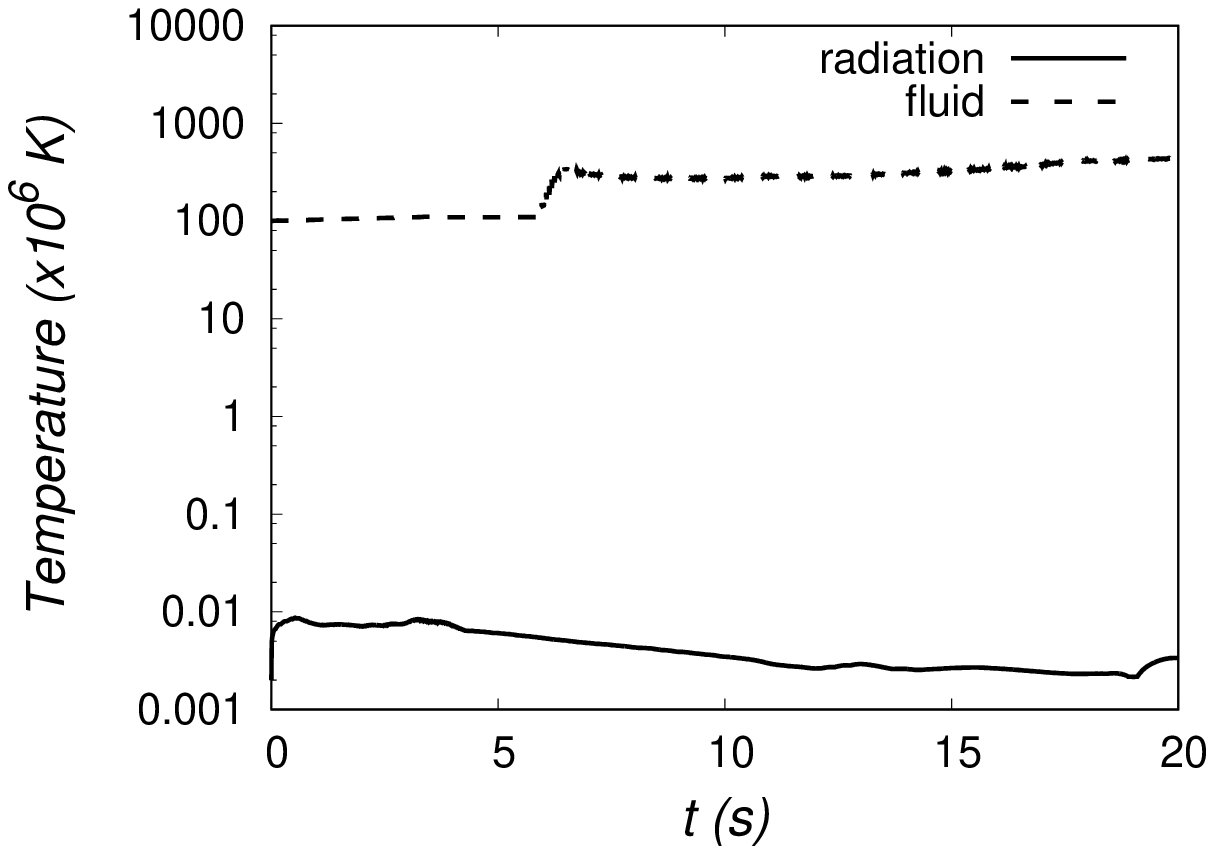}
\caption{Model 5. Top: The LCs measured in two different planes transformed to the observer's frame.  The solid and dotted lines correspond to the perpendicular and inclined detector planes. Bottom: Maximum of both radiation and fluid temperatures. The solid and dotted lines correspond to the radiation and fluid temperatures, respectively, measured in the laboratory frame.} \label{fig:16ti-rad:lum}
\end{figure}

\begin{figure}%[htb]
\includegraphics[width=0.4\textwidth]{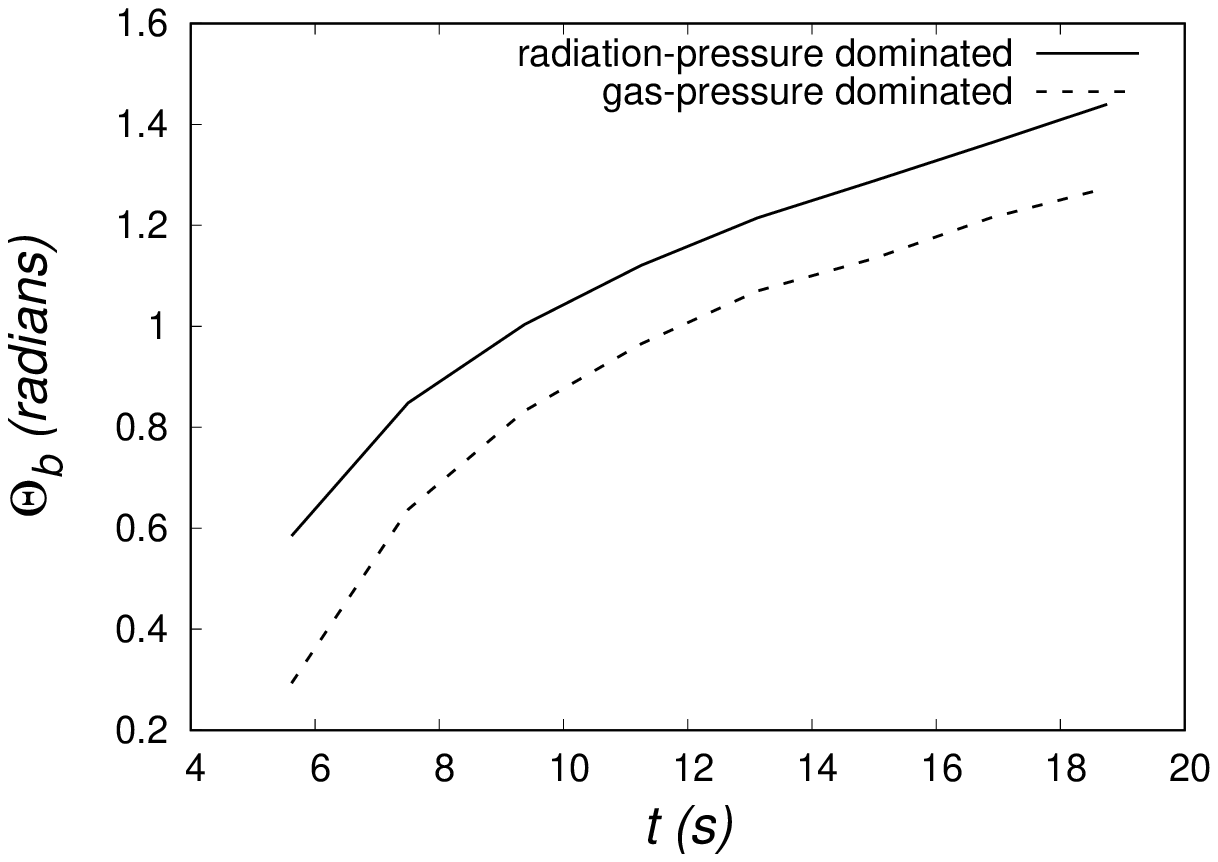}
\caption{Opening angle as a function of time after the jet breakout. The solid and dotted lines correspond to models 5 and 4, respectively, measured in the laboratory frame.} \label{fig:angles}
\end{figure}

The dynamical evolution of model $5$ is similar to that of model $4$. The results are shown in Fig. \ref{fig:16ti-10radrho} for model $5$ at the same instances of times of model $4$. At $t \sim 3 \text{s}$, before the jet reaches the surface of the progenitor star, the reverse shock produced by the jet/stellar-envelope interaction forms a cocoon and the pressure in the cocoon confines the jet. At this time, the only difference with respect to the jet in the gas-pressure dominated scenario is that in this case, the jet is propagating faster.

At $t = 7.5 \ {\rm s}$, we can see a collimated beam after the jet breaks out the progenitor star, and at this time there are two important differences with respect to the model $4$. The first one is that the collimation shock appears at $\sim 2\times 10^{10} \text{cm}$ further ahead on the beam jet. The second  is  the jet opening angle $\Theta_\text{b}$ that we show for models $4$ and  $5$ in Fig. \ref{fig:angles}. This shows that the jet opening angle not only depends on the initial Lorentz factor \cite{MizutaIII} but on other variables as well, such as the radiative energy density.  At $t = 13.12 \ {\rm s}$, a collimated jet continues propagating through a stratified external medium. In this case, the pressure of the cocoon is enough to keep the structure collimated. As time goes on, $t= 18.7 \text{s}$, the matter continues flowing along the forward direction, and the jet remains collimated.

In the bottom  of Fig. \ref{fig:16ti-10radrho}, we show the evolution of the  Lorentz factor along the progenitor and external medium. By $t = 3.75 \ {\rm s}$ the jet is still within the star, and the jet Lorentz factor near the zone of injection grows, whereas the Lorentz factor in the head of the jet decreases. When the jet breaks out the surface of the star, the jet not only continues propagating faster compared to the jet of model $4$ but with a Lorentz factor slightly bigger.

\begin{figure}%[htb]
	\centering
\includegraphics[width=0.2\textwidth]{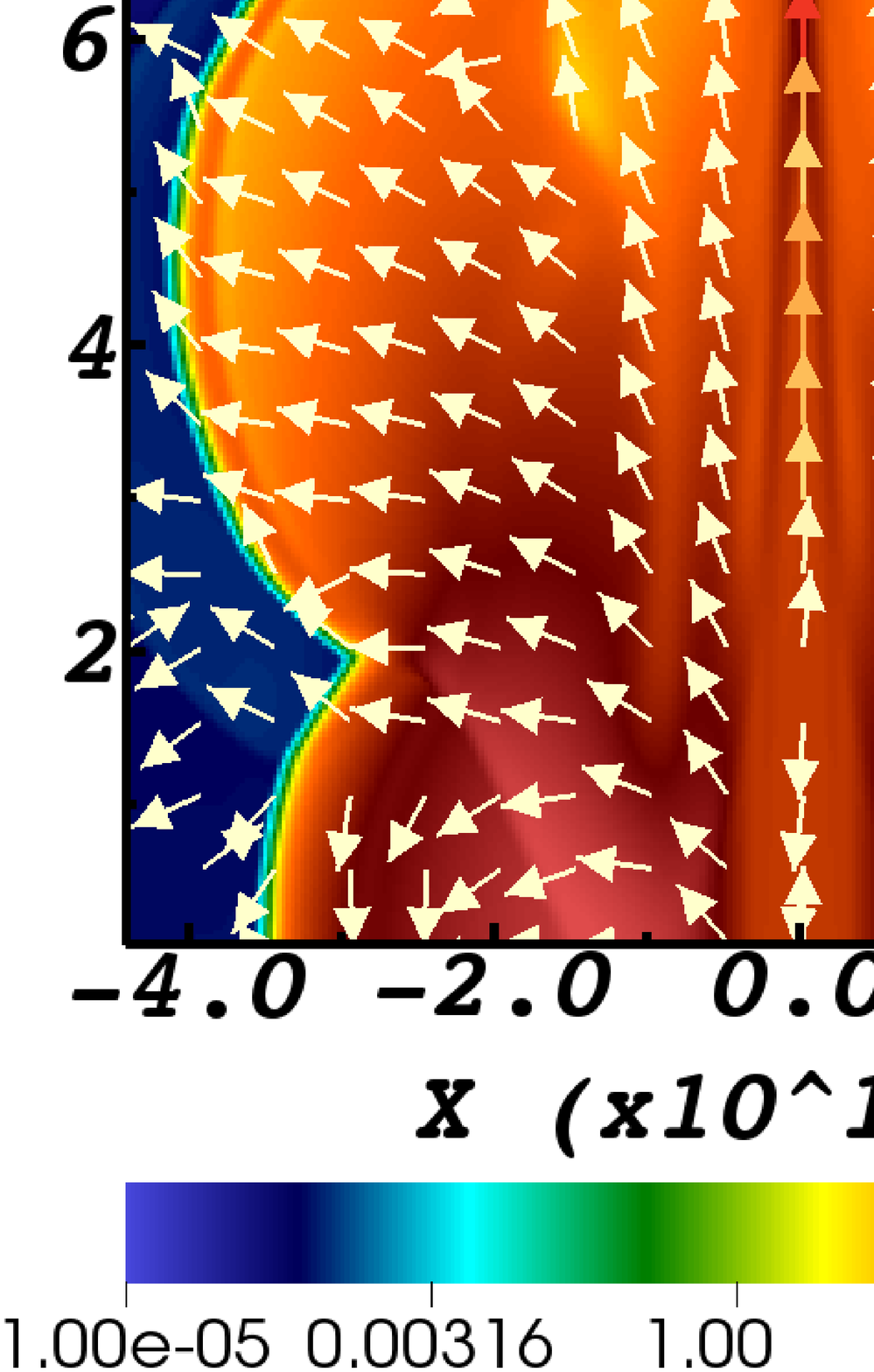}
\includegraphics[width=0.2\textwidth]{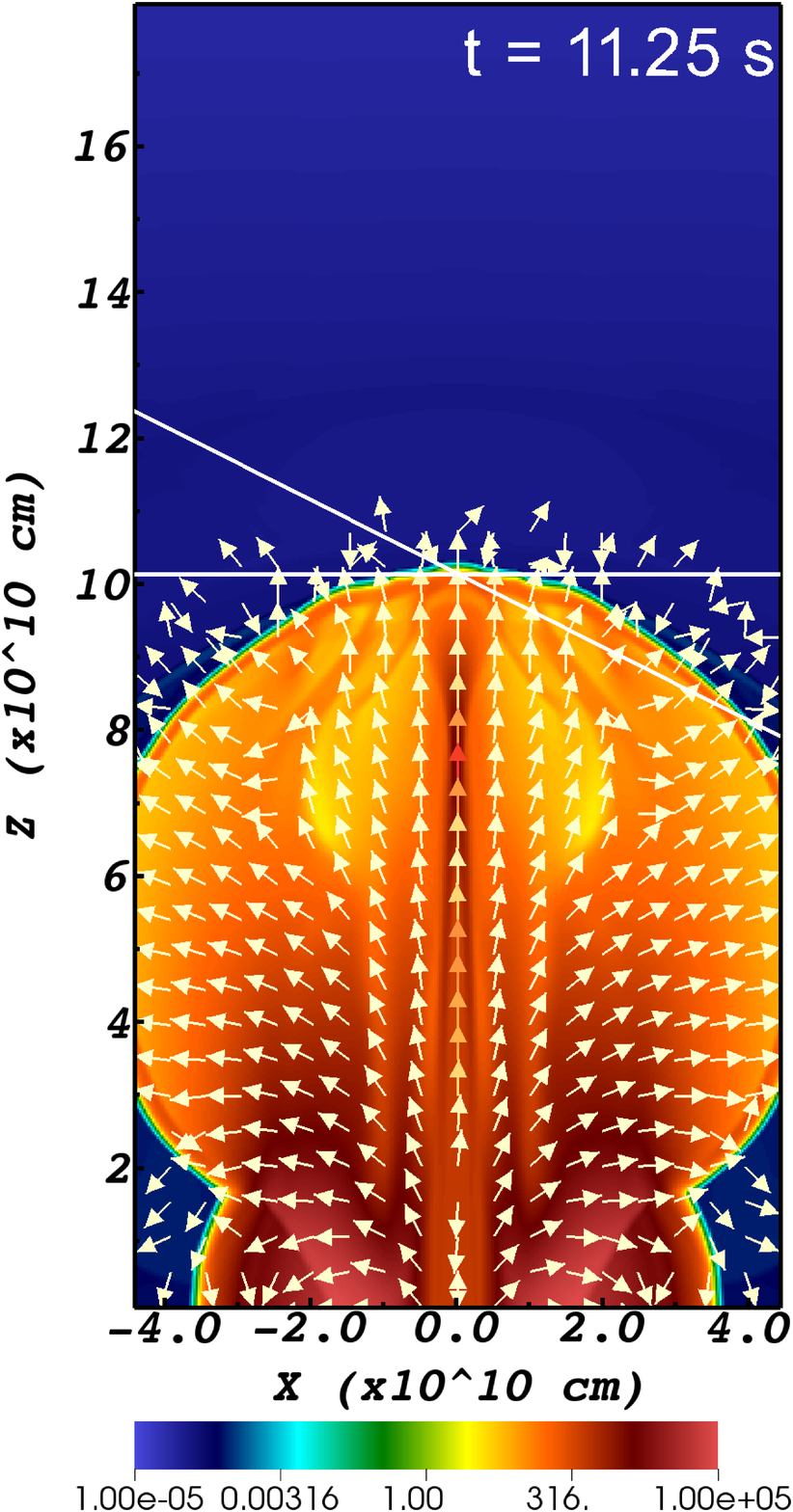}
\caption{Model 5. The pseudo-colour slice represents the rest-mass density of the outflow, and vectors represent the vectorial field of the radiation flux, measured in the laboratory frame. We can see how the radiation flux moves ahead of the GRB jet, and also how it arrives first at the inclined plane. The white lines indicate the position of the perpendicular and inclined detecting planes.} \label{fig:16ti-rad}
\end{figure}

In Fig. \ref{fig:16ti-rad:lum}, we show the LCs for  model $5$. The initial injected energy arrives to the detector with sufficient energy as to obtain the LCs of the order of $10^{48} \ {\rm ergs} \ \text{s}^{-1}$, two orders of magnitude bigger than the LCs of model $4$.  The LC measured by the inclined plane shows a  plateau following the first main peak. Our results suggest that this plateau regime is due to two effects (see Fig. \ref{fig:16ti-rad}): (1) The radiation flux released during the evolution that moves ahead of the GRB jet and interact with the surrounding medium before the outflow drives a shock wave into the external medium and (2) the radiation flux emitted by the GRB jet is detected by an observer in the inclined plane (${\cal O}_1$) before an observer in the perpendicular plane (${\cal O}_2$). Thus, the ${\cal O}_1$ measures a radiation flux, not only before ${\cal O}_2$ but also before the external shock of the jet crosses the inclined plane. This allows  the rarefaction signal to travel inwards towards the jet axis and carry radiation flux that contributes to the amplitude of the LC and thereby slow down the decay of the luminosity.  Consequently, the rarefaction signal seen by ${\cal O}_2$ arrives with a delay  that is enough for the radiation flux to decrease, making ${\cal O}_2$ aware of a deficit of the radiation flux with respect to ${\cal O}_1$, and therefore its LC starts to fall  more quickly. 

\begin{figure}%[htb]
\includegraphics[width=0.4\textwidth]{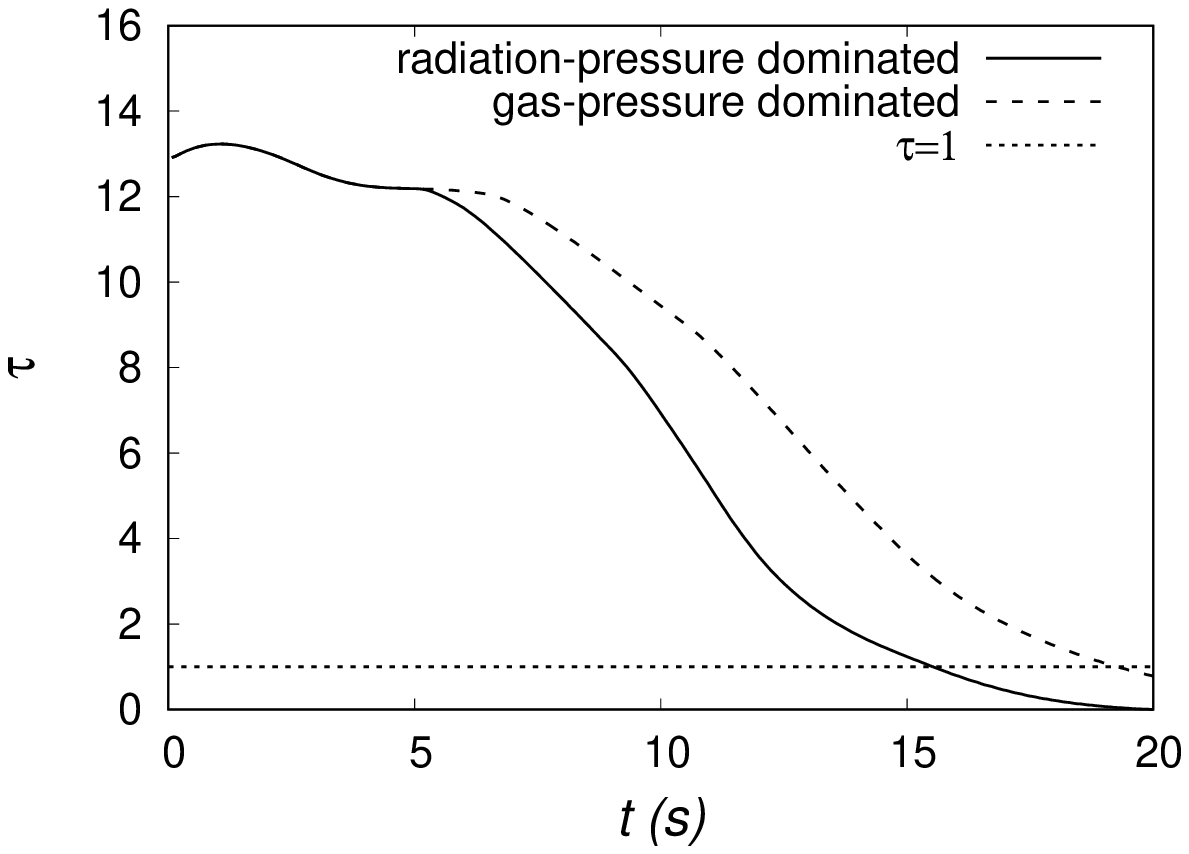}
\caption{Optical depth $\tau$ for models 4 and 5, integrated all along the $z$-axis and measured in the laboratory frame. The solid and dotted lines correspond to the models 5 and 4, respectively. We also plot $\tau=1$ as a threshold for optically thick and thin cases.} \label{fig:tao16ti}
\end{figure}

The bottom of Fig. \ref{fig:16ti-rad:lum} shows the maximum temperatures, in this case the fluid temperature is about $\sim 10^ 8 \ {\rm K}$, whereas the radiation temperature is around $\sim 10^4 \ {\rm K}$, which shows the difference of temperatures during the evolution.
 Finally, in Fig. \ref{fig:tao16ti}, we show the evolution of the optical depth $\tau$ for models 4 and 5, where we can see the transition from optically thick ($\tau >1$) to  optically thin ($\tau <1$) regimes. Notice that the optical depth associated with the radiation-pressure-dominated scenario, becomes optically thin before the gas-pressure-dominated case. This result is in accordance with the fact that the radiation boosts the jet in the radiation dominated model 5.

% --------------------------------------------
% ---------->     SECTION     <----------
% --------------------------------------------

\section{Discussion and conclusions}
\label{sec:conclusions}

We implemented an RRH code in 3D, with the main objective of constructing the LCs produced by jets. The LC is calculated by the integration of the radiation flux, which is fully coupled to the hydrodynamics during the simulations. We present essential tests that validate our code and applications to the jet propagation in stratified toy media and on to a progenitor star model.

As a first application of our code, we considered a model of LGRBs jet evolving through a stratified surrounding medium. For the surrounding medium, we use density and pressure profiles that decrease as power law $\sim r^{-2}$. We also study the dynamics of jets propagating through a progenitor star using the 16TI progenitor model. The simulations assume LTE between the fluid and the radiation field initially. For the definition of the jet, one needs nine parameters: three components for the velocity, the rest-mass density, pressure of matter for hydrodynamics, three components for the radiative flux, and the radiated energy density for the radiation. In particular, we explored the regime in which the jet goes along $z-\text{axis}$ with a highly relativistic velocity. We have combined each one of the jets with values of radiated energy density that created scenarios, where the radiation pressure or gas pressure are dominant, one at a time. Our model is also restricted to a single frequency and opacities associated with free-free, bound-free, bound-bound, and electron-scattering precesses but can be extended to other scenarios.

 In order to close the RRH system of equations, we use a constant adiabatic law EoS for the fluid, while for the radiation field, we use the M1-closure relation. The use of a more realistic EoS for the fluid in stellar models could provide interesting outcomes because the fluid temperature that could be obtained would be different. While in purely hydrodynamical models, the temperature is an auxiliary quantity, for an RRH model, it is essential because it may substantially change the opacities and subsequently the  dynamics  of the GRB jets. On the other hand, we try -as a first approach- to describe the dominant physical processes that contribute to the opacity in stellar interiors, using a classical approximation given by Kramers opacities. This simple approximation provides a qualitative and quantitative description of how important fully coupling the radiation with the hydrodynamics is. 

Under these conditions, we have compared the evolution of jet models with and without the coupling of hydrodynamics to the radiation field and have also shown the effects of gas and radiation-pressure domination. We have found that when gas pressure dominates, the dynamics of the jet is pretty similar to its purely hydrodynamical counterpart. On the other hand, when the radiation pressure is dominant, the effect of radiation is noticeable in comparison to gas pressure and purely hydrodynamics versions because the radiation field acts as a boost accelerating the material density around. This effect is more important in the case of jets propagating across the progenitor density model.

Regarding the luminosity LCs, depending on the combination of radiated energy density, the maximum amplitude of the luminosity lies within the range $\sim [10^{50}-10^{52}] \ {\rm erg} ~ {s}^{-1}$ for the jets propagating along the stratified medium. The scenario with the smallest amplitude is the gas pressure dominated, whereas the biggest amplitude is achieved when the jet is dominated by radiation pressure. This is physically consistent with the fact that the energy injected in the radiation-pressure-dominated scenario is bigger by two orders of magnitude than in the scenario, where the gas pressure dominates.

Additionally, we compute the maximum of the radiation and fluid temperatures for each of the jets, found to be right behind the working surface. For the gas-pressure-dominated scenario the fluid temperature is of $\sim 10^8 \ {\rm K}$, and the radiation temperature is around one order of magnitude smaller. In the case where the radiation pressure dominates, the gas temperature is of order of $10^{9} \ {\rm K}$ and the radiation temperature is $\sim 10^{8} \ {\rm K}$. An important point here to highlight is that during a large part of the evolution, the fluid temperature is bigger than the radiation temperature, which is consistent with the notion that radiation carries energy and acts as a fluid-cooling mechanism.

We also applied our code to evolve jets triggered inside a progenitor. We  verified that the jets propagate inside the progenitor and successfully breakout the surface. For our initial conditions, the LCs peak are of order of $\sim 10^{48} \ {\rm erg} ~ {s}^{-1}$ and $\sim 10^{46} \ {\rm erg} ~ {s}^{-1}$ for the radiation-pressure- and gas-pressure-domination scenarios, respectively. This is comparable to the luminosity of LLGRBs. Similarly to the previous scenarios, the fluid temperature is bigger than radiation temperature, however, in this case, the difference is of  four orders of magnitude. Even though initially the gas and radiation are assumed in thermal equilibrium, during the evolution the temperature difference between the two reaches four orders of magnitude, which indicates how far from thermal equilibrium these components are.

% ----->     ACKNOWLEDGMENTS     <-----

\section*{Acknowledgments}

We appreciate the comments and recommendations from the members of the Valencia Group and  from the anonymous Referee. This research is supported by grants CIC-UMSNH-4.9 and CONACyT 258726 (Fondo Sectorial de Investigaci\'on para la Educaci\'on). Most of the simulations were carried out in the computer farm funded by CONACyT 106466 and the Big Mamma machine of the Laboratory of Artificial Intelligence at the IFM. The authors also acknowledge the computer resources, technical expertise and support provided by the Laboratorio Nacional de Superc\'omputo del Sureste de M\'exico, CONACyT network of national laboratories. We also thank ABACUS Laboratorio de Matem\'aticas Aplicadas y C\'omputo de Alto Rendimiento del CINVESTAV-IPN, grant CONACT-EDOMEX-2011-C01-165873, for providing computer resources.

% -----------------     APPENDIX     ---------------------
\begin{appendix}

\section{Basic tests}
\label{sec:NT}

We only present the standard tests in two dimensions, because the standard 1D tests of our code were presented in \cite{Panchos}. These tests show the performance of our code in both,  optically thick and optically thin regimes. The initial conditions for the tests are the following:
 
\begin{enumerate}
\item \textit{Single beam test}: This test is intended to verify that our code can work properly in optically thin media, where gas and radiation are decoupled ($\kappa_a=\kappa_{total}=0$). This test consists in injecting a simple beam of radiation and checking that the beam does not present any rupture during its evolution. The test was solved in the plane $z=0$ on a $31\times31$ grid. The boundary conditions for all borders are outflow, except in the given region delimited by $y\in[0.4,0.6]$, where the beam is injected with energy density $100$ times larger than that of the environment. The value of the $a_\text{r}$ and adiabatic index of the gas are $1.118\times10^{17}$, code units, and $4/3$, respectively.  

In Fig. \ref{fig:t1}, we can see, at $t=10$, that radiation beam through the whole domain without presenting any rupture. The standard snapshot can be compared with \cite{Sadowski}. 

\begin{figure}%[h]
\centering
\includegraphics[width=8.9cm]{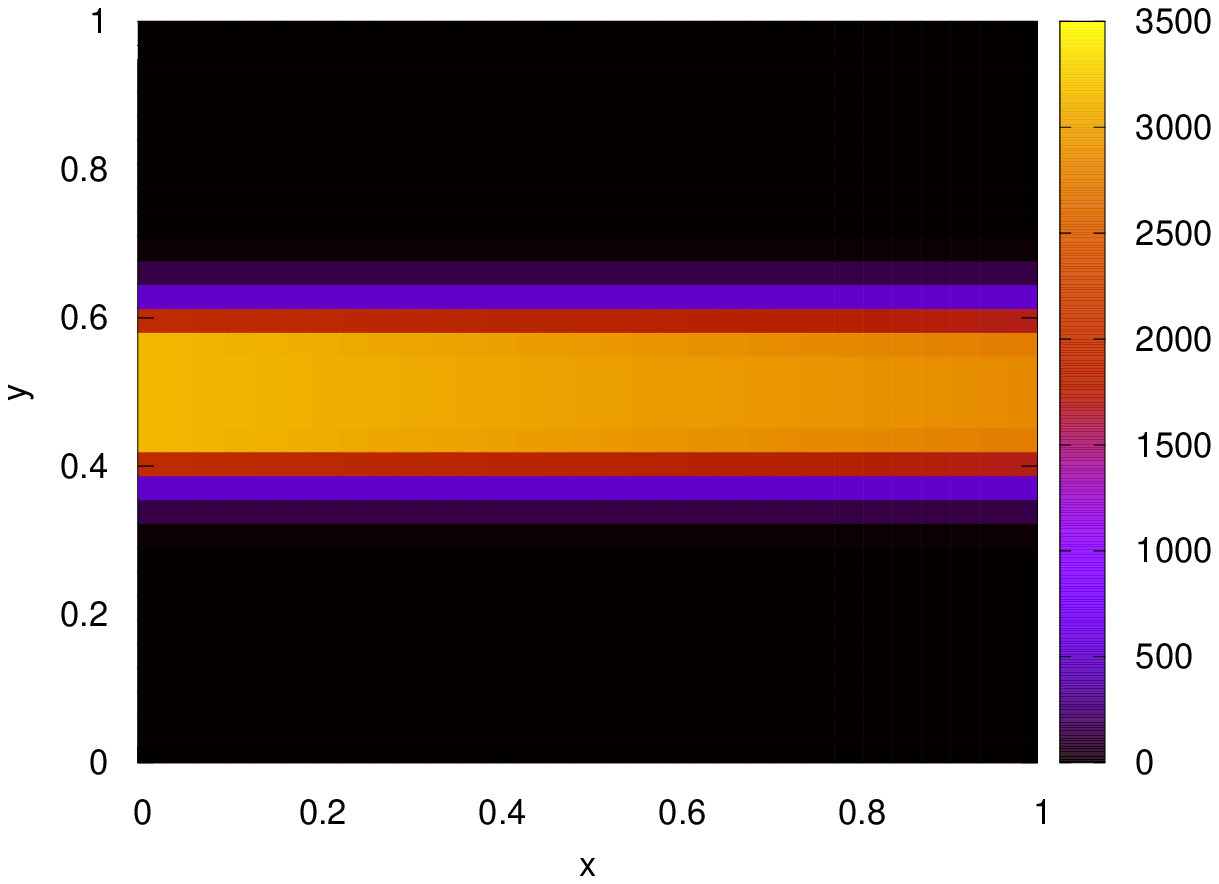}
\caption{\label{fig:t1} Radiation energy density at $t=10$. The radiation beam is located at the left boundary.}
\end{figure}

\item \textit{Shadow}: In order to verify that the M1 approximation works properly, and to illustrate the difference between the M1 and Eddington approximations we solve this problem. The test consists of an optically thick gas lego circle, immersed in an optically thin environment. We solve the test on the plane $z=0$ on a $100\times50$ grid, with a fixed mass density within a lego circle given by  

\begin{equation}
    \rho_0 = \rho_a + (\rho_\text{b} - \rho_a)e^{(-\sqrt{x^2+y^2+z^2}/\omega^2)},
\end{equation}

\noindent where $\rho_a=10^{-4}$, $\rho_\text{b}=10^3$ and $\omega=0.22$. Initially the system is in thermal equilibrium, and has velocities and radiative fluxes equal to zero. The boundary conditions are as follows: inflow at the left border and outflow at the right border. On all other borders, we use periodic boundary conditions. At the border where we imposed the inflow. The values for radiated energy density, radiated flux and gas temperature are $E_L=a_\text{r}T^4_{g,L}$, $F^x=0.99999E_L$, y $T_{g,L}=100T_a$, respectively. The value of $a_\text{r}=351.37$ and $\kappa_a=\kappa_{total}=\rho_0$.

The gas temperature is given such that the pressure is constant throughout the domain,

\begin{equation}
 	T_g = T_a\frac{\rho_a}{\rho_0},
\end{equation}

\noindent with an adiabatic index $\Gamma=1.4$.

In Fig. \ref{fig:t2}, we show the results when the incoming radiation beam passes through the entire domain and reaches a steady state ($t\sim10$). In the upper panel, we show the solution obtained with the Eddington approximation. This approximation treats the radiation field isotropically, as consequence the radiation diffuses rapidly behind the sphere and a shadow cannot be formed. In the lower panel, we show the solution with the M1 approximation, contrary to the Eddington approximation, here we can see that a shadow is formed behind the circle because it is designed to keep moving the flow parallel to itself in optically thin regions for $F_\text{r}\approx E_\text{r}$. The standard snapshot can be compared with  \cite{Sadowski}.
\begin{figure}%[h]	
\centering
\includegraphics[width=8.9cm]{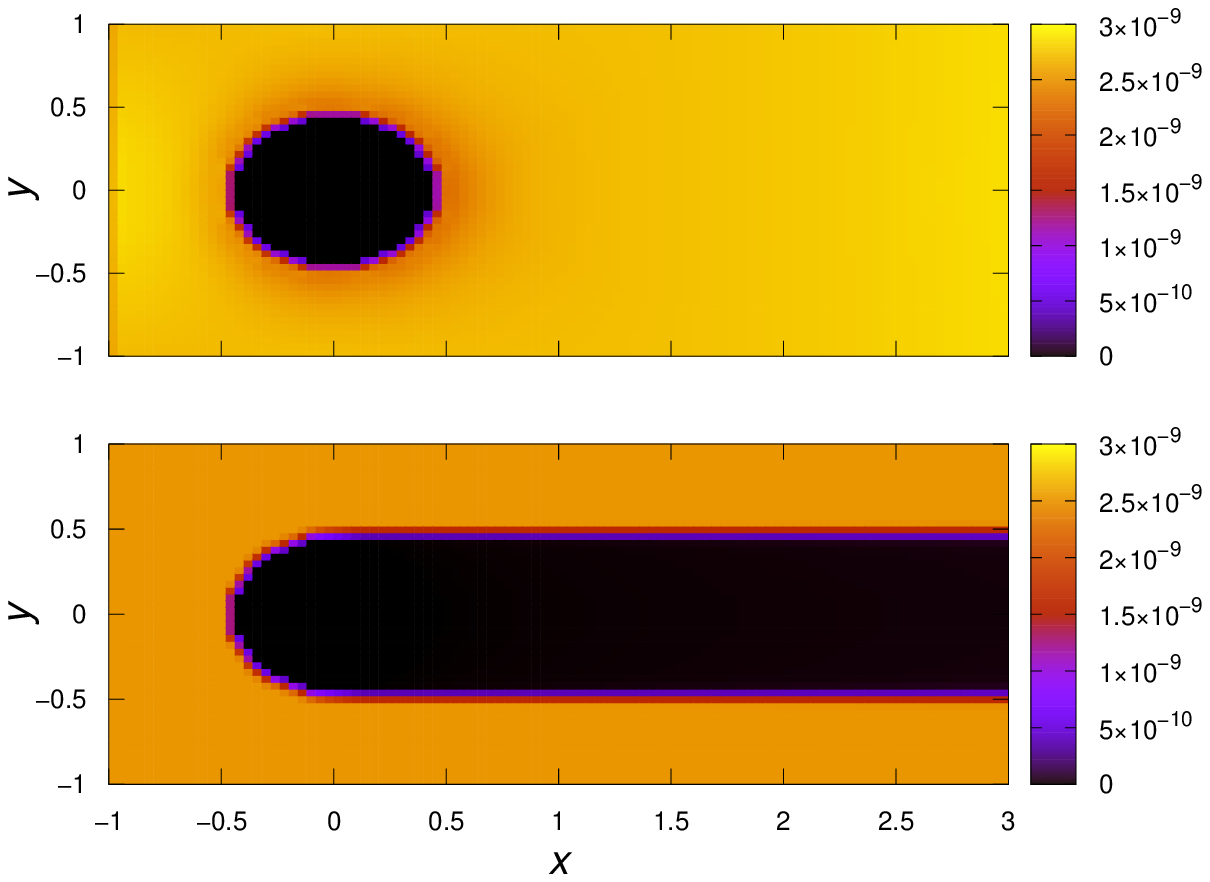}
\caption{\label{fig:t2} Radiation energy density at $t=10$. The source of radiation is located at the left boundary. Top: result corresponding to the Eddington approximation. Bottom: result corresponding to the M1 approximation.}
\end{figure}

\item \textit{Double shadow}: To test the performance and efficiency of our code with multiple sources of light, whose radiative flux is not parallel to the direction of propagation, we implement the double-shadow problem described in \cite{Sadowski}. In this test, a beam of light is injected into a static environment, where the photons move in different directions than the direction of propagation of the beam. The initial conditions for gas and radiation are exactly same as for the simple shadow test, but unlike that test, the beam is injected into the left border for $y>0.3$ with a radiative flux given by $F^x_\text{r}=0.93E_\text{r}$, $F^y_\text{r}=-0.37E_\text{r}$. We also establish a symmetry of reflection in $y=0$. As a consequence, the domain is illuminated by two radiation beams that intersect. In the region near the upper and lower left corners, where the beams do not overlap, the flow direction follows the direction imposed by the boundary conditions. In the region of the overlap, the density of radiative energy increases twice, whereas the flow is purely horizontal because the vertical component is cancelled in this region. The fact that the incident beam is inclined has an effect on the shadow produced behind the lego sphere. On the one hand, we have regions of partial shadow (penumbra) that result from perpendicular photons, while the region of total shadow (umbra) is limited by the edges of the penumbra. This test shows the limits of the M1 approximation that, in principle, does not limit the specific intensity of radiation to a particular direction. But in the case of multiple light sources, it should be used with caution, as seen in Fig. \ref{fig:t3}, the M1 approximation produces an extra horizontal shadow along the x-axis, where the penumbra overlaps. In this region, it is expected to be uniform and without shadow \cite{Jiang}. The standard snapshot, Fig. \ref{fig:t3}, can be compared with  \cite{Sadowski}. 

\begin{figure}%[h]
\centering
\includegraphics[width=8.9cm]{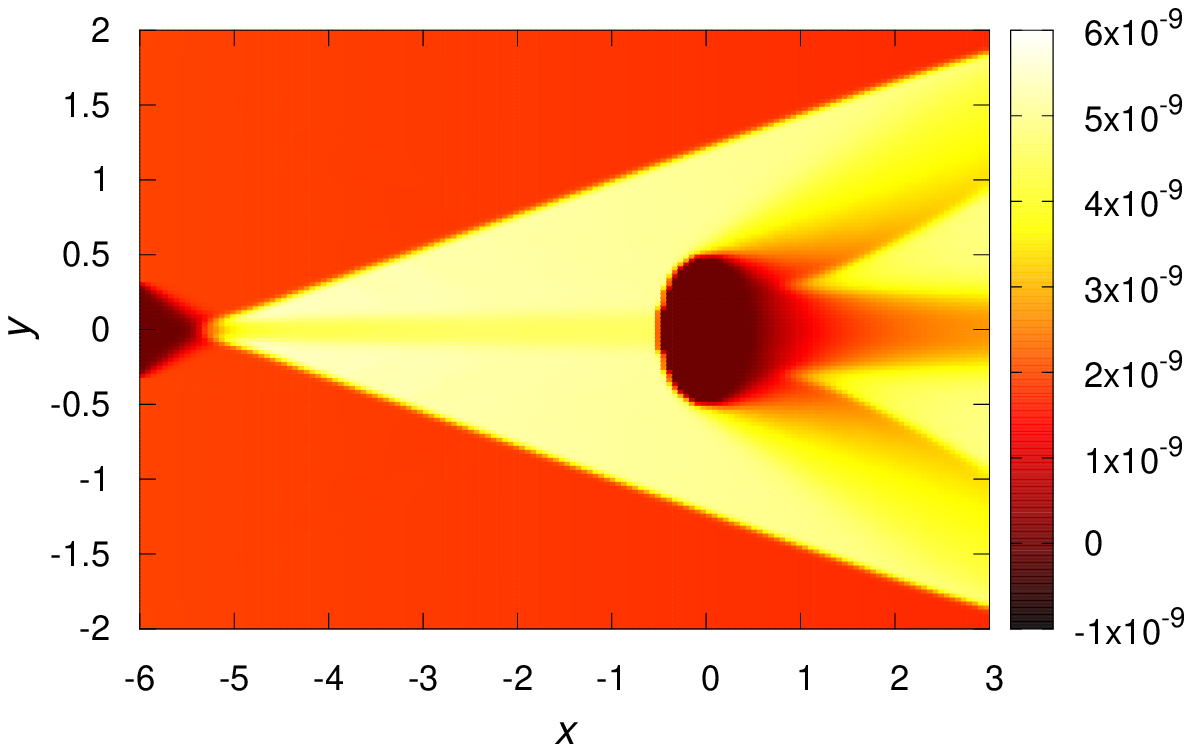}
\caption{\label{fig:t3} Radiation energy density at $t=10$. The source of radiation is located at the left boundary. This result corresponds to the M1-closure approximation.}
\end{figure}

\end{enumerate}

% -----> Subsection <-----
\section{Convergence test}
\label{subsec:rs}
 
In order to test the convergence of the evolution of the jets and select an adequate mesh resolution, we  performed a convergence test. For this we use three different resolutions: 
 low resolution which uses $6$ zones per beam radius ($\Delta x_{1}=1.333\times 10^8 \text{cm}$), 
 medium resolution that uses $8$ zones per beam radius ($\Delta x_{2}=3\Delta x_1/4$), employed in all the simulations listed in Tables \ref{t1} and \ref{t2} and 
high resolution with $10$ zones per beam radius ($\Delta x_{3}=3\Delta x_1/5$), the high resolution. These resolutions were chosen so that the simulations could be done during the time it takes the jet to  travel through the domain.

\begin{figure}%[htb]
\includegraphics[width=0.156\textwidth]{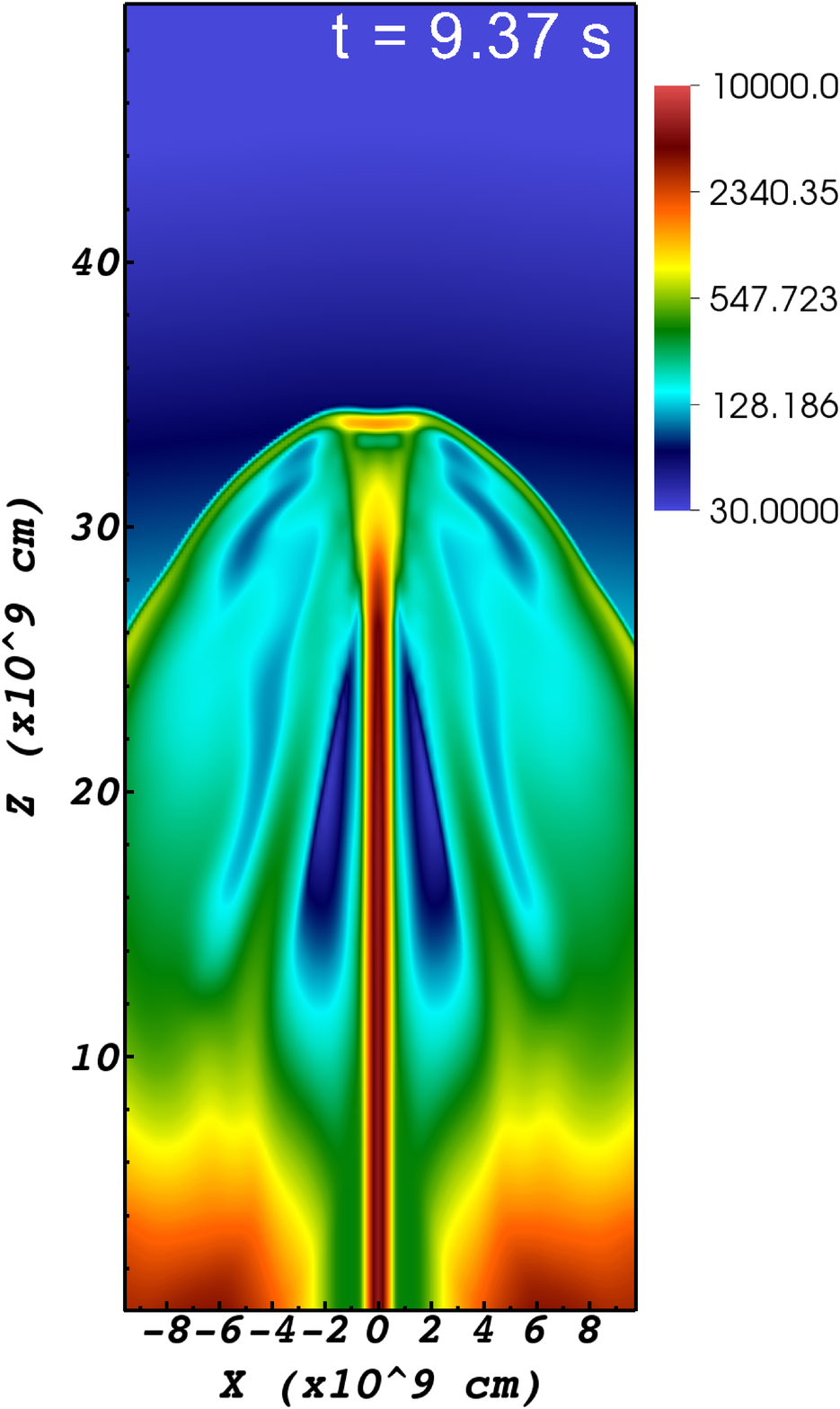}
\includegraphics[width=0.156\textwidth]{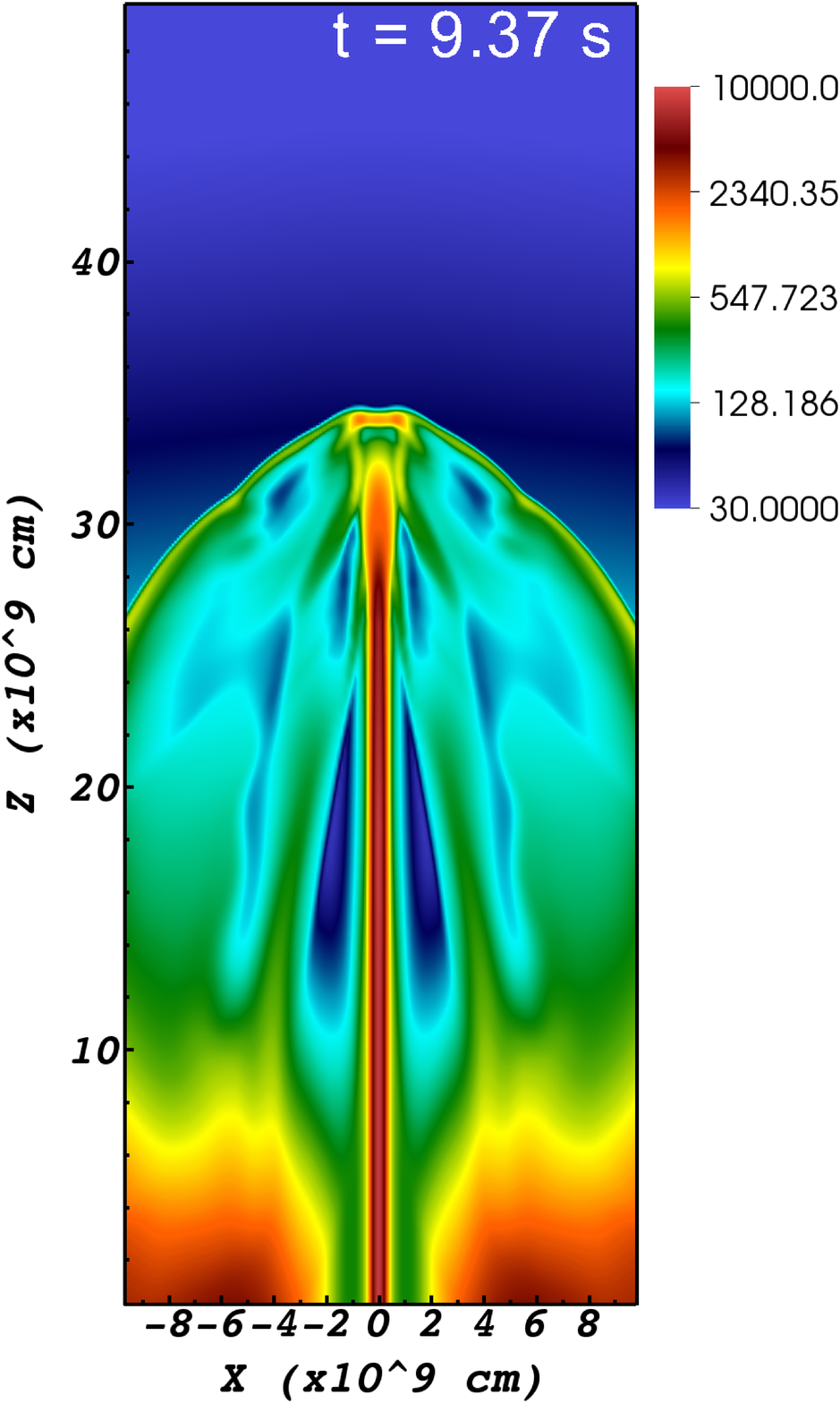}
\includegraphics[width=0.156\textwidth]{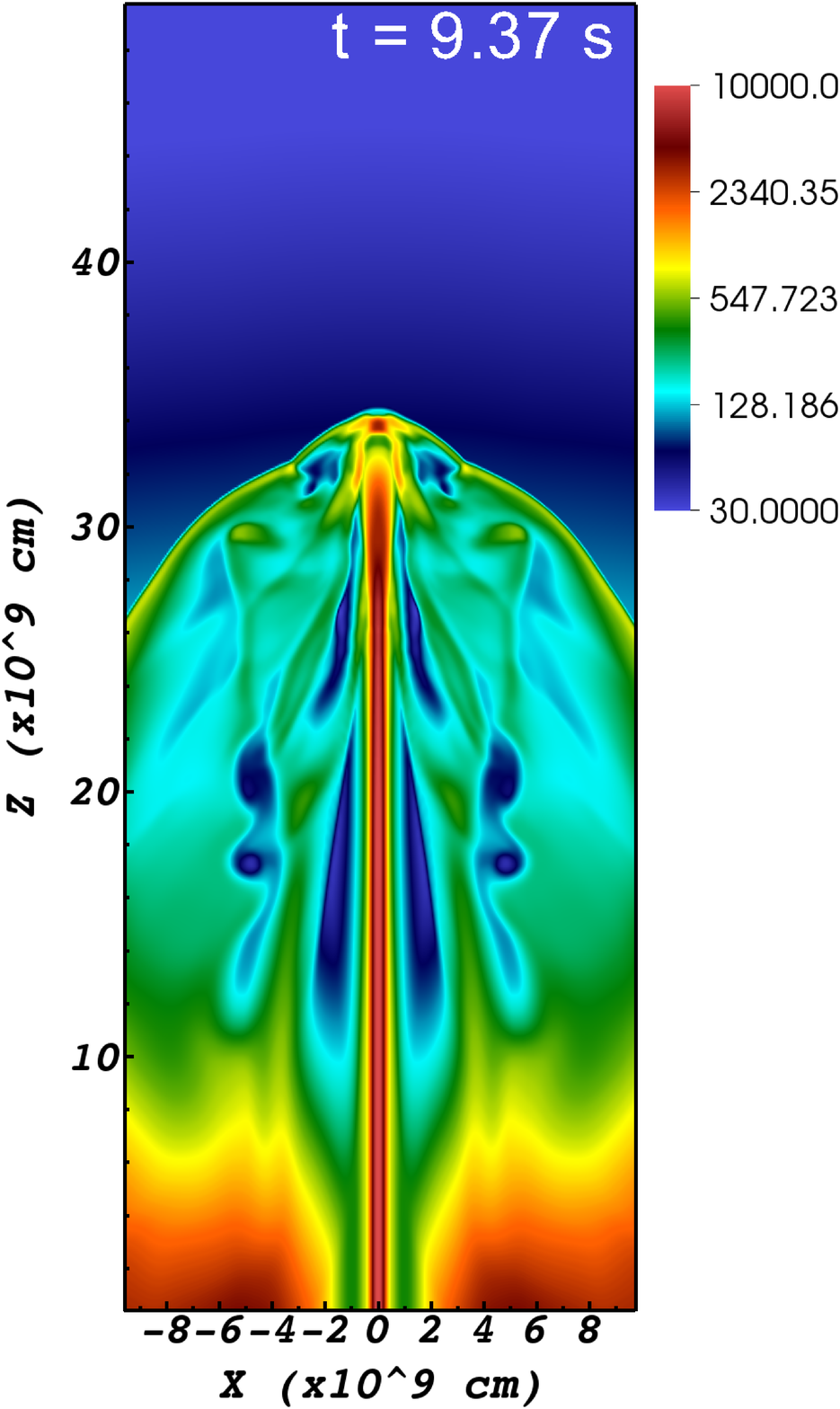}
\caption{Morphology of the rest-mass density for Model $3$  using three different resolutions. From left to right: lower, standard and high resolution.} \label{fig:Jets}
\end{figure}

\begin{figure}%[htb]
\includegraphics[width=0.4\textwidth]{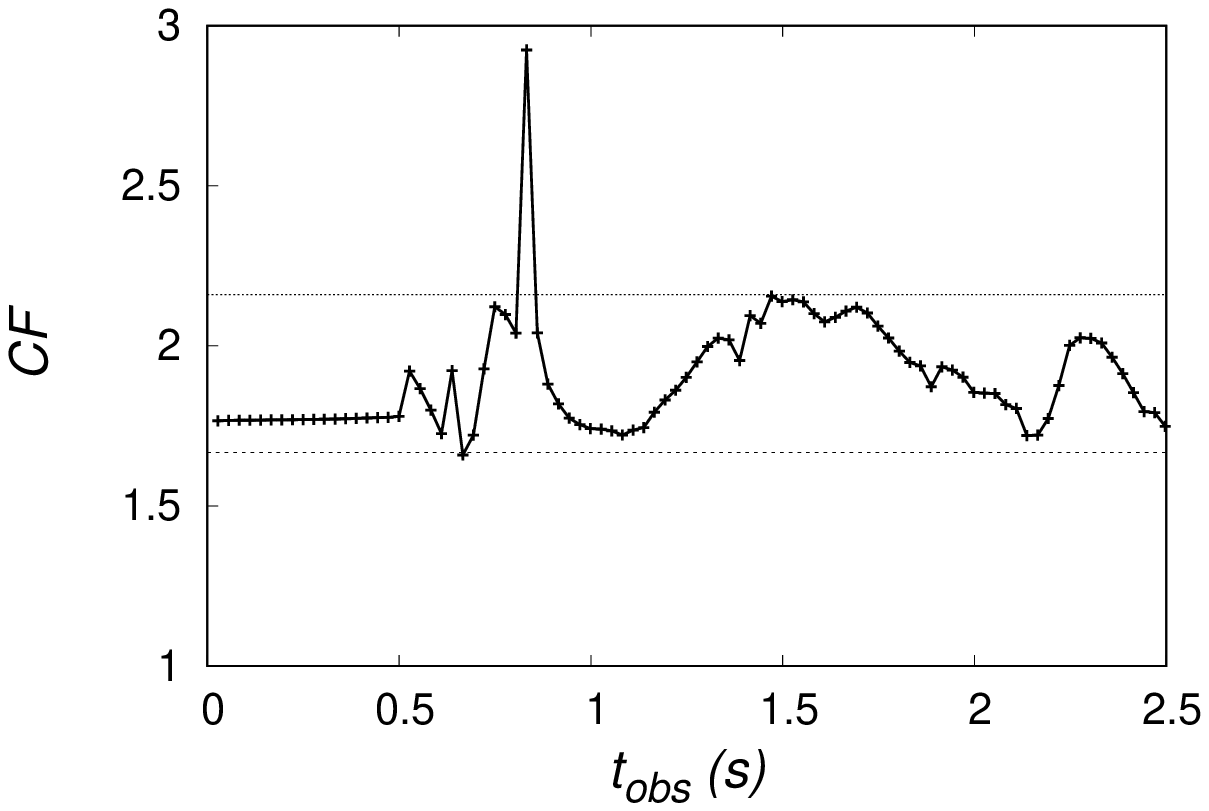}
\caption{Order of self-convergence of the LC measured with a perpendicular detector for model $3$. First order of accuracy is expected for systems with shocks during the evolution, and our sample simulations show a convergence slightly better than first order. The two constant lines indicate the value of $CF$ for first- and second-order convergence for our resolutions, respectively.} \label{fig:fc}
\end{figure}

Fig. \ref{fig:Jets} shows the morphology of the rest mass density for  model $3$ at $t\sim 9.37 \text{s}$. With the three resolutions the morphology is consistent in the sense that the jet head has reached the same position in the z-axis in all the cases. Also, the transverse expansion of the jet is  consistent. However, as expected, higher resolution reveals smaller structures and the exact morphology of the turbulent internal part of the jet is not exactly the same for the different resolutions. The exact details of that region are not expected  to contribute to the thermal emission, which is dominated by the jet/external interaction, nevertheless they have an effect on the LCs which are different for different resolutions,  however their time series should converge. This is the reason why we practice a self-convergence test on the final result of the simulations, namely the LC. Using the three resolutions mentioned $\Delta x_{3} < \Delta x_{2} < \Delta x_{1}$ we calculate the respective LCs $L_1,~L_2,~L_3$ and perform a self convergence test by comparing the differences among them. The convergence factor is given by

\begin{equation}
	CF(L_1,L_2,L_3)=\frac{L_1 - L_2}{L_2 - L_3} \simeq \frac{(\Delta x_{1})^Q - (\Delta x_{2})^Q}{(\Delta x_{2})^Q - (\Delta x_{3})^Q}.
	\label{eq:conver}
\end{equation}

\noindent where $Q$ is the accuracy order of the methods. In our case, the numerical methods used, that is, the piecewise linear reconstructor of variables, the HLLE flux formula and the IMEX integrator, all combined in our simulations, in the presence of shocks are expected to converge within first and second order. The convergence factor for for $Q=1$ is $CF_{Q=1}=5/3$, whereas for $Q=2$ is $CF_{Q=2}=175/81\simeq 2.16$. We show in Fig. \ref{fig:fc} that the Convergence Factor has a value between these two, which shows our results self-converge with the expected accuracy and thus also shows that our simulations use a resolution within a convergence regime,  except for the spikes that usually appear when the numerator or denominator in  (\ref{eq:conver}) decrease with respect to the other.

\end{appendix}

% -------------------------------------------------------
% -----     REFERENCES     ----------
% -------------------------------------------------------

\end{document}